\documentclass{article}
\usepackage{graphicx}  
\usepackage{amsmath}   
\usepackage[compress]{cite}
\usepackage{amssymb}   
\usepackage{bm} 
\usepackage{dcolumn}
\usepackage{color}
\usepackage{mathrsfs}
\usepackage{amsfonts}
\usepackage{varioref}
\usepackage{textcomp}
\usepackage{subcaption}
\usepackage{braket}

\usepackage{cancel}
\RequirePackage[colorlinks,citecolor=blue,urlcolor=magenta,linkcolor=blue]{hyperref}
\allowdisplaybreaks
\def\RN{Reissner-Nordstr\"{o}m }

\def\be{\begin{equation}}
	\def\ee{\end{equation}}
\def\ba{\begin{eqnarray}}
	\def\ea{\end{eqnarray}}

\newcommand{\beq}{\begin{equation}}
\newcommand{\eeq}{\end{equation}}
\newcommand{\bea}{\begin{eqnarray}}
\newcommand{\eea}{\end{eqnarray}}
\newcommand{\bit}{\begin{itemize}}
\newcommand{\eit}{\end{itemize}}
\newcommand{\ben}{\begin{enumerate}}
\newcommand{\een}{\end{enumerate}}

\newcounter{mnotecount}[section]

\renewcommand{\themnotecount}{\thesection.\arabic{mnotecount}}

\newcommand{\mnotex}[1]
{\protect{\stepcounter{mnotecount}}$^{\mbox{\footnotesize
$
\bullet$\themnotecount}}$ \marginpar{
\raggedright\tiny\em
$\!\!\!\!\!\!\,\bullet$\themnotecount:  #1} }
\numberwithin{equation}{section}

\labelformat{section}{Section #1} 
\labelformat{subsection}{Section #1} 
\labelformat{subsubsection}{Section #1}
\labelformat{subsubsubsection}{Section #1}
\labelformat{equation}{Eq.~(#1)} 
\labelformat{figure}{Fig.~#1} 
\labelformat{subfigure}{Fig.~\thefigure#1} 
\labelformat{table}{Tab.~#1} 
\labelformat{appendix}{Appendix #1}

\title{\bf Perturbing the perturbed: Stability of quasi-normal modes in presence of a positive cosmological constant }

\author{Subhodeep Sarkar\footnote{subhodeep.sarkar1@gmail.com}$~^{1a,b}$, Mostafizur Rahman\footnote{mostafizur.r@iitgn.ac.in}$~^{2}$ and Sumanta Chakraborty\footnote{tpsc@iacs.res.in}$~^{3}$\\
\\
$^{1a}${\small{Indian Institute of Information Technology, Allahabad, Prayagraj, Uttar Pradesh - 211015, India}}\\
$^{1b}${\small{Centre for Theoretical Physics, Jamia Millia Islamia, New Delhi - 110025, India}}\\
$^{2}${\small{Indian Institute of Technology, Gandhinagar, Gujarat - 382355, India}}\\
$^{3}${\small{School of Physical Sciences, Indian Association for the Cultivation of Science, Kolkata - 700032, India}}}

\date{ }  
\begin{document}
  
\maketitle
\begin{abstract}

In this work, we wish to address the question --- whether the quasi-normal modes, the characteristic frequencies associated with perturbed black hole spacetimes, central to the stability of these black holes, are themselves stable. Though the differential operator governing the perturbation of black hole spacetimes is self-adjoint, the boundary conditions are dissipative in nature, so that the spectral theorem becomes inapplicable, and there is no guarantee regarding the stability of the quasi-normal modes. We have provided a general method of transforming to the hyperboloidal coordinate system, for both asymptotically flat and asymptotically de Sitter spacetimes which neatly captures the dissipative boundary conditions, and the differential operator becomes non-self-adjoint. Employing the pseudospectrum analysis and numerically implementing the same through Chebyshev's spectral method, we present how the quasi-normal modes will drift away from their unperturbed values under external perturbation of the scattering potential. Intriguingly, for strong enough perturbation, even the fundamental quasi-normal mode, associated with gravitational perturbations, drifts away from its unperturbed position for asymptotically de Sitter black holes, in stark contrast to the case of asymptotically flat black holes. Besides presenting several other interesting results, specifically for asymptotically de Sitter black holes, we also discuss the implications of the instability of the fundamental quasi-normal mode on the strong cosmic censorship conjecture.

\end{abstract}
\section{Introduction}

The historic detection of gravitational waves from the merger of binary black holes \cite{LIGOScientific:2016aoc, LIGOScientific:2017vwq, LIGOScientific:2018mvr}, as well as the observation of the black hole shadow \cite{EventHorizonTelescope:2019dse, EventHorizonTelescope:2021srq}, has made it possible to study the nature of gravity in the strong field regime. These observations have enabled us to probe the physics of gravitational interaction near the photon circular orbits, which are either confined to a spherical surface, known as the photon sphere (for static and spherically symmetric spacetime) or are confined within a certain region of spacetime, known as the photon region (for stationary and axisymmetric spacetime). By and large, these observations suggest that the spacetime geometry till the photon region is well described by black hole (BH) solutions in general relativity. Black holes are the simplest objects in our universe and those in general relativity are characterized by only three hairs --- mass, charge and angular momentum. Gravitational waves, as well as black hole shadow measurements, can not only extract information about these hairs but can also constrain additional hairs for black hole solutions in modified theories of gravity \cite{Cardoso:2016ryw,Banerjee:2022jog,Mishra:2021waw,EventHorizonTelescope:2022xqj,Banerjee:2019nnj,Chakravarti:2019aup,Visinelli:2017bny,Bambi:2019tjh,Gralla:2020pra}. The most important property of a black hole, as far as classical physics is concerned, is its stability under external perturbations. Unlike mechanical systems, the quasi-normal modes of black holes, associated with excitation under external perturbations, have both real and imaginary parts, hence the name quasi-normal modes (QNMs). The real part of the QNMs leads to an oscillatory behaviour, characteristic of the quasi-normal modes, whereas the imaginary part with an appropriate sign, leads to an exponentially damped part, and is the reason behind the stability of black holes \cite{Berti:2009kk,Berti:2015itd,Ghosh:2021mrv}. Astonishingly, the stability of black holes has a one-to-one correspondence with the existence of the photon region, where the effective potential experienced by the perturbing field becomes maximum (these effective potentials are often referred to as scattering potentials). Moreover, the instability time scale associated with the maxima in the effective potential is related to the time scale associated with the exponential decay of the quasi-normal modes. This is why probing the spacetime geometry till the photon region is sufficient to comment on the stability of the black holes. It should be emphasized that, by assuming the spacetime geometries to depict black holes, one tacitly assumes the non-existence of structure beyond the photon region, even though several possibilities regarding the existence of structure beneath the photon region exist and the study of the stability of these non-black hole compact objects is another story altogether (see \cite{Cardoso:2016oxy,Cardoso:2016rao,Abedi:2016hgu,Maggio:2020jml,Maggio:2021ans,Chakraborty:2022zlq,Franzin:2022iai,Dey:2020pth,Dey:2020lhq, Rahman:2021kwb} for further details). The gravitational wave signals from coalescing black hole binaries have neatly captured the fundamental and in some cases, one of the higher overtones, though with a small signal-to-noise ratio, exactly following an oscillatory behaviour modulated by an exponential damping, signalling the observational verification of the stability of black holes under perturbations \cite{Ghosh:2021mrv,Berti:2018vdi,Carullo:2019flw,LIGOScientific:2016lio}. With future generations of ground and space-based gravitational wave detectors, we will be able to detect more overtones with a higher signal-to-noise ratio and we shall acquire a better understanding of the spectroscopy of black hole QNMs \cite{London:2014cma,London:2018gaq,Carullo:2021dui}. 

Having established the fundamental role played by these QNMs in black hole physics, the fact that we hope to observe a large number of these modes in future gravitational wave detectors begs the following question to be asked and answered: how stable are these modes under a small perturbation of the scattering potential from which the modes are themselves generated? The importance of this question lies in the simple fact that no astrophysical object in our universe is isolated, and hence the scattering potential is bound to be perturbed by nearby gravitating objects. Thus, such small perturbations are always present and hence the stability of the QNMs is an important avenue to explore. Such explorations in asymptotically flat spacetimes started from the seminal works in \cite{Aguirregabiria1996ScatteringBB, Nollert:1996rf, Nollert:1998ys}: in \cite{Aguirregabiria1996ScatteringBB} it was shown that all of these modes, excluding the fundamental mode, are unstable\footnote{For a personal recollection of the discovery of quasi-normal modes and comments on this issue in particular, the readers are requested to refer to \cite{vishuIAGRG1996}.} but \cite{Nollert:1996rf, Nollert:1998ys} argued that the instability extended to the fundamental mode as well.
The detection of gravitational waves has sparked a renewed interest in this area, given the central role that quasinormal modes play in black hole spectroscopy \cite{Jaramillo:2020tuu, alsheikh:tel-04116011, Jaramillo:2021tmt, Gasperin:2021kfv, Jaramillo:2022kuv,Jaramillo:2022zvf, Destounis:2021lum, Boyanov:2022ark, Cheung:2021bol, Konoplya:2022pbc}.
In particular, motivated by issues in determining the onset of turbulence in hydrodynamics \cite{tREFETHEN1993/science}, novel methods of finding spectral instability were obtained \cite{Jaramillo:2020tuu,alsheikh:tel-04116011,Jaramillo:2021tmt, Gasperin:2021kfv, Jaramillo:2022kuv} and then applied, for the first time in gravitational systems, to the study of the stability of the QNMs of the P\"{o}schl-Teller potential and (asymptotically flat) Schwarzschild black holes \cite{Jaramillo:2020tuu,alsheikh:tel-04116011}. These set of pioneering works \cite{Jaramillo:2020tuu,alsheikh:tel-04116011,Jaramillo:2021tmt, Gasperin:2021kfv, Jaramillo:2022kuv} were rapidly followed up by exploring the stability of the QNM spectrum of Reissner-Nordstr\"{o}m black holes \cite{Destounis:2021lum} and non-black hole compact objects in asymptotically flat spacetimes \cite{Boyanov:2022ark}.
The key conclusion, in the context of black holes, being that the fundamental mode is stable, whereas the higher overtones are unstable under small perturbations of the scattering potential. No such systematic attempt at determining the stability of these modes for black holes in asymptotically de Sitter spacetime has been made (except for \cite{Yang:2022wlm} where the authors focused exclusively on near-extremal spacetimes) and will form the main focus of this work. The motivation for working with the asymptotically de Sitter spacetime is two-fold, in addition to the fact that our universe is asymptotically de Sitter. First of all, the inclusion of a cosmological constant introduces interesting limiting geometries, e.g., the Nariai solution, which is obtained by taking the limit in which the black hole event horizon almost coincides with the cosmological horizon, and it would interesting to see how these instabilities affect the QNMs of a Nariai spacetime. The inclusion of charge in turn leads to an even richer class of geometries, as it introduces the near-extermal solutions in the picture. 
In the presence of a cosmological constant, a new class of modes also emerges, e.g., besides the well-known photon sphere modes, there are purely imaginary de Sitter modes for Schwazrschild-de Sitter black holes \cite{Lopez-Ortega:2006aal, Cardoso:2017soq, Konoplya:2022xid}. Moreover, Reissner-Nordström-de Sitter black holes exhibit a new mode, known as the near-extremal mode, in addition to the two aforementioned modes, with the inclusion of the charge parameter \cite{Cardoso:2017soq}. It will be worthwhile to study the stability properties of each of these modes under small perturbations. 

Secondly, the presence of a positive cosmological constant greatly modifies the late-time behaviour of the QNMs, and results in a violation of the strong cosmic censorship conjecture, which has no resolution within the purview of classical physics \cite{Cardoso:2017soq, Hollands:2019whz}. It is certainly possible that the introduction of a small perturbation will force the modes to migrate to a different region in the complex frequency plane, such that the strong cosmic censorship conjecture will be restored in a purely classically setting, an exciting prospect that was briefly suggested in \cite{Jaramillo:2020tuu}. As we shall demonstrate, except for some specific situations, the above hope can be realized through the instability of the QNMs.  

The instability of the quasi-normal modes is not confined to black holes alone, rather it exists in a broad range of physical systems, starting from hydrodynamics to quantum mechanics \cite{Ashida:2020dkc}. Just as theoretical estimations regarding the onset of turbulent flow in a hydrodynamic system do not match with experimental outcomes, the eigenspectrum of non-hermitian operators in quantum mechanics does not provide a complete description of such a quantum system. In both cases, the failure of the eigenspectrum to describe the system completely arises from the non-self-adjoint nature of the relevant differential operator \cite{Trefethen1993HydrodynamicSW, DRISCOLL1996125, Krejcirik:2014kaa}. For black holes as well, the boundary conditions imposed on the perturbations are dissipative in nature (black holes absorb everything that falls into it) and hence require a non-self-adjoint operator to describe the eigenvalue problem associated with the quasi-normal modes. Thus the origin of the instability of the modes can be traced back to the existence of a non-self-adjoint operator governing the QNM spectrum. For self-adjoint operators, on the other hand, the spectral theorem asserts that for a small perturbation, the eigenvalues of the self-adjoint operator will also be perturbed, but the perturbed eigenvalues will be confined within a small region, whose size is comparable to the strength of the perturbation. But for non-self-adjoint operators, as is the case for black holes, the QNMs migrate to regions of the complex frequency plane, over a distance that is at least a few orders of magnitude larger than the scale of the perturbation \cite{Jaramillo:2020tuu, Trefethen2005_pseudospectrumbook, Destounis:2021lum}, signaling the existence of instability.
The migration of quasi-normal modes is extremely sensitive to the frequency of perturbations. Particularly, in the case of asymptotically flat black holes, such drifting is present for all higher overtones when subjected to high-frequency (``small-scale") perturbations, potentially rendering these modes unstable. Conversely, these overtones are stable against low-frequency (``large-scale") perturbations. Interestingly, the fundamental mode does not drift away and hence remains stable under external perturbation of the scattering potential, irrespective of the perturbation frequency.\cite{vishuIAGRG1996, Jaramillo:2020tuu}.

In this work, we wish to extend the above result for asymptotically de Sitter black holes and wish to explore if the fundamental mode still remains stable. In case, the fundamental mode demonstrates instability, there will be significant implications for the strong cosmic censorship conjecture, or, in other words, regularity of the perturbation at the Cauchy horizon. If the fundamental QNM drifts in the complex frequency plane towards a smaller imaginary value, the perturbations will no longer be regular at the Cauchy horizon, thus respecting the strong cosmic censorship conjecture. Through this analysis, we also hope to point out other non-trivial features associated with the instability of the QNMs for asymptotically de Sitter black holes.       

The paper is organized as follows: We begin by constructing a hyperboloidal coordinate system for asymptotically de Sitter spacetimes in \ref{sec:hyperboloidal_approach} that is capable of handling the boundary conditions of the scattering problems in a purely geometric manner. Along the way, we also review the situation for asymptotically flat black holes. In the following section, that is, \ref{sec:wave_eqn_details}, we write down the wave equation governing the behavior of the perturbing field in the hyperboloidal coordinate system.  We then proceed to motivate and define a novel probe recently introduced to examine the spectral stability of gravitational systems, the pseudospectrum, and the associated energy norm in \ref{sec:mathematical_def_pseudospectrum_norm}. In \ref{sec:numerical_method}, we have laid down the numerical procedure used in this work, that is, Chebyshev's spectral method. We then embark on a thorough investigation of the spectral stability of asymptotically de Sitter black holes in \ref{sec:spectral_instability_ds_BHs}. We finally end at \ref{sec:disscusion_conclusion} with a brief discussion of our results and concluding remarks.

\emph{Notations and conventions:} We set the fundamental constants $G$ and $c$ to unity. Throughout this paper, we will use mostly positive signature convention, such that the Minkowski spacetime will have the metric $\textrm{diag}(-1,1,1,1)$. 

\section{Hyperboloidal coordinate system for asymptotically de Sitter spacetimes}\label{sec:hyperboloidal_approach}

In this section, we will extensively discuss and delineate a novel and useful coordinate system, known as the hyperboloidal coordinate system \cite{Zenginoglu:2007jw}, through which the boundary conditions, associated with the determination of the QNMs of black holes, get automatically incorporated into the differential equations governing the perturbations \cite{Zenginoglu:2011jz, PanossoMacedo:2018gvw}. The idea of using a hyperboloidal foliation to study QNMs was first suggested by Schmidt \cite{Schmidtgravityessay1993}. While the explicit relationship between the hyperboloidal coordinate transformation and the Regge-Wheeler-Zerilli wavefunctions governing the probe field (and hence its utility in computing the QNMs) was expounded by Zenginoglu in \cite{Zenginoglu:2011jz}. Furthermore, a hyperboloidal foliation constructed using a particular gauge choice (called the \emph{minimal gauge} \cite{Schinkel:2013tka, PanossoMacedo:2019npm}) that is extremely well adapted for the numerical computation of the QNMs of asymptotically flat black holes was introduced and put to use in a set of works by Macedo and his collaborators \cite{Schinkel:2013tka, Schinkel:2013zm, Ansorg:2016ztf, PanossoMacedo:2018hab, PanossoMacedo:2018gvw, PanossoMacedo:2019npm}. Recently, the aforementioned ``numerically efficient" hyperboloidal approach has been used to compute the QNMs of a Kerr black hole \cite{Ripley:2022ypi} and the techniques used therein have also been extended to study the stability of ultra-compact horizonless spacetimes \cite{Zhong:2022jke}. In the context of asymptotically de Sitter spacetimes, certain hyperboloidal coordinate systems have been used, but in a very different setting, e.g., in the study of black hole thermodynamics \cite{Gregory:2017sor, Gregory:2018ghc, Anderson:2020dim, Anderson:2022zuc}; however, these coordinate systems were not constructed keeping in mind the question of investigating spectral stability and calculating QNMs. In this regard, this work presents the first attempt to arrive at the hyperboloidal coordinate system in the minimal gauge, for asymptotically de Sitter spacetimes that is suitable for black hole perturbation theory and gives a precise geometrical context behind its utility in studying QNMs and their stability\footnote{Coordinate systems conceptually related to one we will derive here, have also been used to study the QNMs of pure two-dimensional de Sitter space \cite{Bizon:2020qnd} and \emph{small} Schwarzschild de Sitter black holes \cite{Hintz:2021vfl}.}. It is also worth highlighting the application of other hyperboloidal coordinate systems as well: this includes investigating the dynamics of a Yang-Mills field in various geometries \cite{Bizon:2010mp, Bizon:2014nla}, as well as the study of QNMs in asymptotically anti-de Sitter (AdS) spacetimes, where one can construct a ``regular" coordinate system that shares the usefulness of the hyperboloidal coordinate systems. Even though these regular coordinates are geometrically quite different, such coordinates, known as the Kerr-Star coordinates \cite{Dafermos:2008en}, have been elucidated in \cite{Holzegel:2011uu, Warnick:2013hba} for AdS BHs and, notably, they have been used to address the issue of strong cosmic censorship in Kerr-AdS black holes \cite{Kehle:2020zfg}.

The hyperboloidal foliation of the spacetime is necessary since the differential equation for the perturbations in the standard coordinate system is self-adjoint, it is the dissipative boundary condition at the horizon, which makes the problem non-self-adjoint. Thus to present this issue more explicitly, we need to incorporate the boundary conditions within the differential operator itself, describing the perturbations, such that the non-self-adjoint nature of the operator becomes evident. This is precisely what the hyperboloidal coordinate system achieves.

Therefore, the hyperboloidal coordinate system is a natural one to address the evolution of the perturbations associated with the black hole spacetimes, such that the boundary conditions get automatically incorporated. For asymptotically flat spacetimes, the boundary conditions are imposed on the perturbations at the horizon and at future null infinity. These two null surfaces are connected by a hyperboloid, which is related to the coordinate time by the so-called `height function'. On the other hand, for asymptotically de Sitter spacetimes, the boundary conditions are imposed, on the perturbations, at the black hole horizon, and at the cosmological horizon. Both of these boundary conditions are imposed on null surfaces and hence the hyperboloidal coordinate system is natural for asymptotically de Sitter spacetimes as well. In what follows, we will first demonstrate how such a hyperboloidal coordinate system can be defined, for a generic static and spherically symmetric spacetime. We will subsequently demonstrate how the above analysis will lead to hyperboloidal coordinate system in the context of asymptotically flat spacetimes, before extending it to the case of asymptotically de Sitter spacetimes. We start by writing down the line element for a generic static and spherically symmetric black hole spacetime as,
\begin{equation}
d s^{2} = -f(r) dt^ {2}+  \dfrac{dr^{2}}{f(r)} +r^{2} d\Omega^{2} ,
\end{equation}
where, $d\Omega^{2}\equiv d\theta^{2}+\sin^{2}\theta d\phi^{2}$, is the line element on a unit two-sphere. It is instructive to introduce a dimensionless, as well as compactified coordinate $\sigma$, in place of the radial coordinate $r$, through the following coordinate transformation, 
\begin{align}\label{rsigma}
\dfrac{r}{\lambda}\equiv \dfrac{\rho(\sigma)}{\sigma}~, 
\end{align}
where $\lambda$ is a characteristic length scale associated with the problem of interest. Besides redefining the radial coordinate, we also define the dimensionless null coordinates $v$ and $u$ as,
\begin{align}
v=\dfrac{t+r_{*}}{\lambda}~;
\qquad
u=\dfrac{t-r_{*}}{\lambda}=v-\dfrac{2r_{*}}{\lambda}~;
\qquad
dr_{*}=\dfrac{dr}{f(r)}~,
\end{align}
where $r_{*}$ is the tortoise coordinate, related to the radial coordinate, as in the above expression. In terms of the new coordinates $(v,\sigma,\theta,\phi)$, the above static and spherically symmetric line element becomes,
\begin{align}
ds^{2}=-\lambda^{2}fdv^{2}-2\lambda^{2}\left(\dfrac{\beta(\sigma)}{\sigma^{2}}\right) dvd\sigma+\lambda^{2}\dfrac{\rho^{2}}{\sigma^{2}}d\Omega^{2}~.
\end{align}
Here, we have defined a new function $\beta$ of the compactified radial coordinate $\sigma$ as, 
\begin{align}\label{def_beta}
\beta\equiv \rho-\sigma \rho'~,
\end{align}
where `prime' denotes the derivative with respect to the compactified radial coordinate $\sigma$. Hence we can read off the following non-zero components of the metric and of the inverse metric as follows,
\begin{align}
g_{vv}&=-\lambda^{2}f~;
\qquad
g_{v\sigma}=-\dfrac{\lambda^{2}}{\sigma^{2}}\beta~;
\qquad 
g_{\theta\theta}=\dfrac{g_{\phi \phi}}{\sin^{2}\theta}=\lambda^{2}\dfrac{\rho^{2}}{\sigma^{2}}~,
\\
g^{v\sigma}&=-\dfrac{\sigma^{2}}{\lambda^{2}\beta}~;
\qquad
g^{\sigma \sigma}=\dfrac{\sigma^{4}f}{\beta^{2}\lambda^{2}}~;
\qquad
g^{\theta\theta}=\sin^{2}\theta g^{\phi \phi}=\dfrac{\sigma^{2}}{\lambda^{2}\rho^{2}}~.
\end{align}
These metric elements will be used to raise and lower the indices of any tensorial quantity living in this spacetime. Since the hyperboloidal coordinate system connects two null surfaces, it is natural to connect the null vectors associated with these null surfaces to the hyperboloidal coordinate system. Further, the only null vectors that can be associated with any null surface are the null normals associated with them. In the present context, these two relevant null vectors are given by, 
\begin{align}
\ell_{\alpha}=-A\nabla_{\alpha}u~;
\qquad
k_{\alpha}=-B\nabla_{\alpha}v~,
\end{align}
where $A$ and $B$ are arbitrary normalization factors and these two null vectors must satisfy the following condition, $k_{\alpha}\ell^{\alpha}=-1$. In the $(v,\sigma,\theta,\phi)$ coordinate system, we obtain the following components for the vector $k_{\alpha}$,
\begin{align}
k_{\alpha}=\left(-B,0,0,0\right)~;
\qquad
k^{\alpha}=\left(0,\dfrac{\sigma^{2}}{\lambda^{2}\beta}B,0,0\right)~,
\end{align}
and similarly, the components of the other null vector $\ell_{\alpha}$ read,
\begin{align}
\ell_{\alpha}=\left(-A,-A\dfrac{2\beta}{\sigma^{2}f},0,0\right)~;
\qquad
\ell^{\alpha}=\left(\dfrac{2A}{\lambda^{2}f},-A\dfrac{\sigma^{2}}{\lambda^{2}\beta},0,0\right)~.
\end{align}
Therefore, we obtain the following inner product between these two null vectors: $k_{\alpha}\ell^{\alpha}=-(2AB/\lambda^{2}f)$, which should be set to the value $-1$. Therefore, we immediately obtain,
\begin{align}
AB=\dfrac{\lambda^{2}f}{2}~.
\end{align}
Therefore, the two normalization factors are not independent, rather one can be expressed in terms of the other. Hence the two null vectors take the following form,
\begin{align}
k_{\alpha}&=\zeta\left(-1,0,0,0\right)~;
\qquad
k^{\alpha}=\zeta\dfrac{\sigma^{2}}{\lambda^{2}\beta}\delta^{\mu}_{\sigma}~,
\\
\ell_{\alpha}&=\left(-\dfrac{\lambda^{2}f}{2\zeta},-\dfrac{\lambda^{2}\beta}{\zeta\sigma^{2}},0,0\right)~;
\qquad
\ell^{\alpha}=\dfrac{1}{\zeta}\left(\delta^{\mu}_{v}-\dfrac{\sigma^{2}f}{2\beta}\delta^{\mu}_{\sigma}\right)~,
\end{align}
Let us now make another coordinate transformation, by defining a new time coordinate $\tau$, through the following relation,
\begin{align}
\tau\equiv v+h_{0}(\sigma)~,
\end{align}
where, as of now $h_{0}(\sigma)$ is an arbitrary function of the compactified radial coordinate $\sigma$. In this new coordinate system $(\tau,\sigma,\theta,\phi)$, the line element becomes,
\begin{align}
ds^{2}=-\lambda^{2}fd\tau^{2}+2\lambda^{2}\left(fh_{0}'-\dfrac{\beta}{\sigma^{2}}\right)d\tau d\sigma
+\lambda^{2}\left(2h_{0}'\dfrac{\beta}{\sigma^{2}}-fh_{0}'^{2}\right) d\sigma^{2}+\lambda^{2}\dfrac{\rho^{2}}{\sigma^{2}}d\Omega^{2}
\end{align}
Under this coordinate transformation, to the new coordinate system, the components of the null vectors become,
\begin{align}
\bar{k}^{\alpha}=\left(\dfrac{\sigma^{2}h_{0}'}{\lambda^{2}\beta}\zeta,\dfrac{\sigma^{2}}{\lambda^{2}\beta}\zeta,0,0\right)~;
\qquad
\bar{\ell}^{\alpha}=\left(\dfrac{1}{\zeta}-\dfrac{\sigma^{2}fh_{0}'}{2\zeta\beta},-\dfrac{\sigma^{2}f}{2\zeta\beta},0,0\right)~,
\end{align}
where, $\zeta$ is an arbitrary normalization constant, which we fix by imposing the following condition on the null vector field $k^{\alpha}$: $k^{\alpha}\partial_{\alpha}\tau=1$, and that yields,
\begin{align}
\zeta=\dfrac{\lambda^{2}\beta}{\sigma^{2}h_{0}'}~.
\end{align}
Thus, for the above choice of the normalization parameter $\zeta$, the null vectors become,
\begin{align}\label{nullvectors}
\bar{k}^{\alpha}=\left(1,\dfrac{1}{h_{0}'},0,0\right)=\delta^{\alpha}_{\tau}+\dfrac{1}{h_{0}'}\delta^{\alpha}_{\sigma}~;
\qquad
\bar{\ell}^{\alpha}=\dfrac{\sigma^{2}h_{0}'}{\lambda^{2}\beta}\left(1-\dfrac{\sigma^{2}fh_{0}'}{2\beta}\right)\delta^{\mu}_{\tau}
-\dfrac{\sigma^{4}h_{0}'f}{2\lambda^{2}\beta^{2}}\delta^{\mu}_{\sigma}~.
\end{align}
As of now, we have two unknown functions $\rho(\sigma)$ and $h_{0}(\sigma)$. As we will demonstrate below, the function $h_{0}(\sigma)$ can be determined given $\rho(\sigma)$, since will satisfy a differential equation, depending on the choice of the unknown function $\rho(\sigma)$. For this purpose, we consider $\sigma=\textrm{constant}$ surface, then the norm of the normal to this surface becomes, $g^{\sigma \sigma}\propto \sigma^{4}$. Thus, $\sigma=0$ is a null surface. If we now try to construct a Gaussian null coordinate system around this $\sigma=0$ null surface, then it follows that $\ell^{\tau}\propto (\sigma^{2}/\lambda^{2})$ \cite{Parattu:2015gga,Parattu:2016trq,Chakraborty:2016yna}. Thus from the expression of the null vector $\bar{\ell}^{\alpha}$ in \ref{nullvectors}, it is evident that $h_{0}'$ must satisfy the following algebraic equation,
\begin{align}
\dfrac{h_{0}'}{\lambda^{2}\beta}\left(1-\dfrac{\sigma^{2}fh_{0}'}{2\beta}\right)=C(\sigma)~,
\end{align}
where, $C(\sigma)$ is a function of the coordinate $\sigma$, which becomes constant in the $\sigma \rightarrow 0$ limit. The above algebraic equation can then be expressed as a quadratic equation in $h_{0}'$, which reads,
\begin{align}
\left(\dfrac{\sigma^{2}f}{2\lambda^{2}\beta^{2}}\right)h_{0}'^{2}-\dfrac{h_{0}'}{\lambda^{2}\beta}+C=0~,
\end{align}
with the following solution for $h_{0}'$,
\begin{align}
h_{0}'&=\dfrac{\beta}{\sigma^{2}f}\left[1\pm \sqrt{1-2C\sigma^{2}f\lambda^{2}}\right]~.
\end{align}
Thus, near $\sigma=0$, the function $h_{0}(\sigma)$ satisfies the following first order differential equation,
\begin{align}\label{h0diff}
h_{0}'(\sigma)=\dfrac{2\beta(\sigma)}{\sigma^{2}f(\sigma)}-\beta(\sigma) C(\sigma)\lambda^{2}+\mathcal{O}(\sigma^{2})~.
\end{align}
To solve the above differential equation we need to know the functional form for $\beta(\sigma)$, as well as of $f(\sigma)$, which we determine subsequently, first for the asymptotically flat spacetimes and then for the asymptotically de Sitter spacetimes. 

At this outset, let us introduce a modified height function $h(\sigma)$ by the following coordinate transformation between the old time coordinate $t$ and the new time coordinate $\tau$, and also we introduce a new function $g(\sigma)$, through the tortoise coordinate $r_{*}$ as,
\begin{align}
\tau=\dfrac{t}{\lambda}+h(\sigma)~;
\qquad
\dfrac{r_{*}}{\lambda}=g(\sigma)~.
\end{align}
Such that, the relation between the functions $h_{0}$ and $h$ is given by, 
\begin{align}
h=h_{0}+g~.
\end{align}
Thus given any background spacetime, one can determine the function $f(r)$, from which it is possible to compute the tortoise coordinate $r_{*}$ and hence using \ref{rsigma}, we can determine the function $g(\sigma)$. The function $h_{0}$ can be determined by imposing a regularity condition for the null vectors, which boils down to the differential equation presented in \ref{h0diff}. This in turn requires knowledge about the function $\rho(\sigma)$. We elaborate below on the steps of deriving the function $\rho(\sigma)$ and hence the determination of both $g(\sigma)$ and $h_{0}(\sigma)$, leading to the estimation of the height function $h(\sigma)$.

\subsection{Warm up: Hyperboloidal coordinate for asymptotically flat spacetime}

Let us briefly describe and review the case of asymptotically flat spacetime \cite{PanossoMacedo:2018hab}, which will guide us in generalizing the same to the case of asymptotically de Sitter spacetime. For asymptotically flat spacetime, let us identify the asymptotic point $r=\infty$ with $\sigma=0$ and the event horizon $r=r_{+}$ with $\sigma=1$, such that the region $r\in[r_{+},\infty]$ translates into the compactified region $\sigma\in[1,0]$. Thus we may expand $\beta(\sigma)$ about the asymptotic infinity as,
\begin{align}\label{betasigma}
\beta(\sigma)=\beta_{0}+\beta_{1}\sigma+\beta_{2}\sigma^{2}+\mathcal{O}(\sigma^{3})~,
\end{align}
where, $\beta_{0}$, $\beta_{1}$ and $\beta_{2}$ are constants, to be determined later. Given the above expansion for $\beta(\sigma)$, it is evident from \ref{def_beta} that the function $\rho(\sigma)$ appearing in the definition of the radial function, satisfies the following differential equation,
\begin{align}
\dfrac{d}{d\sigma}\left(\dfrac{\rho}{\sigma}\right)=-\dfrac{\beta_{0}}{\sigma^{2}}-\dfrac{\beta_{1}}{\sigma}-\beta_{2}~.
\end{align}
Integration of the above differential equation yields the following expression for the function $\rho(\sigma)$,
\begin{align}
\rho(\sigma)=\beta_{0}+\rho_{1}\sigma-\beta_{1}\sigma \ln \sigma -\beta_{2}\sigma^{2}~.
\end{align}
Note that, $\rho(\sigma)$ involves a term $\mathcal{O}(\sigma \ln \sigma)$ and hence for $\rho(\sigma)$ to be well-behaved at $\sigma=0$, we must impose the condition $\beta_{1}=0$. Thus, to the leading order, we obtain, 
\begin{align}
\beta(\sigma)=\beta_{0}+\mathcal{O}(\sigma^{2})\equiv \rho_{0}+\mathcal{O}(\sigma^{2})~;
\qquad
\rho(\sigma)=\rho_{0}+\rho_{1}\sigma+\mathcal{O}(\sigma^{2})~,
\end{align}
such that, from \ref{rsigma} it follows that the coordinate transformation between the radial coordinate $r$ and the compactified radial coordinate $\sigma$ becomes,
\begin{align}\label{rsigmaflat}
\dfrac{r}{\lambda}=\dfrac{\rho_{0}+\rho_{1}\sigma}{\sigma}~.
\end{align}
It is to be emphasized that the above choice of the functions $\beta(\sigma)$ and $\rho(\sigma)$ are for asymptotically flat spacetimes, while for asymptotically de Sitter spacetimes the above relation will be different. The fact that $r=\infty$ maps to $\sigma=0$ is evident from \ref{rsigmaflat}, while from \ref{rsigmaflat}, the mapping of the outer horizon $r=r_{+}$, to $\sigma=1$, yields, $r_{+}=\lambda(\rho_{0}+\rho_{1})$. This fixes the constant $\rho_{0}$ to the value: $\rho_{0}=(r_{+}/\lambda)-\rho_{1}$. Thus, finally the relation between the radial coordinate $r$ with the dimensionless coordinate $\sigma$ becomes,
\begin{align}\label{rsigmaflatf}
\dfrac{r}{\lambda}=\dfrac{\sigma_{+}+\rho_{1}\left(\sigma-1\right)}{\sigma}~,
\end{align}
where, for notational convenience, we have defined, $(r_{+}/\lambda)=\sigma_{+}$, which becomes unity for the choice: $\lambda=r_{+}$. As evident, the constant $\rho_{1}$ remains undetermined, fixing of which corresponds to different gauge choices, as discussed in \cite{PanossoMacedo:2018hab}. In what follows, we will keep $\lambda$ arbitrary. 

\subsubsection{Hyperbolic coordinate for Schwarzschild black hole}

\begin{figure}[htb!]
 \centering
  \begin{subfigure}[t]{0.30\textwidth}
    \centering
  \includegraphics[width=\textwidth]
    {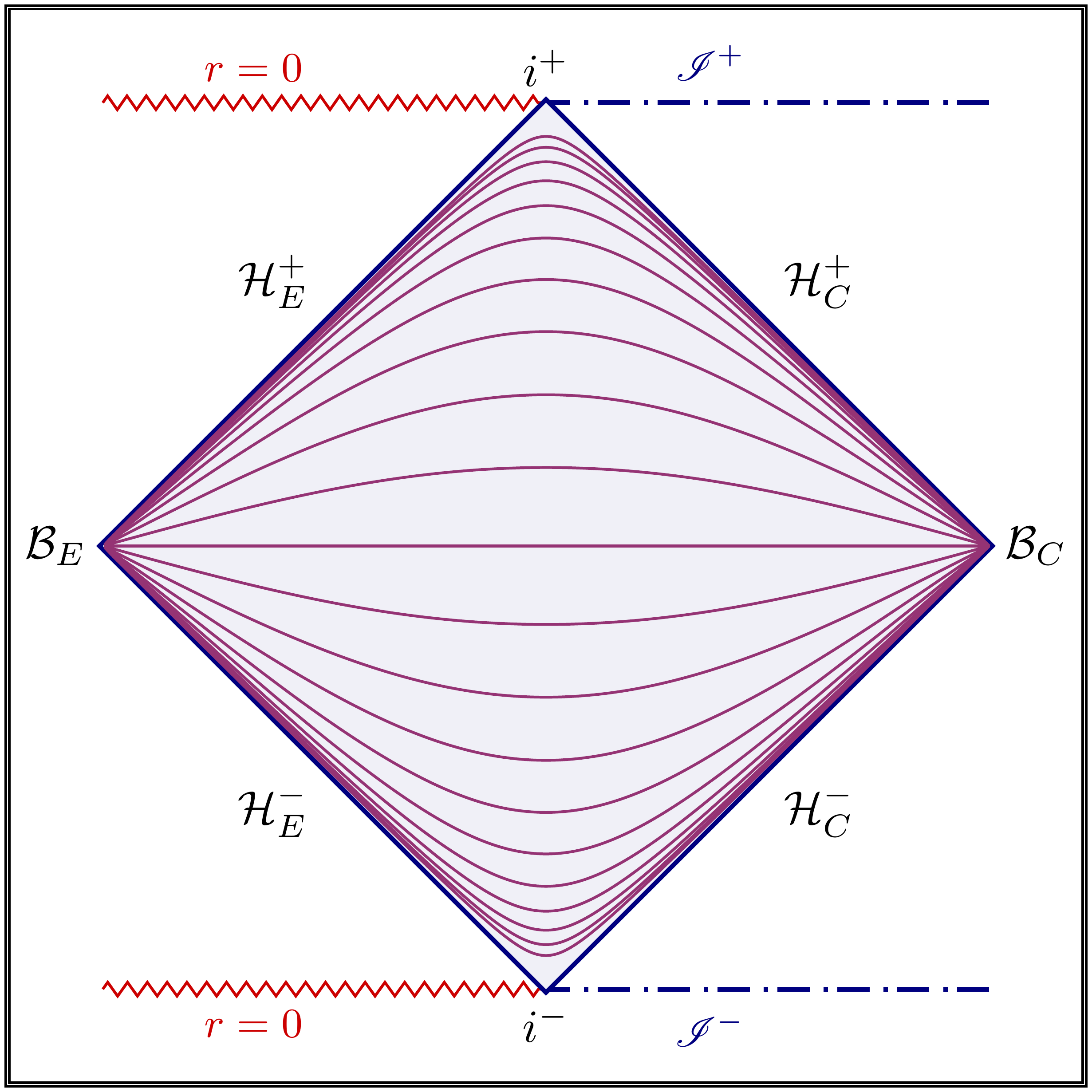}
    \caption{$t =$ constant}
    \label{fig:penrose_const_t}
  \end{subfigure}\hspace{1em}
  \begin{subfigure}[t]{0.30\textwidth}
  \centering
    \includegraphics[width=\textwidth]{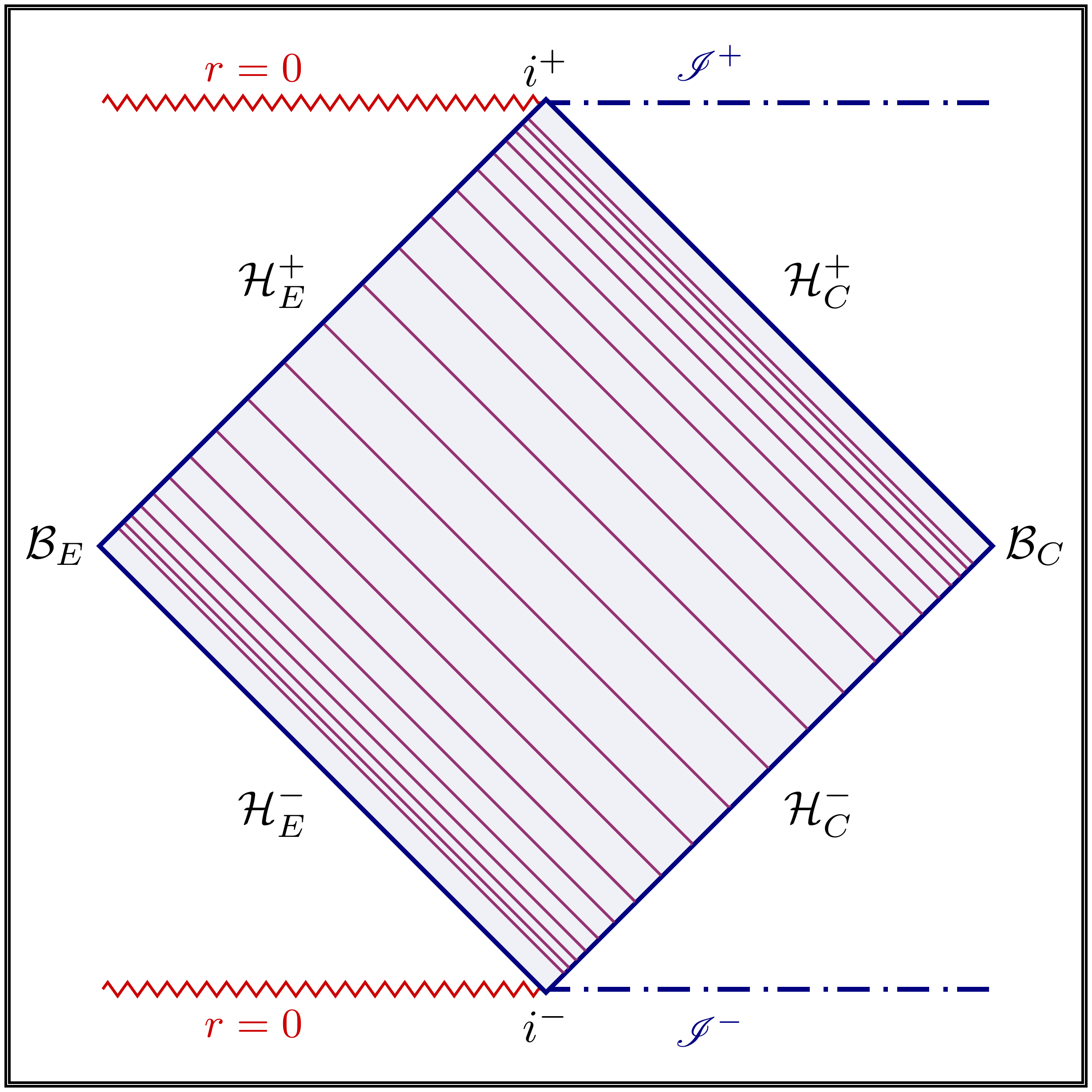}
    \caption{$v =$ constant}
    \label{fig:penrose_const_v}
  \end{subfigure}\hspace{1em}
 \begin{subfigure}[t]{0.30\textwidth}
  \centering
   \includegraphics[width=\textwidth]{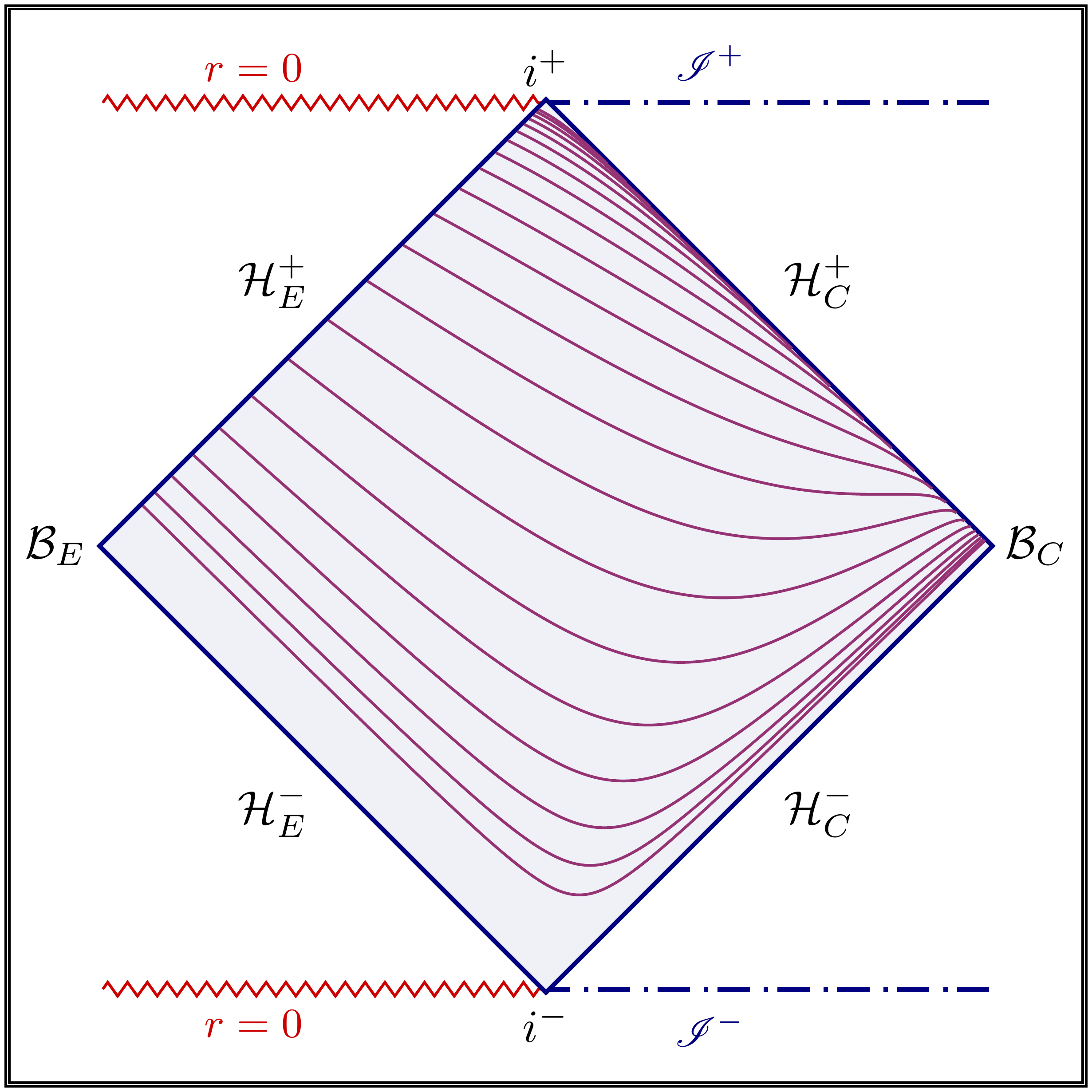}
    \caption{$\tau =$ constant}
    \label{fig:penrose_const_tau}
  \end{subfigure}
  \caption{Conformal diagrams showing constant $t$ (right panel), constant $v$ (middle panel), constant $\tau$ foliations the Schwarzschild de Sitter spacetime with $M = 0.5$ and $\Lambda = 0.2$. In these figures, $\mathcal{H}_E^+$ and $\mathcal{H}_E^-$ are the future and past event horizons, $\mathcal{H}_C^+$ and $\mathcal{H}_C^-$ are the future and past cosmological horizons, $\mathcal{B}_E$ and $\mathcal{B}_C$ are the bifurcation spheres at the (intersections of the two) event horizons and cosmological horizons, $i^+$ and $i^-$ are the future and past timelike infinity, $\mathscr{I}^+$ and $\mathscr{I}^-$ are future and past null infinity, respectively, and $r=0$ represent the curvature singularities. The figure on the extreme right shows the hyperboloidal slicing of a Schwarzschild de Sitter black hole which is emerging as a natural arena for studying perturbations in various black hole spacetimes. The constant time slices were generated using the transformation equations involved in obtaining the conformal compactification of the spacetime metric under consideration\protect\footnotemark. The characteristic length scale $\lambda$ has been set to unity (see. \ref{rsigma}).} 
  \label{fig:pernose}
\end{figure}

As an illustration of the above analysis, let us explicitly determine the functions $h_{0}(\sigma)$ and $g(\sigma)$ for Schwarzschild black hole, which is asymptotically flat. The first step is to solve for the differential equation of $h_{0}(\sigma)$, which requires casting the metric function $f(r)$ in terms of the compactified radial coordinate $\sigma$. Using \ref{rsigmaflatf}, we obtain, 
\begin{align}
f(\sigma)&=1-\dfrac{2M}{\lambda}\dfrac{\sigma}{\rho(\sigma)}
=1-\dfrac{\sigma\sigma_{+}}{\rho_{0}+\rho_{1}\sigma}
\simeq 1-\dfrac{\sigma_{+}}{\rho_{0}}\sigma+\mathcal{O}(\sigma^{2})~,
\end{align}
where, $\sigma_{+}\equiv (2M/\lambda)$. In the subsequent calculations, we shall ignore all the $\mathcal{O}(\sigma^2)$ terms as in the above expression. This corresponds to making a gauge choice which has been called the minimal gauge in \cite{Schinkel:2013tka, PanossoMacedo:2019npm}. Geometrically speaking, there are several possible hyperboloidal slices (in fact, any arbitrary line) connecting the future event horizon $\mathcal{H}_E^+$ and the future cosmological horizon $\mathcal{H}_C^+$ (see \ref{fig:penrose_const_tau}{\footnote{The final figures were drawn using the TikZ package in LaTeX after the data was generated in Python and is adapted from a publicly available code snippet shared by An{\i}l Zengino\u{g}lu for asymptotically flat spacetimes \cite{Zenginoglu:2011jz}.}}), the minimal gauge is one such choice which has been demonstrated to be extremely well suited for numerical computations \cite{Schinkel:2013tka, Schinkel:2013zm, Ansorg:2016ztf, PanossoMacedo:2018hab, PanossoMacedo:2018gvw, PanossoMacedo:2019npm}. Given the above expression for the metric function in terms of the compactified radial coordinate $\sigma$, from \ref{h0diff}, we obtain the following differential equation satisfied by $h_{0}(\sigma)$,
\begin{align}
h_{0}'(\sigma)=\dfrac{2\rho_{0}}{\sigma^{2}}\left(1+\dfrac{\sigma_{+}}{\rho_{0}}\sigma\right)+\mathcal{O}(1)~,
\end{align}
which integrates to,
\begin{align}\label{h0gensol}
h_{0}(\sigma)=-\dfrac{2\rho_{0}}{\sigma}+2\sigma_{+}\ln\sigma~.
\end{align}
Thus, we have derived the modified height function $h_{0}(\sigma)$ and let us now determine the function $g(\sigma)$. For that purpose, we first write down the tortoise coordinate in terms of the radial coordinate $r$ and then convert the same to the dimensionless and compactified coordinate $\sigma$ as,
\begin{align}
\dfrac{r_{*}}{\lambda}&=\dfrac{r}{\lambda}+\dfrac{2M}{\lambda}\ln\Big|\dfrac{r}{2M}-1\Big|
\nonumber
\\
&=\dfrac{\rho_{0}+\rho_{1}\sigma}{\sigma}+\sigma_{+}\ln \Big|\dfrac{\rho_{0}+\rho_{1}\sigma}{\sigma \sigma_{+}}-1\Big|~.
\end{align}
Substituting, $\rho_{0}=\sigma_{+}-\rho_{1}$, we obtain the following functional dependence of the function $g(\sigma)$ as,
\begin{align}
g(\sigma)&=\rho_{1}+\dfrac{\sigma_{+}-\rho_{1}}{\sigma}+\sigma_{+}\ln \Big|\dfrac{\sigma_{+}-\rho_{1}+(\rho_{1}-\sigma_{+})\sigma}{\sigma \sigma_{+}}\Big|
\nonumber
\\
&=\rho_{1}+\dfrac{\sigma_{+}-\rho_{1}}{\sigma}+\sigma_{+}\ln (\sigma_{+}-\rho_{1})+\sigma_{+}\ln (1-\sigma)-\sigma_{+}\ln \left(\sigma \sigma_{+}\right)
\nonumber
\\
&\simeq \dfrac{\sigma_{+}-\rho_{1}}{\sigma}+\sigma_{+}\ln (1-\sigma)-\sigma_{+}\ln \sigma + \mathcal{O}(1)
\end{align}
Having derived both $h_{0}(\sigma)$ and $g(\sigma)$, the height function $h(\sigma)$ becomes, 
\begin{align}
h(\sigma)=g(\sigma)+h_{0}(\sigma)=\sigma_{+}\ln \sigma - \dfrac{\sigma_{+}-\rho_{1}}{\sigma} +\sigma_{+}\ln (1-\sigma)~,
\end{align}
where, we have ignored all the terms which are independent of the compactified radial coordinate $\sigma$. This is because, as we will see later, only derivative of these functions with respect to $\sigma$ will be of relevance for the purpose of this work. Using these two functions $h(\sigma)$ and $g(\sigma)$, provides the necessary transformation from the Schwarzschild coordinates to hyperboloidal coordinates. We will now demonstrate the same for the Reissner-Nordstr\"{o}m spacetime, another asymptotically flat spacetime. 

\subsubsection{Hyperbolic coordinate for Reissner-Nordstr\"{o}m black hole}

The second example of deriving the hyperboloidal coordinate for asymptotically flat spacetime corresponds to that of a charged black hole, also known as the Reissner-Nordstr\"{o}m black hole (for details, see \cite{PanossoMacedo:2018hab}). In this case, the metric function can be expressed in terms of the radial coordinate as,
\begin{align}\label{metric_RN}
f(r)=1-\dfrac{2M}{r}+\dfrac{Q^{2}}{r^{2}}=\dfrac{r^{2}-2Mr+Q^{2}}{r^{2}}=\dfrac{(r-r_{+})(r-r_{-})}{r^{2}}~,
\end{align}
where, $r_{\pm}=M\pm\sqrt{M^{2}-Q^{2}}$ are the event and the Cauchy horizons respectively. Transforming the radial coordinate to the dimensionless compactified coordinate $\sigma$, we obtain,
\begin{align}
f(\sigma)&=\left(1-\dfrac{r_{+}}{\lambda}\dfrac{\sigma}{\sigma_{+}-\rho_{1}+\rho_{1}\sigma} \right) 
\left(1-\dfrac{r_{-}}{\lambda}\dfrac{\sigma}{\sigma_{+}-\rho_{1}+\rho_{1}\sigma} \right) 
\nonumber
\\
&=\dfrac{\left\{\sigma_{+}-\rho_{1}+\left(\rho_{1}-\sigma_{+}\right)\sigma \right\}\left\{\sigma_{+}-\rho_{1}+\left(\rho_{1}-\dfrac{r_{-}}{\lambda}\right)\sigma \right\}}{\left(\sigma_{+}-\rho_{1}+\rho_{1}\sigma\right)^{2}}
\nonumber
\\
&\simeq 1-\left(\dfrac{r_{+}+r_{-}}{\lambda(\sigma_{+}-\rho_{1})}\right)\sigma+\mathcal{O}(\sigma^{2})~.
\end{align}
Note that the $\mathcal{O}(\sigma)$ term does not depend on the electric charge and the effect of the charge only arises in the $\mathcal{O}(\sigma^{2})$ term. Thus at the leading order, the solution for $h_{0}$ is given by \ref{h0gensol}, irrespective of the existence of electric charge in the spacetime. Expressed in detail, we obtain,
\begin{align}\label{h0RN}
h_{0}(\sigma)=-\dfrac{2(\sigma_{+}-\rho_{1})}{\sigma}+2\left(\sigma_{+}+\dfrac{r_{-}}{\lambda}\right) \ln\sigma~.
\end{align}
The corrections due to the presence of electric charge will keep pouring in as we write down the tortoise coordinate in terms of the compactified radial coordinate $\sigma$ and hence the function $g(\sigma)$ will be significantly modified. Given the metric function of the Reissner-Nordstr\"{o}m black hole in \ref{metric_RN}, the tortoise coordinate becomes,
\begin{align}
r_{*}=r-\dfrac{1}{2|\kappa_{-}|}\ln \left( \dfrac{r}{r_{-}}-1 \right) +\dfrac{1}{2\kappa_{+}}\ln \left(\dfrac{r}{r_{+}}-1\right)~,
\end{align}
where, $\kappa_{+}$ and $\kappa_{-}$ are the surface gravities at the event and the Cauchy horizon, respectively, with the following expressions,
\begin{align}
\kappa_{+}=\dfrac{1}{2}f'(r_{+})=\dfrac{r_{+}-r_{-}}{2r_{+}^{2}}~;
\qquad
\kappa_{-}=\dfrac{1}{2}f'(r_{-})=-\dfrac{r_{+}-r_{-}}{2r_{-}^{2}}~.
\end{align}
Note that $\kappa_{+}$ is positive, but $\kappa_{-}$ is a negative quantity. Transforming the tortoise coordinate, as presented above, to the compactified radial coordinate $\sigma$, we obtain the function $g(\sigma)$, central to the hyperboloidal coordinate as,
\begin{align}
g(\sigma)&=\dfrac{\sigma_{+}-\rho_{1}+\rho_{1}\sigma}{\sigma}-\dfrac{\sigma_{+}}{2r_{+}|\kappa_{-}|}\ln \left(\dfrac{\sigma_{+}-\rho_{1}+\rho_{1}\sigma}{\sigma \sigma_{+}}\dfrac{r_{+}}{r_{-}}-1\right)+\dfrac{\sigma_{+}}{2r_{+}\kappa_{+}}\ln \left(\dfrac{\sigma_{+}-\rho_{1}+\rho_{1}\sigma}{\sigma \sigma_{+}}-1\right)~.
\nonumber
\\
&=\rho_{1}+\dfrac{\sigma_{+}-\rho_{1}}{\sigma}+\dfrac{\sigma_{+}}{2r_{+}\kappa_{+}}\Big[\ln \left(\sigma_{+}-\rho_{1}\right)+\ln \left(1-\sigma\right) -\ln \sigma- \ln \sigma_{+}\Big]
\nonumber
\\
&\qquad -\dfrac{\sigma_{+}}{2r_{+}|\kappa_{-}|}\ln \left(\dfrac{\sigma_{+}-\rho_{1}+\rho_{1}\sigma}{\sigma\sigma_{+}}\dfrac{r_{+}}{r_{-}}-1\right)~.
\end{align}
In the above form, the limit of the function $g(\sigma)$ to the Schwarzschild case is not a straightforward one. For this purpose, we define, 
\begin{align}
q^{2}=\dfrac{r_{-}}{r_{+}}~.
\end{align}
In terms of this quantity $q^{2}$, we can express a few geometrical entities as: $r_{+}\kappa_{+}=\{(1-q^{2})/2\}$, and $r_{+}\kappa_{-}=-\{(1-q^{2})/2q^{4}\}$. Therefore, the function $g(\sigma)$ becomes,
\begin{align}
g(\sigma)&=\rho_{1}+\dfrac{\sigma_{+}-\rho_{1}}{\sigma}+\dfrac{\sigma_{+}}{1-q^{2}}\Big[\ln \left(\sigma_{+}-\rho_{1}\right)+\ln \left(1-\sigma\right) -\ln \sigma -\ln \sigma_{+}\Big]
-\dfrac{\sigma_{+} q^{4}}{1-q^{2}}\ln \left(\dfrac{\sigma_{+}-\rho_{1}+\rho_{1}\sigma}{q^{2}\sigma \sigma_{+}}-1\right)
\nonumber
\\
&=\rho_{1}+\dfrac{\sigma_{+}-\rho_{1}}{\sigma}+\dfrac{\sigma_{+}\ln \left(\sigma_{+}-\rho_{1}\right)}{1-q^{2}}+\dfrac{\sigma_{+}\ln \left(1-\sigma\right)}{1-q^{2}} 
-\left(1+q^{2}\right)\sigma_{+}\ln \left(\sigma \sigma_{+}\right)
\nonumber
\\
&\qquad -\dfrac{\sigma_{+}q^{4}}{1-q^{2}}\ln \left[\sigma_{+}-\rho_{1}+\left(\rho_{1}-\sigma_{+}q^{2}\right)\sigma\right]
+\dfrac{\sigma_{+}q^{4}}{1-q^{2}}\ln q^{2}
\nonumber
\\
&=\dfrac{\sigma_{+}-\rho_{1}}{\sigma}+\dfrac{\sigma_{+}\ln \left(1-\sigma\right)}{1-q^{2}} 
-\left(1+q^{2}\right)\sigma_{+}\ln \sigma-\dfrac{\sigma_{+}q^{4}}{1-q^{2}}\ln \left[\sigma_{+}-\rho_{1}+\left(\rho_{1}-\sigma_{+}q^{2}\right)\sigma\right]~.
\end{align}
Here also we have ignored terms which are independent of the compactified radial coordinate $\sigma$. Note that the parameter $\rho_{1}$, present in the above expression for $g(\sigma)$, needs to be fixed. There are two possible choices for this parameter, depending on the possible location of the Cauchy horizon. Since our main interest is in asymptotically de Sitter spacetime, we will not discuss the above gauge choices any further.  Finally, using $h_{0}(\sigma)$ from \ref{h0RN}, we obtain, the height function to yield, 
\begin{align}
h(\sigma)&=-\dfrac{\sigma_{+}-\rho_{1}}{\sigma}+\dfrac{\sigma_{+}\ln \left(1-\sigma\right)}{1-q^{2}} 
+\sigma_{+}\left(1+q^{2}\right)\ln \sigma-\dfrac{\sigma_{+}q^{4}}{1-q^{2}}\ln \left[\sigma_{+}-\rho_{1}+\left(\rho_{1}-\sigma_{+}q^{2}\right)\sigma\right]~.
\end{align}
Thus, we have derived the necessary functions to transform the Reissner-Nordstr\"{o}m coordinate system to the hyperboloidal coordinate system. Having demonstrated the procedure for asymptotically flat spacetime, we will now derive the respective functions for asymptotically de Sitter spacetimes. 
\subsection{Hyperbolic coordinate for asymptotically de Sitter spacetime}

In the case of asymptotically de Sitter spacetime, the region of interest is bounded by two finite values of the radial coordinate, the cosmological horizon $r_{\rm c}$ and the outer event horizon $r_{+}$. In this case if we want to convert the above range of radial coordinates to the following range of $\sigma$: $\sigma \in [0,1]$. Then we need to set $\beta_{0}=0$ in \ref{betasigma} (note that well-behaved $\rho(\sigma)$ has already demanded $\beta_{1}=0$ as well), and hence we will obtain, 
\begin{align}
\beta(\sigma)=\beta_{2}\sigma^{2}~;
\qquad
\rho(\sigma)=\rho_{1}\sigma-\beta_{2}\sigma^{2}~,
\end{align}
such that the transformation between the radial coordinate $r$, with the compactified and dimensionless radial coordinate $\sigma$ reads,
\begin{align}\label{rsigmadS}
\dfrac{r}{\lambda}=\dfrac{\rho_{1}\sigma-\beta_{2}\sigma^{2}}{\sigma}=\rho_{1}-\beta_{2}\sigma~.
\end{align}
As evident from \ref{rsigmadS}, if we want to map the cosmological horizon to $\sigma=0$, then we will obtain, $r_{c}=\lambda \rho_{1}$. Similarly, if we map the outer event horizon to $\sigma=1$, then we obtain another condition: $r_{+}=\lambda(\rho_{1}-\beta_{2})$. Defining, as in the case of asymptotically flat spacetime: $\sigma_{+}=(r_{+}/\lambda)$, we must have from the above conditions: $\rho_{1}=\sigma_{+}+\beta_{2}$ and $\rho_{1}=\sigma_{+}(r_{c}/r_{+})$. Note that, unlike the case of asymptotically flat spacetime, here both the parameters $\rho_{1}$ and $\beta_{2}$ get fixed. Thus, finally the relation between radial coordinate $r$ with the dimensionless coordinate $\sigma$ yields,
\begin{align}\label{rsigmadS_2}
\dfrac{r}{\lambda}=\sigma_{+}\sigma+\dfrac{r_{c}}{r_{+}}\sigma_{+}\left(1-\sigma\right)~.
\end{align}
We would like to emphasize that the above transformation is very different from that of asymptotically flat spacetimes. Further, given the above expression for $\beta(\sigma)$, with $\beta_{2}=\rho_{1}-\sigma_{+}$, from \ref{h0diff}, the differential equation satisfied by $h_{0}$ reads,
\begin{align}\label{h0diffdS}
h_{0}'(\sigma)=\dfrac{2\left(\rho_{1}-\sigma_{+}\right)}{f(\sigma)}+\mathcal{O}(\sigma^{2})~.
\end{align}
Thus, given the metric function $f(r)$, expressed in terms of the compactified radial coordinate $\sigma$, the above differential equation can be solved yielding the function $h_{0}(\sigma)$. This in turn will help in determining the height function $h(\sigma)$ through the tortoise coordinate. 

\subsubsection{Hyperbolic coordinate for Schwarzschild de Sitter black hole}


We start our examples of the asymptotically de Sitter spacetimes from the Schwarzschild de Sitter black hole, for which the metric function $f(r)$, in the standard radial coordinate reads, 
\begin{align}
f(r)=1-\dfrac{2M}{r}-\dfrac{\Lambda r^{2}}{3}~;
\qquad
\dfrac{\Lambda}{3}=\dfrac{1}{L_{\rm dS}^{2}}~.
\end{align}
Here, $L_{\rm dS}$ corresponds to the de Sitter length scale, which becomes larger, the smaller the cosmological constant is. The above metric function can further be expressed in the following manner,
\begin{align}
f(r)=-\dfrac{1}{rL_{\rm dS}^{2}}\left(r^{3}-L_{\rm dS}^{2}r+2ML_{\rm dS}^{2}\right)
=-\dfrac{1}{rL_{\rm dS}^{2}}\left(r-r_{+}\right)\left(r-r_{\rm c}\right)\left(r+r_{0}\right)~,
\end{align}
where, $r_{+}$ is the event horizon, $r_{\rm c}$ is the cosmological horizon and $r_{0}=r_{+}+r_{\rm c}$. We need to transform the above functional dependence of the metric function on the radial coordinate $r$, to the compactified radial coordinate $\sigma$, which yields,
\begin{align}
f(\sigma)&=-\dfrac{\lambda^{2}}{(r/\lambda)L_{\rm dS}^{2}}\left(\dfrac{r}{\lambda}-\dfrac{r_{+}}{\lambda}\right)\left(\dfrac{r}{\lambda}-\dfrac{r_{\rm c}}{\lambda}\right)\left(\dfrac{r}{\lambda}+\dfrac{r_{0}}{\lambda}\right)
\nonumber
\\
&=-\dfrac{\lambda^{2}}{(\rho_{1}-\beta_{2}\sigma)L_{\rm dS}^{2}}\left(\rho_{1}-\beta_{2}\sigma-\dfrac{r_{+}}{\lambda}\right)\left(\rho_{1}-\beta_{2}\sigma-\dfrac{r_{\rm c}}{\lambda}\right)\left(\rho_{1}-\beta_{2}\sigma+\dfrac{r_{0}}{\lambda}\right)
\nonumber
\\
&=-\dfrac{\lambda^{2}}{\rho_{1}L_{\rm dS}^{2}}
\left(\rho_{1}-\dfrac{r_{+}}{\lambda}\right)\left(\rho_{1}-\dfrac{r_{\rm c}}{\lambda}\right)\left(\rho_{1}+\dfrac{r_{0}}{\lambda}\right)
\nonumber
\\
&\qquad \qquad \qquad  
\left[1-\beta_{2}\sigma\left(\dfrac{1}{\left(\rho_{1}-\dfrac{r_{+}}{\lambda}\right)}
+\dfrac{1}{\left(\rho_{1}-\dfrac{r_{\rm c}}{\lambda}\right)}
+\dfrac{1}{\left(\rho_{1}+\dfrac{r_{0}}{\lambda}\right)}-\dfrac{1}{\rho_{1}}\right)+\mathcal{O}(\sigma^{2})\right]~.
\end{align}
For the choices of the parameters presented above, $\rho_{1}=\sigma_{+}(r_{\rm c}/r_{+})$, and $\beta_{2}=\rho_{1}-\sigma_{+}$, we observe that the term involving $\{\rho_{1}-(r_{\rm c}/\lambda)\}$, identically vanishes and hence the metric function will be proportional to the compactified radial coordinate $\sigma$, such that,
\begin{align}
f(\sigma)&=\beta_{2}\dfrac{r_{+}\sigma_{+}\lambda^{2}}{r_{\rm c}L_{\rm dS}^{2}}\sigma
\left(\dfrac{r_{\rm c}}{r_{+}}-1\right)\left(\dfrac{r_{\rm c}}{r_{+}}+\dfrac{r_{0}}{r_{+}}\right)
\nonumber
\\
&=\dfrac{r_{+}\sigma_{+}^{2}\lambda^{2}}{r_{\rm c}L_{\rm dS}^{2}}
\left(\dfrac{r_{\rm c}}{r_{+}}-1\right)^{2}\left(2\dfrac{r_{\rm c}}{r_{+}}+1\right)\sigma~.
\end{align}
The same expansion can also be obtained from the direct substitution of $r(\sigma)$ in the metric function, as one can check in a straightforward manner, by using the results that $f(r_{c})=0=f(r_{+})$ and $\lambda\rho_{1}=r_{c}$.

Having derived the metric function in terms of the re-scaled and compactified radial coordinate $\sigma$, the differential equation for $h_{0}$, as presented in \ref{h0diffdS}, takes the following form,
\begin{align}
h_{0}'(\sigma)=\dfrac{2r_{\rm c}\sigma_{+}L_{\rm dS}^{2}}{r_{+}^{3}}
\dfrac{1}{\left(\dfrac{r_{\rm c}}{r_{+}}-1\right)\left(2\dfrac{r_{\rm c}}{r_{+}}+1\right)}\dfrac{1}{\sigma}~.
\end{align}
The above differential equation for $h_{0}(\sigma)$ can be integrated, yielding,
\begin{align}
h_{0}(\sigma)=\dfrac{2r_{\rm c}\sigma_{+}L_{\rm dS}^{2}}{r_{+}^{3}}
\dfrac{1}{\left(\dfrac{r_{\rm c}}{r_{+}}-1\right)\left(2\dfrac{r_{\rm c}}{r_{+}}+1\right)}\ln \sigma~,
\end{align}
where, any constant contributions to the function $h_{0}(\sigma)$ have been ignored. This result will be used in determining the height function subsequently.

It turns out that the above expression for $h_{0}(\sigma)$, as well as the expression for the tortoise coordinate simplifies significantly if expressed in terms of the surface gravities associated with these horizons. Given the above metric function, the following three surface gravity expressions can be obtained,
\begin{align}
\kappa_{+}&=-\dfrac{1}{2r_{+}L_{\rm dS}^{2}}\left(r_{+}-r_{\rm c}\right)\left(r_{+}+r_{0}\right)
=\dfrac{1}{2r_{+}L_{\rm dS}^{2}}\left(r_{\rm c}-r_{+}\right)\left(2r_{+}+r_{\rm c}\right)~,
\\
\kappa_{\rm c}&=-\dfrac{1}{2r_{\rm c}L_{\rm dS}^{2}}\left(r_{\rm c}-r_{+}\right)\left(r_{\rm c}+r_{0}\right)
=-\dfrac{1}{2r_{\rm c}L_{\rm dS}^{2}}\left(r_{\rm c}-r_{+}\right)\left(2r_{\rm c}+r_{+}\right)~,
\\
\kappa_{0}&=\dfrac{1}{2r_{0}L_{\rm dS}^{2}}\left(r_{0}+r_{+}\right)\left(r_{0}+r_{\rm c}\right)
=\dfrac{1}{2r_{0}L_{\rm dS}^{2}}\left(r_{\rm c}+2r_{+}\right)\left(r_{+}+2r_{\rm c}\right)~.
\end{align}
Note that both $\kappa_{+}$ and $\kappa_{0}$ are positive quantities, while $\kappa_{c}$ is negative. In terms of these expressions for surface gravity, the simplified expression for $h_{0}(\sigma)$ becomes,
\begin{align}
h_{0}(\sigma)=\dfrac{\sigma_{+}}{|\kappa_{c}|r_{+}}\ln \sigma~.
\end{align}
The next task, in determining the hyperboloidal coordinate system, is to express the tortoise coordinate in terms of the compactified radial coordinate $\sigma$. To this end, we first express the tortoise coordinate in terms of the radial coordinate $r$ as, 
\begin{align}
r_{*}=\dfrac{1}{2\kappa_{+}}\ln \left(\dfrac{r}{r_{+}}-1\right)-\dfrac{1}{2|\kappa_{c}|}\ln \biggr|\dfrac{r}{r_{\rm c}}-1\biggr|+\dfrac{1}{2\kappa_{0}}\ln \left(\dfrac{r}{r_{0}}+1\right)~,
\end{align}
and then express the radial coordinate $r$ in terms of the compactified coordinate $\sigma$ using \ref{rsigmadS_2}. Therefore, the function $g(\sigma)$, which is simply the tortoise coordinate expressed in terms of the compactified coordinate $\sigma$ reads,
\begin{align}\label{tortoise_sds}
g(\sigma)&=\dfrac{\sigma_{+}}{2\kappa_{+}r_{+}}\ln \left[\sigma+\dfrac{r_{c}}{r_{+}}\left(1-\sigma\right)-1\right]
-\dfrac{\sigma_{+}}{2|\kappa_{c}|r_{+}}\ln \left[\left(\sigma+\dfrac{r_{c}}{r_{+}}\left(1-\sigma\right)\right)\dfrac{r_{+}}{r_{\rm c}}-1\right]
\nonumber
\\
&\qquad +\dfrac{\sigma_{+}}{2\kappa_{0}r_{+}}\ln \left[\left(\sigma+\dfrac{r_{c}}{r_{+}}\left(1-\sigma\right)\right)\dfrac{r_{+}}{r_{0}}+1\right]
\nonumber
\\
&=\dfrac{\sigma_{+}}{2\kappa_{+}r_{+}}\ln \left(1-\sigma\right)
-\dfrac{\sigma_{+}}{2|\kappa_{c}|r_{+}}\ln \sigma
+\dfrac{\sigma_{+}}{2\kappa_{0}r_{+}}\ln \left[\left(1+\dfrac{r_{c}}{r_{0}}\right)+\sigma
\left(\dfrac{r_{+}}{r_{0}}-\dfrac{r_{c}}{r_{0}}\right)\right]~.
\end{align}
Here we have discarded all the terms which are independent of $\sigma$ since we shall only be concerned with the derivatives of $g(\sigma)$ and $h(\sigma)$ in this work. 
Hence the height function $h(\sigma)$ in the Schwarzschild de Sitter black hole spacetime becomes,
\begin{align}\label{height_sds}
h(\sigma)&=\dfrac{\sigma_{+}}{2\kappa_{+}r_{+}}\ln \left(1-\sigma\right)
+\dfrac{\sigma_{+}}{2|\kappa_{c}|r_{+}}\ln \sigma
+\dfrac{\sigma_{+}}{2\kappa_{0}r_{+}}\ln \left[\left(1+\dfrac{r_{c}}{r_{0}}\right)+\sigma
\left(\dfrac{r_{+}}{r_{0}}-\dfrac{r_{c}}{r_{0}}\right)\right]~.
\end{align}
Note that unlike the case of asymptotically flat spacetime, where terms $\mathcal{O}(1/\sigma)$ were present in the height function, for asymptotically de Sitter spacetime, the divergences in the height function solely arises from the logarithmic terms. Moreover, there are no free parameters in the above expressions for the height function $h(\sigma)$ and the tortoise coordinate $g(\sigma)$, implying the absence of any residual gauge freedom in the choice of hyperbolic coordinates in the minimal gauge. We will now demonstrate that this feature is a generic one and appears in the presence of electric charge as well. 
\subsubsection{Hyperbolic coordinate for Reissner-Nordstr\"{o}m-de Sitter black hole}

In this final example, we will derive the height function necessary to make a transition to hyperboloidal coordinate system for Reissner-Nordstr\"{o}m-de Sitter black hole. This will be the generalization of the Reissner-Nordstr\"{o}m black hole to asymptotically de Sitter spacetimes. The metric function associated with the above spacetime reads, 
\begin{align}
f(r)=1-\dfrac{2M}{r}+\dfrac{Q^{2}}{r^{2}}-\dfrac{\Lambda r^{2}}{3}~;
\qquad
\dfrac{\Lambda}{3}=\dfrac{1}{L_{\rm dS}^{2}}~,
\end{align}
where, the last relation defines the de Sitter length scale $L_{\rm dS}$, as in the previous section. The metric element can be expressed as, 
\begin{align}
f(r)=-\dfrac{1}{r^{2}L_{\rm dS}^{2}}\left(r^{4}-L_{\rm dS}^{2}r^{2}+2ML_{\rm dS}^{2}r-Q^{2}L_{\rm dS}^{2}\right)
=-\dfrac{1}{r^{2}L_{\rm dS}^{2}}\left(r-r_{+}\right)\left(r-r_{-}\right)\left(r-r_{\rm c}\right)\left(r+r_{0}\right)
\end{align}
where, $r_{+}$ is the event horizon, $r_{-}$ is the Cauchy horizon and $r_{\rm c}$ is the cosmological horizon, with $r_{0}=r_{+}+r_{\rm c}+r_{-}$. In this case, with the same parameter choices as in the Schwarzschild de Sitter black hole, we obtain,
\begin{align}
f(\sigma)&=\dfrac{r_{+}^{4}}{r_{\rm c}^{2}L_{\rm dS}^{2}}
\left(\dfrac{r_{\rm c}}{r_{+}}-1\right)^{2}\left(2\dfrac{r_{\rm c}}{r_{+}}+\dfrac{r_{-}}{r_{+}}+1\right)\left(\dfrac{r_{\rm c}}{r_{+}}-\dfrac{r_{-}}{r_{+}}\right)\sigma
\end{align}
Having derived the metric function in terms of the re-scaled radial coordinate $\sigma$, the differential equation for $h_{0}$ takes the following form,
\begin{align}
h_{0}'(\sigma)=\dfrac{2\sigma_{+}r_{\rm c}^{2}L_{\rm dS}^{2}}{r_{+}^{4}}
\dfrac{1}{\left(\dfrac{r_{\rm c}}{r_{+}}-1\right)\left(2\dfrac{r_{\rm c}}{r_{+}}+\dfrac{r_{-}}{r_{+}}+1\right)\left(\dfrac{r_{\rm c}}{r_{+}}-\dfrac{r_{-}}{r_{+}}\right)}\dfrac{1}{\sigma}
\end{align}
which integrates to,
\begin{align}
h_{0}(\sigma)=\dfrac{2\sigma_{+}r_{\rm c}^{2}L_{\rm dS}^{2}}{r_{+}^{4}}
\dfrac{1}{\left(\dfrac{r_{\rm c}}{r_{+}}-1\right)\left(2\dfrac{r_{\rm c}}{r_{+}}+\dfrac{r_{-}}{r_{+}}+1\right)\left(\dfrac{r_{\rm c}}{r_{+}}-\dfrac{r_{-}}{r_{+}}\right)}\ln \sigma
\end{align}
This will be used in determining the height function. 

Next task is to find out the tortoise coordinate. For which we need to define the following three surface gravity expressions,
\begin{align}
\kappa_{+}&=-\dfrac{1}{2r_{+}^{2}L_{\rm dS}^{2}}\left(r_{+}-r_{\rm c}\right)\left(r_{+}+r_{0}\right)\left(r_{+}-r_{-}\right)
=\dfrac{1}{2r_{+}^{2}L_{\rm dS}^{2}}\left(r_{\rm c}-r_{+}\right)\left(2r_{+}+r_{\rm c}+r_{-}\right)\left(r_{+}-r_{-}\right)
\\
\kappa_{-}&=-\dfrac{1}{2r_{-}^{2}L_{\rm dS}^{2}}\left(r_{-}-r_{\rm c}\right)\left(r_{-}+r_{0}\right)\left(r_{-}-r_{+}\right)
=-\dfrac{1}{2r_{-}^{2}L_{\rm dS}^{2}}\left(r_{\rm c}-r_{-}\right)\left(2r_{-}+r_{\rm c}+r_{+}\right)\left(r_{+}-r_{-}\right)
\\
\kappa_{\rm c}&=-\dfrac{1}{2r_{\rm c}^{2}L_{\rm dS}^{2}}\left(r_{\rm c}-r_{+}\right)\left(r_{\rm c}+r_{0}\right)\left(r_{\rm c}-r_{-}\right)
=-\dfrac{1}{2r_{\rm c}^{2}L_{\rm dS}^{2}}\left(r_{\rm c}-r_{+}\right)\left(2r_{\rm c}+r_{+}+r_{-}\right)\left(r_{\rm c}-r_{-}\right)
\\
\kappa_{0}&=\dfrac{1}{2r_{0}^{2}L_{\rm dS}^{2}}\left(r_{0}+r_{+}\right)\left(r_{0}+r_{\rm c}\right)\left(r_{0}+r_{-}\right)
=\dfrac{1}{2r_{0}^{2}L_{\rm dS}^{2}}\left(r_{\rm c}+2r_{+}+r_{-}\right)\left(r_{+}+2r_{\rm c}+r_{-}\right)\left(r_{\rm c}+r_{+}+2r_{-}\right)
\end{align}
In terms of these expressions for surface gravity, the expression for $h_{0}(\sigma)$ can be further simplified, yielding,
\begin{align}
h_{0}(\sigma)=-\dfrac{\sigma_{+}}{\kappa_{c}r_{+}}\ln \sigma
\end{align}
Note that both $\kappa_{+}$ and $\kappa_{0}$ are positive quantities, while $\kappa_{c}$ and $\kappa_{-}$ are negative, therefore, $h_{0}(\sigma)$ reads,
\begin{align}
h_{0}(\sigma)=\dfrac{\sigma_{+}}{|\kappa_{c}|r_{+}}\ln \sigma
\end{align}
Thus, we obtain, the tortoise coordinate as, 
\begin{align}
r_{*}=\dfrac{1}{2\kappa_{+}}\ln \left(\dfrac{r}{r_{+}}-1\right)+\dfrac{1}{2\kappa_{c}}\ln \biggr|\dfrac{r}{r_{\rm c}}-1\biggr| +\dfrac{1}{2\kappa_{0}}\ln \left(\dfrac{r}{r_{0}}+1\right)+\dfrac{1}{2\kappa_{-}}\ln \left(\dfrac{r}{r_{-}}-1\right)
\end{align}
Therefore, $g(\sigma)$ reads,
\begin{align}
g(\sigma)&=\dfrac{\sigma_{+}}{2\kappa_{+}r_{+}}\ln \left[\sigma+\dfrac{r_{c}}{r_{+}}\left(1-\sigma\right)-1\right]
-\dfrac{\sigma_{+}}{2|\kappa_{c}|r_{+}}\ln \left[\left(\sigma+\dfrac{r_{c}}{r_{+}}\left(1-\sigma\right)\right)\dfrac{r_{+}}{r_{\rm c}}-1\right]
\nonumber
\\
&\qquad +\dfrac{\sigma_{+}}{2\kappa_{0}r_{+}}\ln \left[\left(\sigma+\dfrac{r_{c}}{r_{+}}\left(1-\sigma\right)\right)\dfrac{r_{+}}{r_{0}}+1\right]
-\dfrac{\sigma_{+}}{2|\kappa_{-}|r_{+}}\ln \left[\left(\sigma+\dfrac{r_{c}}{r_{+}}\left(1-\sigma\right)\right)\dfrac{r_{+}}{r_{-}}-1\right]
\nonumber
\\
&=\dfrac{\sigma_{+}}{2\kappa_{+}r_{+}}\ln \left[\left(1-\sigma\right)\left(1-\dfrac{r_{c}}{r_{+}}\right)\right]
-\dfrac{\sigma_{+}}{2|\kappa_{c}|r_{+}}\ln \left[\sigma \left(\dfrac{r_{+}}{r_{c}}-1\right)\right]
\nonumber
\\
&\qquad +\dfrac{\sigma_{+}}{2\kappa_{0}r_{+}}\ln \left[\left(1+\dfrac{r_{c}}{r_{0}}\right)+\sigma \left(\dfrac{r_{+}}{r_{0}}-\dfrac{r_{c}}{r_{0}}\right)\right]
-\dfrac{\sigma_{+}}{2|\kappa_{-}|r_{+}}\ln \left[\left(\dfrac{r_{c}}{r_{-}}-1\right)+\sigma \left(\dfrac{r_{+}}{r_{-}}-\dfrac{r_{c}}{r_{-}}\right)\right]
\nonumber
\\
&=\dfrac{\sigma_{+}}{2\kappa_{+}r_{+}}\ln \left(1-\sigma\right)
-\dfrac{\sigma_{+}}{2|\kappa_{c}|r_{+}}\ln \sigma 
+\dfrac{\sigma_{+}}{2\kappa_{0}r_{+}}\ln \left[\left(1+\dfrac{r_{c}}{r_{0}}\right)+\sigma \left(\dfrac{r_{+}}{r_{0}}-\dfrac{r_{c}}{r_{0}}\right)\right]
\nonumber
\\
&\qquad 
-\dfrac{\sigma_{+}}{2|\kappa_{-}|r_{+}}\ln \left[\left(\dfrac{r_{c}}{r_{-}}-1\right)+\sigma \left(\dfrac{r_{+}}{r_{-}}-\dfrac{r_{c}}{r_{-}}\right)\right]~.
\end{align}
Here, we have ignored terms which are independent of the compactified radial coordinate $\sigma$. Therefore, using this expression for $g(\sigma)$ and the one for $h_{0}(\sigma)$, derived earlier, the height function becomes,
\begin{align}
h(\sigma)&=\dfrac{\sigma_{+}}{2\kappa_{+}r_{+}}\ln \left(1-\sigma\right)
+\dfrac{\sigma_{+}}{2|\kappa_{c}|r_{+}}\ln \sigma
+\dfrac{\sigma_{+}}{2\kappa_{0}r_{+}}\ln \left[\left(1+\dfrac{r_{c}}{r_{0}}\right)+\sigma \left(\dfrac{r_{+}}{r_{0}}-\dfrac{r_{c}}{r_{0}}\right)\right]
\nonumber
\\
&\qquad 
-\dfrac{\sigma_{+}}{2|\kappa_{-}|r_{+}}\ln \left[\left(\dfrac{r_{c}}{r_{-}}-1\right)+\sigma \left(\dfrac{r_{+}}{r_{-}}-\dfrac{r_{c}}{r_{-}}\right)\right]~.
\end{align}
Again, the above expression for the height function does not depend on any free parameters, except for $\sigma_{+}$, and hence there are no residual gauge freedom associated with the above expressions, unlike the case of asymptotically flat spacetimes. 

Having extended the derivation of the hyperboloidal coordinate system for asymptotically de Sitter spacetime, we have also applied them in two examples, namely Schwarzschild de Sitter and Reissner-Nordstr\"{o}m de Sitter spacetimes. In the subsequent sections, we will express the wave equations satisfied by perturbations in these backgrounds in the hyperboloidal coordinates, which will help in studying the pseudo-spectrum of the QNM frequencies.   

\section{Wave equation for perturbations of asymptotically de Sitter spacetimes}\label{sec:wave_eqn_details}

Having introduced the hyperboloidal coordinate system for asymptotically de Sitter spacetimes, let us now determine the wave equation governing the perturbations in this coordinate system\footnote{In the ensuing general discussion about the wave operator describing perturbations, we will closely follow the notations and conventions introduced in \cite{Jaramillo:2020tuu} before specializing to the de Sitter case from \ref{eigenvalue_eq}.}. In general, owing to the static and spherical symmetry of the background spacetimes, any perturbation can be decomposed into a factor of $e^{-i\omega t}$, a radial function $\phi_{\ell m}(r)$ and angular functions. The angular functions are given by $Y_{\ell m}(\theta,\phi)$ for scalar perturbation, and derivatives of $Y_{\ell m}(\theta,\phi)$ for electromagnetic and gravitational perturbations. The radial function $\phi_{\ell m}(r)$, in general, satisfy the following wave equation:
\begin{align}
\left(\dfrac{\partial^{2}}{\partial t^{2}}-\dfrac{\partial^{2}}{\partial r_{*}^{2}}+V_{\ell}\right) \phi_{\ell m}=0~,
\end{align}
where, $V_{\ell}$ will depend on the nature of the perturbation, the frequency $\omega$, the angular indices, and the geometry of the background spacetime. The above wave equation is in the $(t,r,\theta,\phi)$ coordinate system. For our purpose, we need to transform the above to the hyperboloidal coordinate system devised in the previous section. For asymptotically flat spacetimes, transforming the above wave equation to the hyperboloidal coordinate system has already been performed, here we do the same, but for asymptotically de Sitter spacetimes. First, we discuss the general result and then shall specialize on both the examples discussed above.

To start with, we transform the above wave equation to the hyperboloidal $(\tau,\sigma,\theta,\phi)$ coordinate system, using the transformations: $\tau=(t/\lambda)+h(\sigma)$ and $(r_{*}/\lambda)=g(\sigma)$, where $\lambda$ is an arbitrary length scale associated with the problem. Therefore, we obtain the following results, using the chain rule for partial differentiation,
\begin{align}
\dfrac{\partial}{\partial t}=\dfrac{1}{\lambda}\dfrac{\partial}{\partial \tau}~;
\qquad
\dfrac{\partial}{\partial r_{*}}=\dfrac{1}{\lambda g'}\dfrac{\partial}{\partial \sigma}
+\dfrac{h'}{\lambda g'}\dfrac{\partial}{\partial \tau}~.
\end{align}
Note that, the time evolution vector $(\partial/\partial t)$ is simply scaled by $\lambda$ under this coordinate transformation, or, in other words, $(\partial/\partial \tau)$ also generates time translation. Hence one can define the quasi-normal mode frequency as conjugate to the redefined time coordinate $\tau$ as well, except for the fact that the quasi-normal mode frequencies defined using $\tau$ will scale by $\lambda$ compared to the quasi-normal mode frequencies defined using $t$. This is another reason for the introduction of the hyperboloidal coordinate since it simply rescales the quasi-normal mode frequencies by a constant factor. Substituting the above relations between $(\partial/\partial t)$ and $(\partial/\partial r_{*})$ with $(\partial/\partial \tau)$ and $(\partial/\partial \sigma)$, the above wave equation for the radial part of the perturbation reduces to,
\begin{align}
\dfrac{1}{\lambda^{2}}\ddot{\phi}_{\ell m}-\left(\dfrac{1}{\lambda g'}\dfrac{\partial}{\partial \sigma}
+\dfrac{h'}{\lambda g'}\dfrac{\partial}{\partial \tau}\right)\left(\dfrac{1}{\lambda g'}\dfrac{\partial \phi_{\ell m}}{\partial \sigma}+\dfrac{h'}{\lambda g'}\dfrac{\partial \phi_{\ell m}}{\partial \tau}\right)+V_{\ell}\phi_{\ell m}=0~,
\end{align}
where `prime' denotes the derivative with respect to the re-scaled radial coordinate $\sigma$ and `dot' denotes the derivative with respect to $\tau$. Simplifying further, we obtain,
\begin{align}
\dfrac{1}{\lambda^{2}}\left(1-\dfrac{h'^{2}}{g'^{2}}\right)\ddot{\phi}_{\ell m}
-\left[\dfrac{2}{\lambda^{2}}\dfrac{h'}{g'^{2}}\partial_{\sigma}+\dfrac{1}{\lambda^{2}g'}\partial_{\sigma}\left(\dfrac{h'}{g'}\right)\right]\dot{\phi}_{\ell m}
-\dfrac{1}{\lambda^{2}}\dfrac{1}{g'}\partial_{\sigma}\left(\dfrac{1}{g'}\partial_{\sigma}\phi_{\ell m}\right)
+V_{\ell}\phi_{\ell m}=0~.
\end{align}
Multiplying the above equation throughout by $\lambda^{2}$, as well as by the combination $\{g'^{2}/(g'^{2}-h'^{2})\}$ along with an overall negative sign, the above wave equation can be rewritten as,
\begin{align}
-\ddot{\phi}_{\ell m}
&+\left[2\left(\dfrac{h'}{g'^{2}-h'^{2}}\right)\partial_{\sigma}+\left(\dfrac{g'}{g'^{2}-h'^{2}}\right)\partial_{\sigma}\left(\dfrac{h'}{g'}\right)\right]\dot{\phi}_{\ell m}
\nonumber
\\
&\qquad \qquad +\left(\dfrac{g'}{g'^{2}-h'^{2}}\right)\partial_{\sigma}\left(\dfrac{1}{g'}\partial_{\sigma}\phi_{\ell m} \right)-\left(\dfrac{g'^{2}}{g'^{2}-h'^{2}}\right)\lambda^{2}V_{\ell}\phi_{\ell m}=0~.
\end{align}
Note that, the above equation is invariant under the transformation $g'\rightarrow -g'$, and hence we can replace every $g'$ by $|g'|$. Therefore, it follows that, we can re-express the above equation as,
\begin{align}\label{diffeq1}
-\ddot{\phi}_{\ell m}+\dfrac{1}{w(\sigma)}\Big[2\gamma(\sigma)\partial_{\sigma}+\left(\partial_{\sigma}\gamma \right)\Big]\dot{\phi}_{\ell m}+\dfrac{1}{w(\sigma)}\Big[\partial_{\sigma}\left\{p(\sigma)\partial_{\sigma} \right\}-q_{\ell}(\sigma)\Big]\phi_{\ell m}=0~,
\end{align}
where, we have introduced the following four functions of the re-scaled radial coordinate $\sigma$:
\begin{align}\label{definition1}
w(\sigma)\equiv \dfrac{g'^{2}-h'^{2}}{|g'|}~;
\qquad
\gamma(\sigma)\equiv \dfrac{h'}{|g'|}~;
\qquad
p(\sigma)\equiv \dfrac{1}{|g'|}~;
\qquad
q_{\ell}(\sigma)\equiv \lambda^{2}|g'|V_{\ell}~.
\end{align}
It turns out that at the boundaries, located at $\sigma=\pm 1$, $|g'|$ diverges and hence the quantity $p(\sigma)$ identically vanishes. On the other hand, all the other quantities, namely $w(\sigma)$, $\gamma(\sigma)$, and $q_{\ell}(\sigma)$ remain finite and positive at the black hole and cosmological horizons. Moreover, it follows that $\partial_{\sigma}p$ is positive at the cosmological horizon (located at $\sigma=0$), but becomes negative at the event horizon (located at $\sigma=1$). As a consequence, the above differential equation will have no $\partial_{\sigma}^{2}$ term. The vanishing of the coefficient $\partial_\sigma^2$ term at the boundaries renders the above equation to be a singular second-order differential equation, one in which no boundary conditions are allowed if we demand regular solutions. This amounts to the fact that the outgoing/ingoing boundary conditions are already incorporated into the (bulk) operator.

Further, we can rewrite the differential equation as presented in \ref{diffeq1}, by introducing $\psi_{\ell m}=\dot{\phi}_{\ell m}$ and constructing a second rank column vector using $\phi_{\ell m}$ and $\psi_{\ell m}$ as, 
\begin{align}\label{diffeq2}
\dot{u}_{\ell m}=iLu_{\ell m}~;
\qquad 
u_{\ell m}=
\begin{pmatrix}
\phi_{\ell m} \\
\psi_{\ell m}
\end{pmatrix}~,
\end{align}
where, $L$ is a $(2\times 2)$ square matrix, whose entries are functions and differential operators involving $\sigma$:
\begin{align}\label{non_self_adjoint_opertor}
L=\dfrac{1}{i}
\begin{pmatrix}
0     & 1 \\
L_{1} & L_{2}
\end{pmatrix}~;
\quad
L_{1}=\dfrac{1}{w(\sigma)}\Big[\partial_{\sigma}\left\{p(\sigma)\partial_{\sigma} \right\}-q_{\ell}(\sigma)\Big]~;
\quad
L_{2}=\dfrac{1}{w(\sigma)}\Big[2\gamma(\sigma)\partial_{\sigma}+\left(\partial_{\sigma}\gamma \right)\Big]~.
\end{align}
Since the differential operator $L$ is solely dependent on the re-scaled radial function $\sigma$, it follows that one can integrate \ref{diffeq2} and obtain $u_{\ell m}(\tau,\sigma)=\exp(iL\tau)u_{\ell m}(0,\sigma)$. Thus the operator $L$ is like the time evolution operator, generating time translation along $\sigma$. Moreover, the evolution by the time coordinate $t$ and $\tau$ are related by a constant re-scaling factor $\lambda$, and hence it follows that we can expand the mode function as $\phi(\tau,\sigma)=\exp(i\omega \tau)\phi(\sigma)$, for each of the quasi-normal modes. Thus the determination of the QNMs reduces the following eigenvalue problem: 
\begin{align}\label{eigenvalue_eq}
Lu_{n,\ell m}=\omega_{n,\ell m}u_{n,\ell m}~.
\end{align}
The spectral theorem would guarantee the stability of the eigenvalues $\omega_{n,\ell m}$, provided the differential operator $L$ is hermitian. However, as evident from our previous analysis, it follows that $p(\sigma=0)=0=p(\sigma=1)$, i.e., the function $p(\sigma)$ vanishes at the boundaries. As a consequence, the Sturm–Liouville operator $L_{1}$ becomes singular at the boundaries and the operator $L$ ceases to be self-adjoint. Therefore, the stability of the QNMs is not guaranteed, since the spectral theorem does not apply.  In order to cure this singular behavior certain regularity conditions for the perturbation must be imposed on the eigenmodes $u_{n,\ell m}$, which in turn ensures that appropriate boundary conditions are taken at the two boundaries. This is how the hyperboloidal coordinate system makes the non-self-adjoint nature of the eigenvalue problem for black hole perturbation theory explicit. 

Having spelled out all the necessary ingredients to turn the wave equation into an eigenvalue problem for the QNMs, we will now determine the functions defined in \ref{definition1} for Schwarzschild de Sitter spacetime. We will not present an explicit form of these functions for the Reissner-Nordstr\"{o}m de Sitter spacetime due to the complicated nature of these functions. Using the expressions for the height function and the tortoise coordinate for Schwarzschild de Sitter spacetime from \ref{height_sds} and \ref{tortoise_sds}, respectively, we obtain:
\begin{align}
p(\sigma)&=\left(\dfrac{2r_{+}^{2}\kappa_{+}|\kappa_{c}|}{\sigma_{+}\kappa_{0}r_{0}}\right)\dfrac{\sigma \left(1-\sigma\right)\left[\left(\dfrac{2r_{\rm c}+r_{+}}{r_{c}-r_{+}}\right)-\sigma\right]}{1-\left(\dfrac{r_{c}-r_{+}}{r_{c}}\right)\sigma}
\\
\gamma(\sigma)&=1-\dfrac{2r_{+}|\kappa_{c}|}{\kappa_{0}^{2}r_{0}}\left[\dfrac{\sigma}{1-\left(\dfrac{r_{c}-r_{+}}{r_{c}}\right)\sigma} \right]\left[\kappa_{+}+\kappa_{0}\left(\dfrac{2r_{\rm c}+r_{+}}{r_{c}-r_{+}}\right)-\left(\kappa_{0}+\kappa_{+}\right)\sigma \right]
\\
w(\sigma)&=\dfrac{2\sigma_{+}}{\kappa_{0}^{2}r_{0}}\dfrac{\left[\kappa_{+}+\kappa_{0}\left(\dfrac{2r_{\rm c}+r_{+}}{r_{c}-r_{+}}\right)-\left(\kappa_{0}+\kappa_{+}\right)\sigma \right]}{1-\left(\dfrac{r_{c}-r_{+}}{r_{c}}\right)\sigma}
\\
q_{\ell}&=\lambda^{2}\left(\dfrac{\sigma_{+}\kappa_{0}r_{0}}{2r_{+}^{2}\kappa_{+}|\kappa_{c}|}\right)\dfrac{1-\left(\dfrac{r_{c}-r_{+}}{r_{c}}\right)\sigma}{\sigma \left(1-\sigma\right)\left[\left(\dfrac{2r_{\rm c}+r_{+}}{r_{c}-r_{+}}\right)-\sigma\right]}V_{\ell}
\end{align}
where, the potential $V_{\ell}(r)$ reads, 
\begin{align}
V_{\ell}(r)=f(r)\left[\dfrac{\ell(\ell+1)}{r^2} + (1-s^2) \left(\dfrac{2M}{r^3} -\dfrac{4-s^2}{6}\Lambda \right)\right]~.
\end{align}
Here, $s=0$ corresponds to scalar perturbations, $s=1$ is associated with the electromagnetic perturbations, and the axial gravitational perturbations are connected to $s=2$. For $\Lambda=0$, we get back the potential experienced by the perturbations in Schwarzschild spacetime. One striking difference between the asymptotically de Sitter spacetimes with asymptotically flat spacetimes must be noted here. For both Schwarzschild and Reissner-Nordstr\"{o}m spacetimes, the function $p(\sigma)$ had a double root at $\sigma=0$ and a single root at $\sigma=1$, which translates to the existence of an essential singularity at future null infinity and a removable singularity at the horizon. While in the case of Schwarzchild-de Sitter as well as for Reissner-Nordstr\"{o}m-de Sitter spacetimes, the function $p(\sigma)$ vanishes at $\sigma=0$ and at $\sigma=1$ linearly. Thus in asymptotically de Sitter spacetimes, both the singularities of the Sturm–Liouville operator $L_{1}$ are removable. We will now move on to the discussion involving the stability of the QNMs under external perturbations.

\section{Pseudospectrum and energy norm}
\label{sec:mathematical_def_pseudospectrum_norm}
The spectral stability of an eigenvalue problem is tightly linked with the nature of the underlying system, in particular, whether the system is conservative or not. Conservative systems are generically associated with self-adjoint operators whose eigenvalues are real, and the eigenfunctions form a complete orthonormal basis. Since we will need properties of such operators, it is a good place to pause for a while and provide a formal definition of such operators. For this purpose, we may consider a Hilbert space $\mathcal{H}$ with a scalar product $\langle \cdot,\cdot\rangle$ and norm $||\cdot||$. If $L$ is a bounded and linear operator acting on vectors in the Hibert space $\mathcal{H}$, then Riesz’s representation theorem guarantees the existence of a unique bounded operator $L^{\dagger}$ called the adjoint operator of $L$, which satisfies the relation: $\langle L^{\dagger} u,v\rangle=\langle u, L v\rangle$ for all $u,v\in \mathcal{H}$ \cite{axler}. The existence of an adjoint operator does not guarantee that the eigenfunctions of this operator $L$ will form a complete orthogonal basis, for which $L$ must be \emph{normal}, i.e., it must commute with its adjoint ($[L, L^{\dagger}]=0$) \cite{Trefethen2005_pseudospectrumbook}. 
The self-adjoint operators are a sub-class of the normal operators, satisfying the condition $L^{\dagger}=L$. For normal operators, there exists a \textit{spectral theorem} that ensures that the spectrum of $L$, which corresponds to the set of all possible eigenvalues of $L$, and denoted by $\sigma(L)$ --- is stable against perturbation of the operator $L$. By stability, we mean that the shift in the spectrum of $L$ due to the perturbation of the operator $L$ is bounded by the order of the perturbation, see \cite{Trefethen2005_pseudospectrumbook,Jaramillo:2020tuu} for more details.  

For non-conservative or, open systems, the situation becomes complicated. Consider black holes as an example, for which any energy going through the event horizon cannot be extracted by any classical process, which is also true for the energy crossing future null infinity. Thus energy is dissipated into the event horizon and at infinity, and hence operators associated with black holes must be non-self-adjoint, which is characteristic of any open system. As a consequence, their eigenfunctions are referred to as quasi-normal modes with complex eigenvalues. The complex eigenvalues are as such not a problem, actually, the existence of such complex eigenvalues stabilizes black holes under external perturbations. The problematic feature being the eigenfunctions are, in general, not orthogonal and do not form a complete set \cite{RevModPhys.70.1545}. Consequently, a small perturbation of the operator $L$ can produce an unbounded shift in the spectrum, thereby rendering the spectrum unstable. In other words, the spectral analysis fails to capture the whole picture of the spectral problem associated with non-self-adjoint operators. In the following, we discuss methods to circumvent these difficulties associated with non-self-adjoint operators and an appropriate definition of a norm in this context. 

\subsection{Pseudospectrum}

In this section, we will briefly review the concept of the pseudospectrum, a powerful mathematical tool to study the spectral problem associated with the non-self-adjoint operators. Consider a non-self-adjoint operator $L \in M_n(\mathbb{C})$ (the space of complex $n\times n$ matrices), whose spectrum is given by $\sigma(L)$.  Note that $\sigma(L)$ is a set of complex numbers.  If we consider now a real number $\epsilon>0$, then the set of all complex numbers that are eigenvalues of the perturbed operator $L+\delta L$ for some $||\delta L||<\epsilon$ is referred to as the \emph{$\epsilon$-pseudospectrum} and is denoted by $\sigma^{\epsilon}(L)$. Mathematically, the pseudospectrum is defined as \cite{Trefethen2005_pseudospectrumbook,Jaramillo:2020tuu}:
\begin{equation}
\sigma^{\epsilon}(L)=\left\{\omega\in \mathbb{C}~, ~\exists ~\delta L \in M_n(\mathbb{C}),~ ||\delta L||<\epsilon~: \omega\in \sigma(L+\delta L)\right\}~.
\end{equation}
For a stable spectral configuration, it is expected that the perturbed spectrum $\sigma^{\epsilon}(L)$ resides in close vicinity (of order $\sim \epsilon$ ) to the unperturbed spectrum. However, if the perturbed spectra drift away from the unperturbed ones at a distance that is a few orders of magnitude larger than $\epsilon$, it signals spectral instability, as the eigenvalues of $L$ are extremely sensitive to external perturbations. Moreover, the pseudospectrum associated with different $\epsilon$ values are just nested sets around the spectrum $\sigma(L)$ in the complex frequency plane, such that $\sigma^{\epsilon_1}(L)\subseteq \sigma^{\epsilon_2}(L)$ for $0<\epsilon_1\leq \epsilon_2$.

A mathematically equivalent definition of the pseudospectrum can be arrived at by considering the resolvent of the operator $L$, defined as: $R_L(\omega)=(\omega \mathbf{I}-L)^{-1}$, which becomes singular when $\omega$ is an eigenvalue of $L$. Intuitively, it is expected that when $\omega$ is close to the spectrum $\sigma(L)$, the norm of the resolvent $||R_L(\omega)||$ will be large. However, for the non-self-adjoint operators, the resolvent norm can be large even when $\omega$ is far away from the eigenvalue. This is because the resolvent satisfies the following relation, $||R_L(\omega)||_2\leq\kappa/\textrm{dist.}(\omega,\sigma(L))$, where the subscript `2' signifies that the above is a $L^{2}$ norm, $\kappa$ is the condition number and $\textrm{dist.}(\omega,\sigma(L))$ is the distance between a point $\omega$ to the set $\sigma(L)$ in the complex plane \cite{Trefethen2005_pseudospectrumbook}. For a normal operator $\kappa=1$; thus, the resolvent norm is large only in the close vicinity of the spectrum $\sigma(L)$. However, for the non-normal operators, the condition number can be very large and hence the resolvent norm can have large values even at distances far away from the spectrum. Therefore, the region in the complex plane, where the resolvent $||R_L(\omega)||$ is large provides another definition of the pseudospectrum $\sigma^{\epsilon}(L)$ of the non-self-adjoint operator $L$ under perturbation as \cite{Trefethen2005_pseudospectrumbook, Jaramillo:2020tuu, Destounis:2021lum}:
\begin{equation}
\sigma^{\epsilon} (L) =\left\{\omega\in \mathbb{C}: ||R_L(\omega)||\equiv||\left(\omega \mathbf{I}-L\right)^{-1}||>1/\epsilon\right\}~.
\end{equation}
In words, the above mathematical statement implies that the pseudospectrum is the region of the complex plane, which is bounded by the $1/\epsilon$ level curve (contour lines) of the norm of the resolvent. Thus one can conclude that determining the resolvent is the best way of checking the stability of the spectrum of any non-self-adjoint operator $L$. If one plots the resolvent norm as a function of frequency on the complex plane, it will demonstrate how the perturbation in the operator results in the drifting of the QNM frequencies. In particular, if the pseudospectrum $\sigma^{\epsilon} (L)$ extends far away from $\sigma(L)$, it signals strong non-normality of the operator $L$ and poor analytic behavior of the resolvent $R_L(\omega)$ \cite{Jaramillo:2020tuu}. 

\subsection{Energy norm}

The stability of the QNM frequencies of black holes is related to the shift in the spectrum of the perturbation operator $L$, defined in \ref{non_self_adjoint_opertor}, in the complex frequency plane under the perturbation $\delta L$ (induced by the perturbation of the potential $q_\ell$). A large shift in the spectrum under a small perturbation indicates spectral instability. However, it is important to specify what is meant by ``small'' or ``large'' perturbations, and for this purpose, we need to provide a choice for the operator norm $||\cdot||$, along with the choice for the scalar product $\langle\cdot,\cdot\rangle$ \cite{Gasperin:2021kfv}. It may be plausible that the perturbation perceived as small for a given choice of operator norm can become large for a different choice. In that regard, the energy norm provides a natural way to define the scale of the perturbation.
Note that \ref{diffeq1} describes the dynamics of a perturbing field mode $\phi_{\ell m}$ (the perturbation can be of scalar, electromagnetic, or, gravitational origin) with scattering potential $V_\ell$ in $(1+1)$ dimensions. Following \cite{Jaramillo:2020tuu, Gasperin:2021kfv}, we define the energy norm in terms of the energy associated with the field modes as,
\begin{equation}\label{energy_norm}
\begin{aligned}
    ||u||_E^2=\left\lVert\begin{pmatrix} \phi  \\ \psi  \end{pmatrix}\right\rVert_E^2 :=E(\phi,\psi)=\dfrac{1}{2}\int_{0}^{1}d\sigma\left[w(\sigma) |\psi|^2+p(\sigma) |\partial_\sigma \phi|^2+q_\ell(\sigma) |\phi|^2\right]~.
    \end{aligned}
\end{equation}
Here, we have dropped the subscripts $\ell m$ in $\phi_{\ell m}$ and $\psi_{\ell m}$ for brevity and improved readability. The integration limit in the above equation corresponds to the boundary of the re-scaled spatial parameter $\sigma\in [0,1]$. In this setting, the operator energy norm of the perturbed operator  $||\delta L||_E$ is related to the energy concentrated near the peaks of the perturbation energy distribution \cite{Gasperin:2021kfv}. With the definition of energy norm in \ref{energy_norm}, we can introduce the definition of the energy scalar product as \cite{Jaramillo:2020tuu, Gasperin:2021kfv}
\begin{equation}\label{inner_product}
\begin{aligned}
  \langle u_1,u_2\rangle_{\rm E}= \left\langle\begin{pmatrix} \phi_{1}  \\ \psi_{1}  \end{pmatrix},\begin{pmatrix} \phi_{2}  \\ \psi_{2}  \end{pmatrix}\right\rangle_E
 =
\dfrac{1}{2}\int_{0}^{1}d\sigma\left[w(\sigma) \bar \psi_1 \psi_2+p(\sigma) \partial_\sigma \bar \phi_1 \partial_\sigma  \phi_2+q_\ell(\sigma) \bar \phi_1 \phi_2\right]~.
    \end{aligned}
\end{equation}
Note that, for $u_1=u_2$, \ref{inner_product} reduces to \ref{energy_norm}, as it should. With the definition of the scalar product, we can calculate the adjoint operator $L^{\dagger}$ \cite{Jaramillo:2020tuu}, which turns out to be: $L^{\dagger}=L+L^{\delta}$, with 
\begin{equation}\label{adjoint}
\begin{aligned}
 L^{\delta}=\begin{pmatrix} 0 & 0 \\ 0 & L_2^{\delta} \end{pmatrix}~,
    \end{aligned}
\end{equation}
where $L_2^{\delta}$ is given by the following expression 
\begin{equation}\label{part_ad}
\begin{aligned}
 L_2^{\dagger}=2\dfrac{\gamma(\sigma)}{w(\sigma)}\left[\delta(\sigma-0)+\delta(\sigma-1)\right]~.
    \end{aligned}
\end{equation}
Here, $\delta(\sigma)$ denotes the Dirac delta function. The above expression explicitly indicates the non-self-adjointness of the operator $L$ defined in \ref{non_self_adjoint_opertor}, since $L\neq L^{\dagger}$ and the difference occurs precisely at the boundaries with dissipative boundary conditions. In the following we elaborate on the numerical techniques for determining the resolvent using the energy norm, defined above, which in turn will tell us about the stability of the QNM frequencies for asymptotically de Sitter spacetime.

\section{Numerical Implementation: Chebyshev's spectral method}
\label{sec:numerical_method}

In the previous section, we have elaborated on the mathematical definitions of pseudospectrum and the associated energy norm suitable for our purpose. In this section we will provide the numerical technique to determine the pseudospectrum arising from the differential operator $L$ governing the behaviour of the perturbation with the black hole event horizon and the cosmological horizon. We start by employing the Chebyshev's spectral method to discretize the differential operator $L$, defined in \ref{non_self_adjoint_opertor} and convert the task of computing the (pseudo)spectrum into a linear algebra problem \cite{trefethenMATLAB10.5555/357801, Jaramillo:2020tuu, Dias:2015nua} that can be implemented on a computer in a straightforward manner. We begin by sampling $N+1$ points from a Chebyshev-Gauss-Lobatto (CGL) grid,
\begin{equation}
    \sigma_j = \dfrac{1}{2} + \dfrac{1}{2}\cos\left(\dfrac{j\pi}{N}\right), ~ j = 0, 1, \cdots,N~, \label{cgl_grid}
\end{equation}
where $\sigma_j \in [0, 1]$\footnote{Note that this convention gives us $\sigma_0 = 1$ and $ \sigma_N=0$, corresponding to the locations of the event and cosmological horizons, respectively, and the grid points are ordered in reverse fashion.}. We then approximate the differential operators using the corresponding Chebyshev differentiation matrices \cite{trefethenMATLAB10.5555/357801, Boyd2001-kt, Markakis:2014nja, Markakis:2023pfh}. This leaves us with a square matrix whose dimension is $2(N+1)$. We can now compute the QNM frequencies directly by finding out the eigenvalues of the aforementioned matrix. 

The reason behind using the spectral method on a CGL grid is motivated by the fact that our problem involves a non-periodic domain, and we would like to sample more points near the horizons to avoid Runge's phenomenon. Furthermore, if the eigenfunctions are smooth, then the Chebyshev collocation method is guaranteed to converge exponentially. However this approach yields dense matrices which makes the problem much more difficult to solve unless one employs suitable matrix decomposition schemes. 

We also have to accurately compute matrix norms in order to calculate the pseudospectra and study explicit perturbations to the operator  $L$. Therefore, we have to use the discretized version of the energy scalar product to evaluate various inner products. The expression of the inner product \cite{Jaramillo:2020tuu} is the following
\begin{equation}
    \braket{u | v}_{\rm E} =(u^{\dagger})^i G^{\rm E}_{ij} v^{j} = u^{\dagger}\cdot G^{\rm E} \cdot v, ~~ u,v \in \mathbb{C}^{N+1}~,
\end{equation}
where $u^{\dagger}$ is the hermitian conjugate of $u$, and $G^{\rm E}$ is the Gram matrix. The adjoint of the operator $L$ is then given by \cite{Jaramillo:2020tuu}
\begin{equation}
    L^{\dagger} = (G_{\rm E})^{-1}\cdot L^{*} \cdot G_{\rm E}~.
\end{equation}
The computation of the gram matrix $G^{\rm E}$ is rather subtle and one must employ a suitable interpolation scheme to ensure the accuracy of the scalar product whenever the order of the product $\braket{u|v}_{\rm E}$ becomes greater than the resolution of the collocation grid $N$. We have employed the strategy detailed in Appendix A of \cite{Jaramillo:2020tuu}.

Having discussed the basic ingredients for the numerical scheme, let us turn our attention to the determination of the pseudospectrum. For this purpose, it is useful to employ the following characterization for the pseudospectrum in the energy norm,
\begin{equation}
\sigma_{\rm E}^{\epsilon}(L) = \{ \omega \in \mathbb{C}:s^{\mathrm{min}}_{\rm E}(\omega \mathbf{I}-L)<\epsilon \}~, 
\label{pseudospectra_def_energy_norm}
\end{equation}
where $s^{\mathrm{min}}_{\rm E}$ is the smallest generalized singular value obtained by performing a singular value decomposition which takes into account the energy norm, viz.,
\begin{equation}
s^{\mathrm{min}}_{\rm E}(M) = \mathrm{min} \{  \sqrt{\omega}: \omega  \in \sigma(M^{\dagger}M)\}, ~~ M = \omega \mathbf{I} -L~.
\end{equation}
However, it is important to note that this path involves computing the full eigenspectrum of $(\omega \mathbf{I}-L)$ over the entire region of the complex plane under investigation. This step is computationally expensive since the time complexity of the algorithm grows as $\mathcal{O}(n^3)$ at each point of the complex plane, where $n=2(N+1)$ \cite{lui1997cont, Wright2001}. We, therefore, take advantage of the fact that the gram matrix $G_{\rm E}$ is a positive-definite Hermitian matrix, i.e., $G^{*}_{\rm E}=G_{\rm E}$, and perform a Cholesky decomposition to obtain
\begin{equation}
G_{\rm E} = W^{*} \cdot W~, 
\label{cholesky}
\end{equation}
where $W$ is an upper-triangular matrix. We can then compute the energy norm of a vector $u$ as follows,
\begin{equation}
|| u ||_{\rm E}^2 = \braket{u|W^{*}W|u} = \braket{Wu|Wu} = ||Wu||^2_2~,
\end{equation}
where $||.||_{\rm E}$ is the energy norm and $||.||_2$ is the $L^{2}$ norm. The above decomposition, as in  \ref{cholesky} further enables us to write the adjoint $L^{\dagger}$ of the operator $L$ as,
\begin{equation}
L^{\dagger} = (W^{*}\cdot W)^{-1} \cdot L^{*}  \cdot (W^{*}\cdot W)~. 
\label{adj_w}
\end{equation}
Following the above result, in terms of the upper-triangular matrix $W$, it follows that we can introduce another operator $\tilde{L}$, such that $\tilde{L}\equiv  W\cdot L\cdot W^{-1}$ and following \cite{trefethen_1999}, we obtain the following connection between the energy norm and the $L^{2}$ norm,
\begin{equation}
||f(L)||_{\rm E}=||f(\tilde{L})||_2~,
\end{equation}
for any function $f$. The above equation, namely \ref{cholesky} enables us to perform subsequent computations in the $L^{2}$ norm using the matrix $\tilde{L}$, rather then the energy norm using the matrix $L$. In particular, with the $L^{2}$ norm we can compute the standard singular value decomposition to determine the minimum singular value, and hence, the pseudospectrum. In fact, before performing the singular value decomposition, one can transform the matrix $\tilde{L}$ into an upper Hessenberg matrix $H$ \cite{lui1997cont, Wright2001} so that the singular value calculation can make use of more efficient algorithms. It follows that all the entries below the first sub-diagonal of $H$ are zero by definition and hence $H$ has the same eigenspectrum as $L$ and $\tilde{L}$. Note that, we can also write: $\tilde{L}^{*} = W\cdot L^{\dagger}\cdot W^{-1}$. Hence the minimum singular value $s_2^{\mathrm{min}}$ of $H$ can be obtained using,
\begin{equation}
s_2^{\rm min}(\tilde{M}) = \mathrm{min} \{  \sqrt{\omega}: \omega  \in \sigma(\tilde{M}^{*}\tilde{M})\}, ~~ \tilde{M} = \omega \mathbf{I} -H~,
\end{equation}
and construct the pseudospectrum through,
\begin{equation}
\sigma_2^{\epsilon}({H}) = \{ \omega \in \mathbb{C}:s^{\mathrm{min}}_2(\omega \mathbf{I}- {H})<\epsilon \}~,
\end{equation}
which is equivalent to \ref{pseudospectra_def_energy_norm}. Since we are interested in estimating the minimum singular value, we can use Arnoldi iteration (or, rather Lanczos iteration since $\tilde{M}^{*}\tilde{M}$ is Hermitian). Assuming that the cost of performing one-time Cholesky and Hessenberg decompositions are negligible compared to performing the full singular value decomposition, the task of computing the pseudospectrum can now be performed using an algorithm with a time complexity $\mathcal{O}(n^2)$ at each point of the complex plane \cite{lui1997cont,Wright2001}, thereby providing a significant speed-up. There is one notable aspect related to the computation of (pseudo)spectra of black holes that deserves mentioning: the operator $L$ is highly non-normal and hence if the computation is carried out using machine precision, then the (pseudo)spectrum is likely to be contaminated by spurious eigenvalues resulting from round-off errors in floating point operations. Such numerical artifacts can severely hinder one from drawing tangible conclusions using numerical studies. Hence it is essential to carry out intermediate computations using extended precision. We have typically used $10$ times the machine precision, thereby keeping track of $ \sim 160$ digits in the intermediate calculations. We shall also discuss issues related to numerical convergence in greater detail in the next section. Lastly, the computation of the pseudospectra is well suited for parallelization since the calculation of the smallest singular value at each point of the two-dimensional grid is independent of one another. We found \texttt{Wolfram Mathematica} to be a suitable computational tool for our purpose since it can easily implement the numerical requirements discussed so far. 

\section{(In)stability of the QNM spectra of asymptotically de Sitter spacetimes}
\label{sec:spectral_instability_ds_BHs}

 \begin{figure}[htbp!]
 \centering
  \begin{subfigure}[t]{0.45\textwidth}
    \centering
    \includegraphics[width=\textwidth]
    {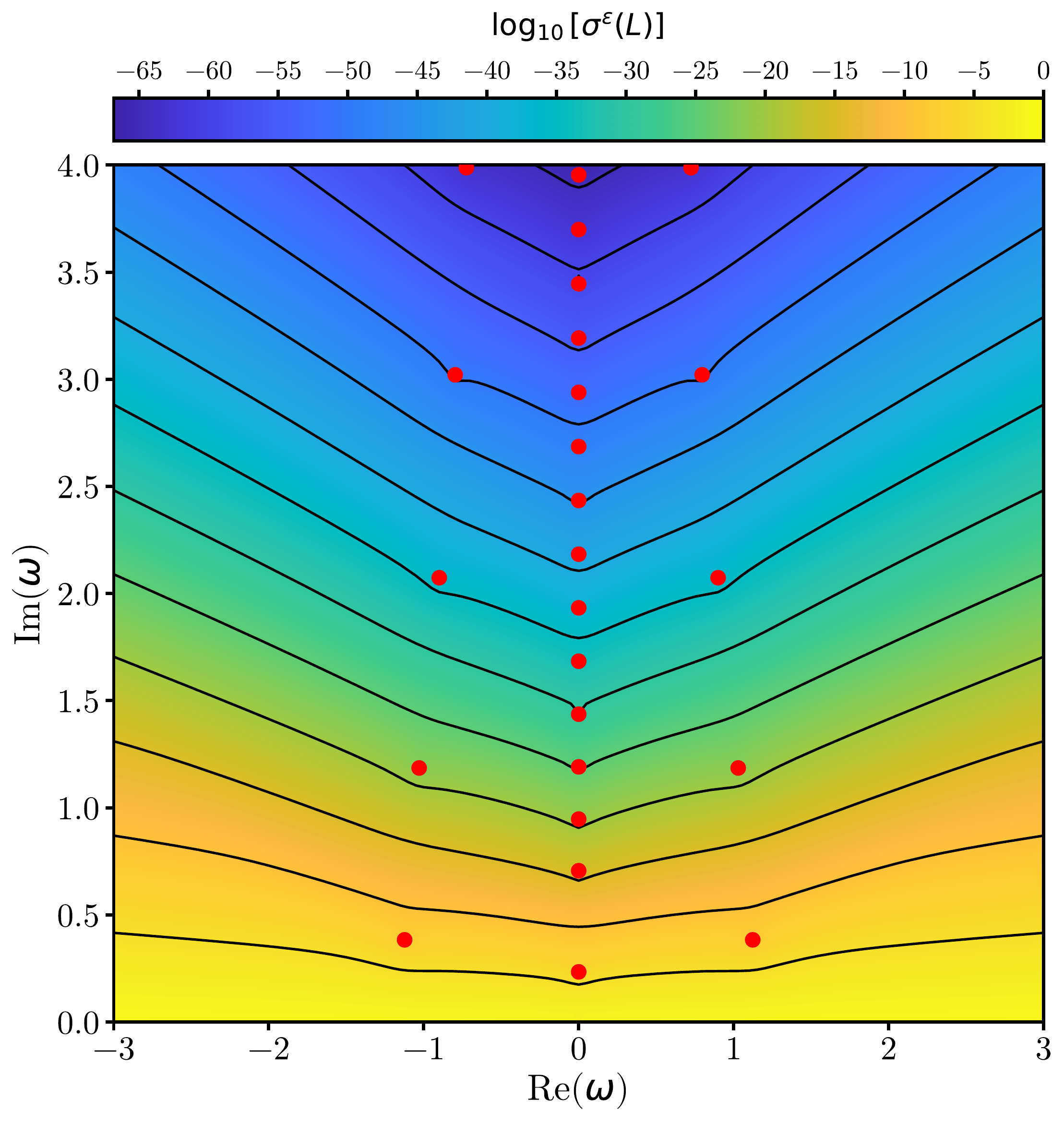}
    \caption{The $\ell=1$ \emph{scalar} $\epsilon$ pseudospectrum. The lowest-lying QNM being a de Sitter mode.}
    \label{fig:pseudospectra_scalar_l1_color_contour}
  \end{subfigure}\hspace{1em} 
  \begin{subfigure}[t]{0.45\textwidth}
  \centering
    \includegraphics[width=\textwidth]{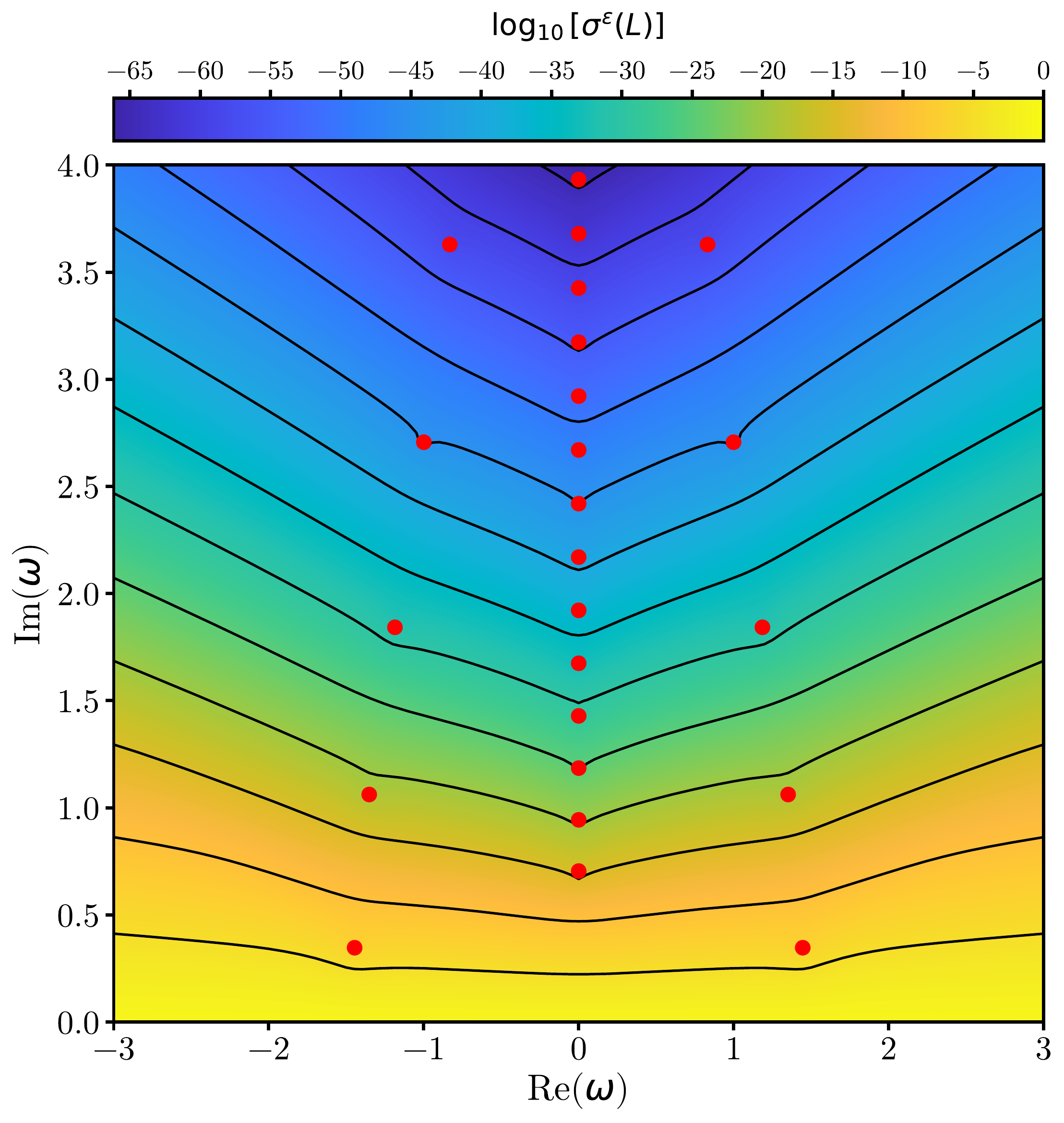}
    \caption{The $\ell=2$ \emph{gravitational} $\epsilon$ pseudospectrum. The lowest-lying QNM is a photon sphere mode.}
    \label{fig:pseudospectra_grav_l2_color_contour}
  \end{subfigure}
  \caption{The $\epsilon-$ pseudospectra of a Schwarzschild de Sitter black hole with $M=1$ and $\Lambda=0.01$ have been presented for scalar and gravitational perturbations. The unperturbed QNMs (filled red circles) for the scalar and gravitational perturbations have been indicated for reference. The solid black contour levels correspond to various choices of $\log_{10}\epsilon$, ranging from $-65$ (top level) to $-5$ (bottom level) in steps of $5$ in both the figures. The calculations have been carried out using $N=230$ collocation points with an internal precision corresponding to roughly $10 \times$ machine  precision, and a step size of $0.05$ along both the axes on the complex frequency plane. The characteristic length scale in the problem has been set to $\lambda = 2 r_+$.}
  \label{fig:pseudospectra_CC_0.01_color_contour}
\end{figure}

After providing all the necessary details regarding the construction of the hyperboloidal coordinate system as well as the numerical techniques to be used, we devote this section to the analysis of the QNM instability for asymptotically de Sitter black holes. The $\epsilon-$pseudospectrum of a black hole can be read as a topographic map that characterizes the stability of the quasi-normal mode spectrum. As we have already stressed, if the spectrum were stable against external perturbations then the pseudospectrum would have had a ``flat" structure, and the contour lines around a particular eigenvalue would then resemble circles of radius $\epsilon$, corresponding to the ``strength" of the perturbation. One could interpret the stability of an eigenvalue as being indicated by extremely steep throats around that particular eigenvalue which rapidly decay and become flat as one moves away from the eigenvalue further into the complex plane. Therefore, any non-trivial topographic structure extending far into the complex plane as one moves away from the eigenspectrum would be the hallmark of instability. The pseudospectra of asymptotically flat black holes \cite{Jaramillo:2020tuu, Destounis:2021lum} and exotic compact objects \cite{Boyanov:2022ark} show highly non-trivial topographic structure, indicating instability of the corresponding QNM spectra. It turns out that asymptotically de Sitter black holes are no exceptions, as evident from \ref{fig:pseudospectra_CC_0.01_color_contour}, the pseudospectra of these black holes also depict similar \emph{inverted ridge-like} topographic structures. However, unlike the case of asymptotically flat black holes, in the present situation the contour lines associated with the pseudospectrum of the lowest lying QNMs open up into the complex plane even for modest values of $\epsilon$ (see \ref{fig:pseudospectra_CC_0.01_color_contour}). Furthermore, \ref{fig:pseudospectra_CC_0.01_color_contour} also demonstrates that for the overtones the contours open up for very small values of $\epsilon$, indicating that instability increases as one ``moves up" the spectrum. We would like to emphasize that these features are generic and hold for both scalar and (axial) gravitational perturbations.

The predictive power of the $\epsilon-$pseudospectra has been highlighted in \cite{Jaramillo:2020tuu}: the key idea being that the pseudospectrum of a black hole may give us an opportunity estimate how the QNM spectrum of a black hole could possibly change under the influence of a perturbation $\delta V$ to the scattering potential experienced by a perturbing field on the black hole background. In particular, one must determine the strength of the perturbation $||\delta V||\lesssim \epsilon$ and locate the corresponding $\log_{10}{\epsilon}$ contour line on the pseudospectrum, like the ones shown in \ref{fig:pseudospectra_CC_0.01_color_contour}. The predicted change in the spectrum would then be the following: the portion of the spectra below  $\log_{10}\epsilon$ contour will remain untouched, but there is a possibility that the eigenvalues lying above this particular contour line could be significantly modified (depending upon the nature of the perturbing source). If the nature of the perturbation is indeed capable of modifying the spectrum then, what is truly remarkable is that (most of) the new perturbed eigenvalues (which we shall refer to as the \emph{perturbed QNMs}) will settle along one of the contour lines lying close to (but always above) the $\log_{10}\epsilon$ contour \footnote{We stress that the pseudospectrum can only inform us whether the spectrum of a non-conservative system like a black hole is inherently unstable. The pseudospectrum cannot tell us what kind of perturbations to the black hole potential can lead to an instability of the QNM spectrum. Hence, to understand the possible physical origin of the QNM instability, the pseudospectrum analysis should be complemented by studying explicit perturbations to the potential, a task that is explored in detail in the subsequent section of the present work. Furthermore, before assessing the predictive power of the pseudospectra, there are certain functional analysis issues that need to be thoroughly addressed (we refer the reader to \cite{Jaramillo:2020tuu} for further discussion).}. To put it differently, the perturbed QNMs will \emph{migrate to new branches following the boundaries of the pseudospectrum}. These branches have been referred to as the Nollert-Price branches \cite{Jaramillo:2020tuu} since it was first observed in \cite{Nollert:1996rf, Nollert:1998ys}. Alternatively, if we are able to detect several (perturbed) overtones along with the fundamental mode, and discover that the detected modes are spread in an orderly fashion that closely follows the contours of the pseudospectrum then we can gain valuable insight into the nature of the perturbing source based on the strength of the perturbation (that can be read off the pseudospectrum). The notion of the pseudospectra is therefore a definitive and robust culmination of the idea that was first foreseen in \cite{Aguirregabiria1996ScatteringBB}, viz., black hole spectroscopy will be able to probe the environment in the vicinity of the black hole; depending on the severity of the instability, it can also in principle help us probe near horizon quantum effects. Besides holding information about the ambient environment, our findings indicate that the pseudospectrum encodes information about the asymptotic structure of the black hole spacetime as well, i.e., the behavior of the perturbed QNMs of asymptotically flat and de Sitter black holes have crucial differences.

To put the above claim on firmer grounds and  draw conclusions that are relevant to asymptotically de Sitter black holes, we shall now explicitly study the perturbed QNM spectra of a Schwarzschild de Sitter black hole. For this purpose, let us briefly discuss the structure of the perturbation $\delta L$ that we add to the operator $L$ by modifying the same. The perturbation $\delta L$ is taken to have the following form:
\begin{equation}
    \label{e:delta_L_operator}
\delta L =\dfrac{1}{i}\!
\left(
\begin{array}{c|c}
	0 & 0 \\
	\hline 
	 \lambda^2 \delta V(\sigma)/w(\sigma) & 0
\end{array}
\right)~.
\end{equation}
The operator $\delta L$ should also be normalized with respect to the energy norm such that $||\delta L|| = \epsilon$. Such a choice ensures two things --- (a) the structure of $\delta L$ inflicts modifications exclusively on the black hole potential since $\delta V(\sigma)/w(\sigma)$ is a diagonal matrix by design and it is added to the sub-matrix of $L$ that contains $L_1$ (c.f. \ref{non_self_adjoint_opertor}), and (b) by specifying the norm we can apply specific meaning to the ``size" of the perturbation. The functional form of $\delta V(\sigma)$ depends on the physical phenomena we are trying to model. For example, to simulate both high and low-frequency perturbations, we can choose,
\begin{equation}
\delta V_d(\sigma) \sim \cos(2 \pi k \sigma)~,
\end{equation}
where $k$ is the inverse length scale associated with the perturbation; in \cite{Jaramillo:2020tuu} such generic oscillatory perturbations were labeled as ``deterministic perturbations".

\begin{figure}[tbh!]
	\centering
	\minipage{0.33\textwidth}
	\includegraphics[width=\linewidth]{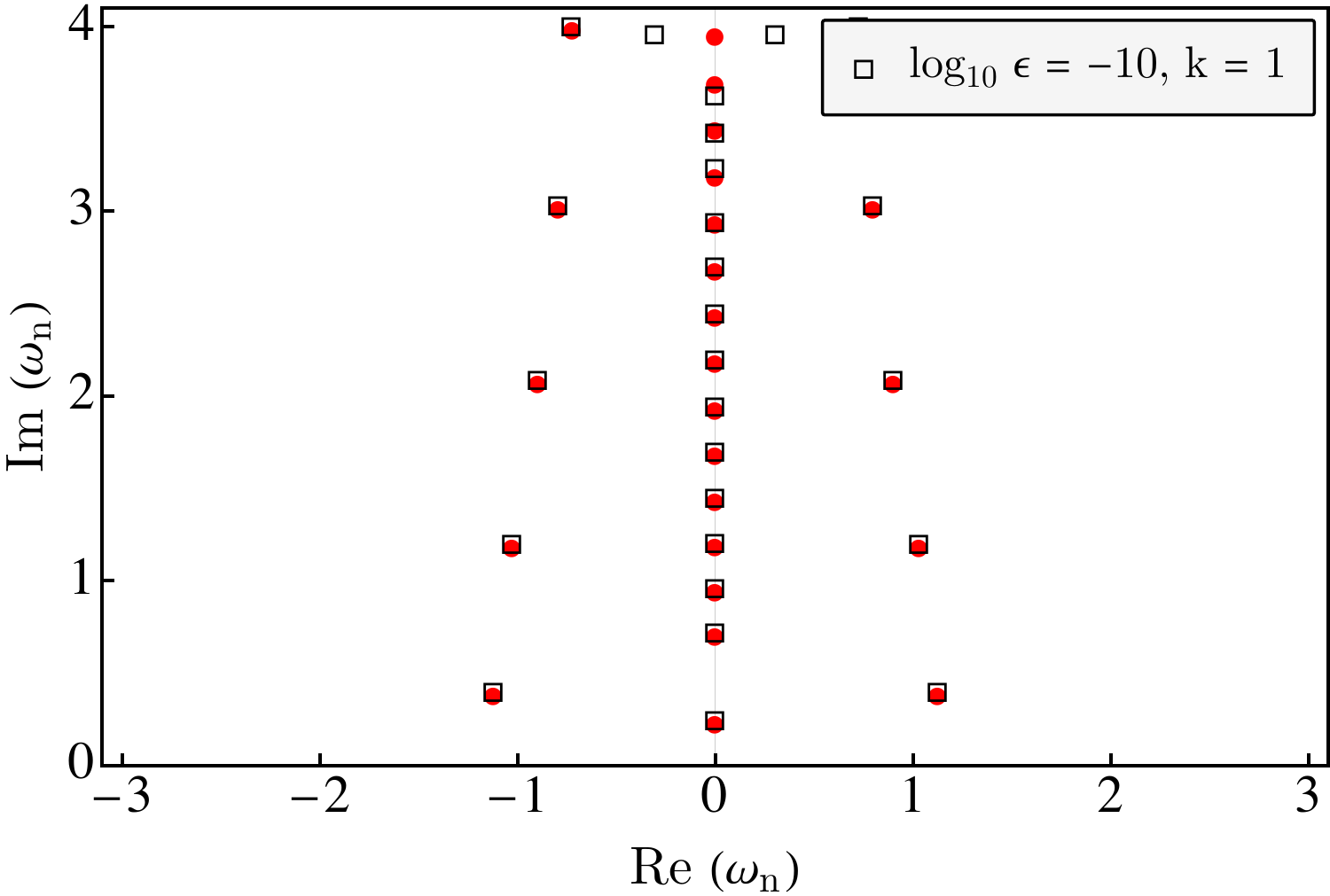}
	\endminipage\hfill
	\minipage{0.33\textwidth}
	\includegraphics[width=\linewidth]{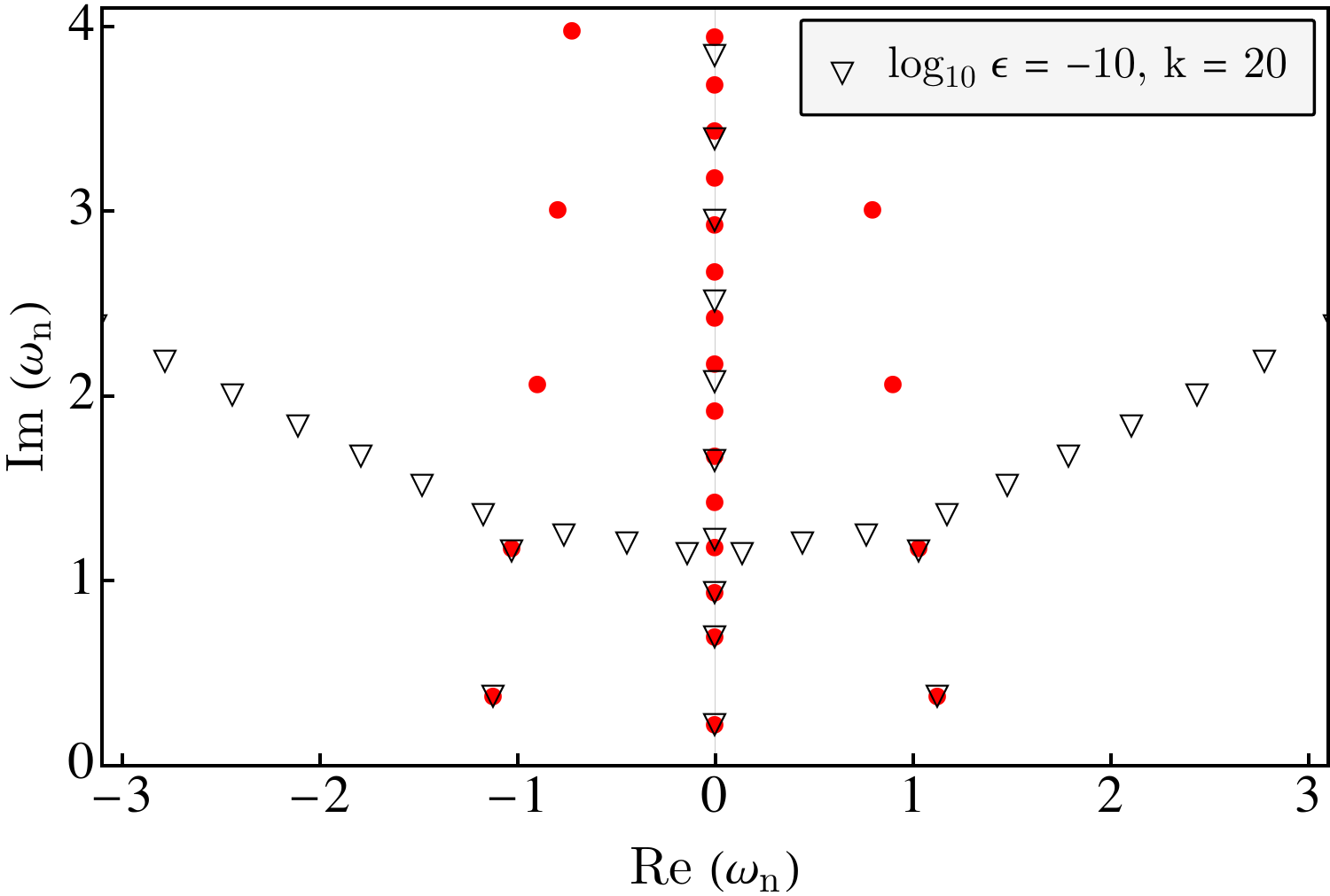}
	\endminipage\hfill
	\minipage{0.33\textwidth}
	\includegraphics[width=\linewidth]{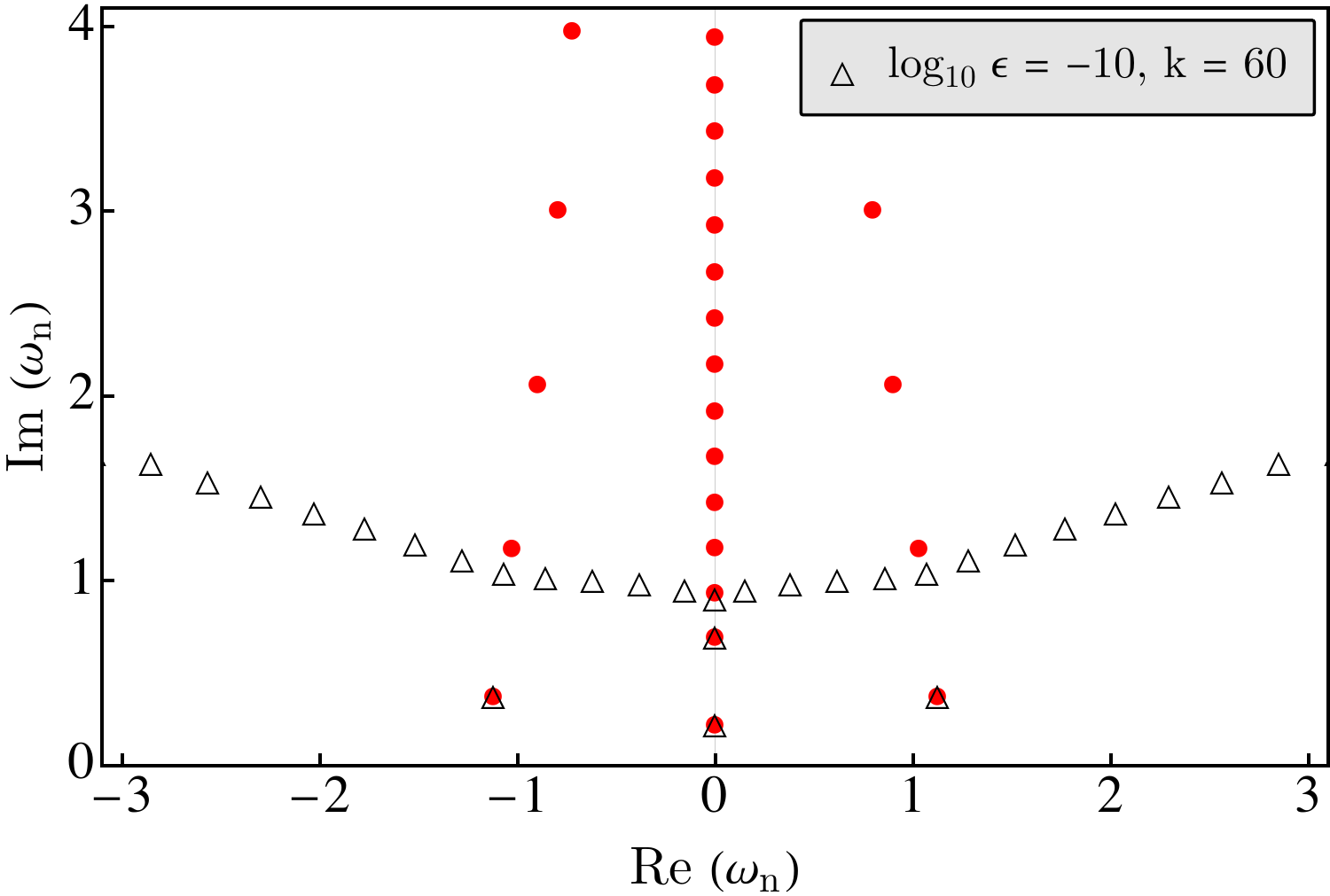}
	\endminipage\hfill
 \minipage{0.33\textwidth}
	\includegraphics[width=\linewidth]{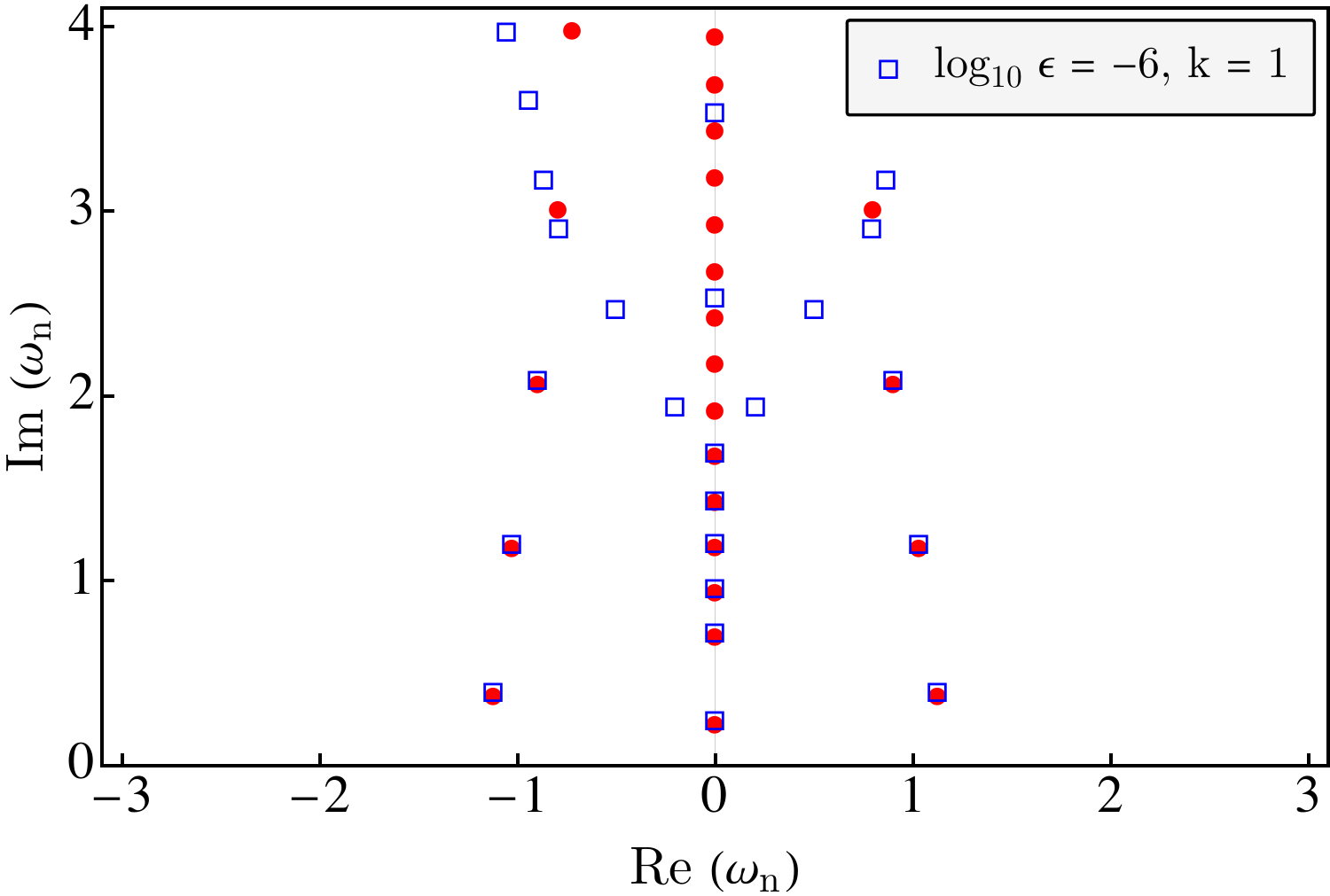}
	\endminipage\hfill
	\minipage{0.33\textwidth}
	\includegraphics[width=\linewidth]{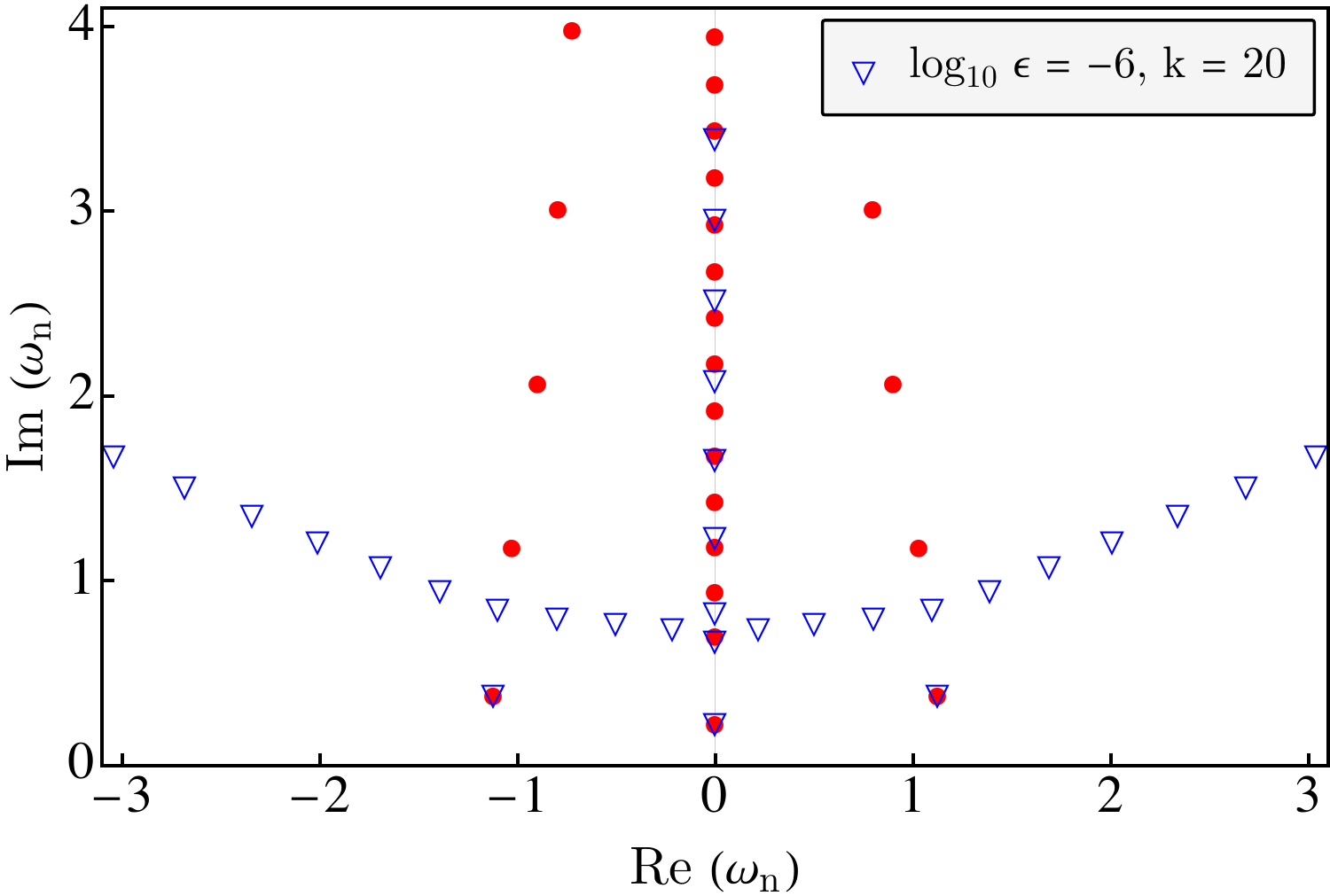}
	\endminipage\hfill
	\minipage{0.33\textwidth}
	\includegraphics[width=\linewidth]{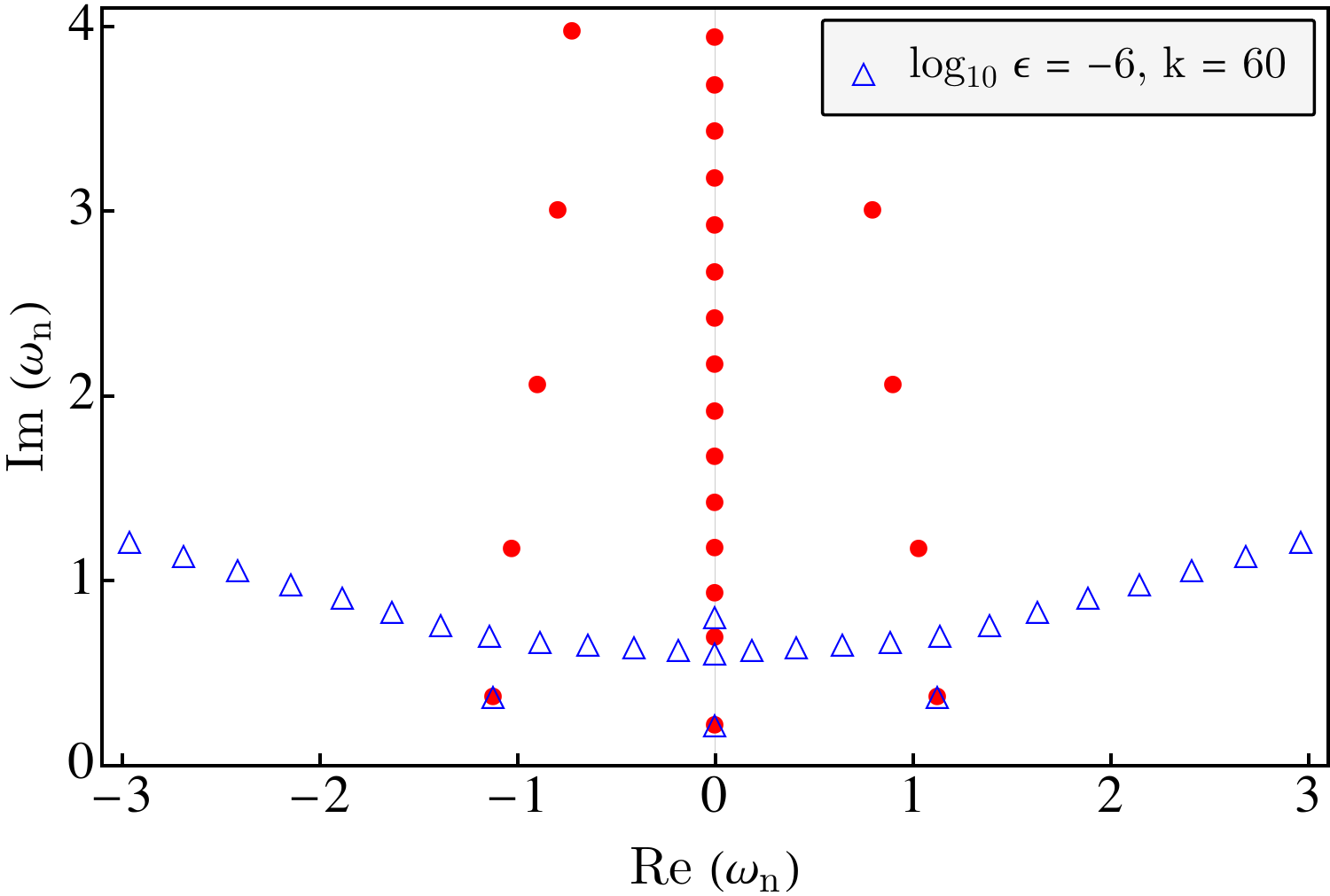}
	\endminipage\hfill
 \minipage{0.33\textwidth}
	\includegraphics[width=\linewidth]{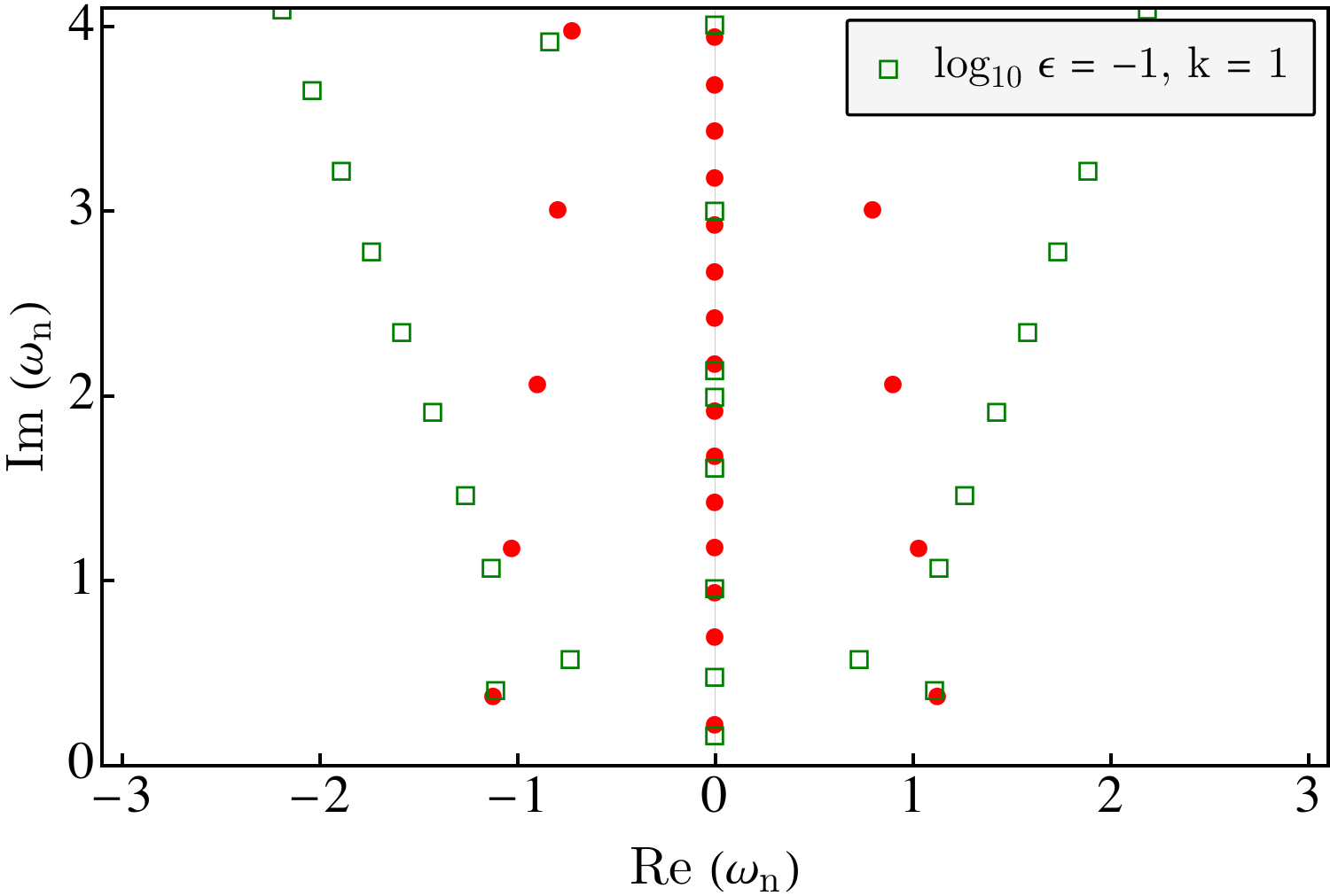}
	\endminipage\hfill
	\minipage{0.33\textwidth}
	\includegraphics[width=\linewidth]{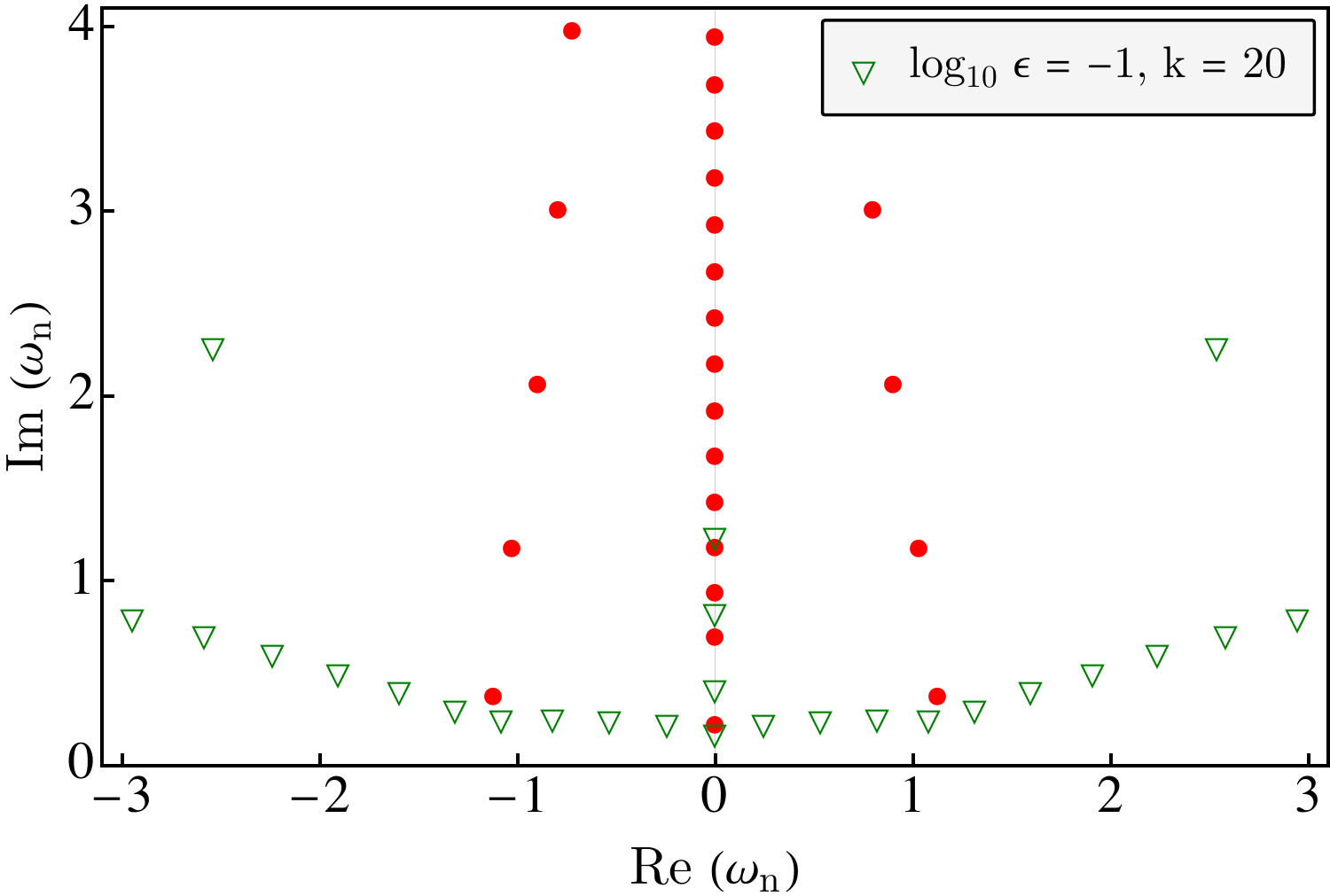}
	\endminipage\hfill
	\minipage{0.33\textwidth}
	\includegraphics[width=\linewidth]{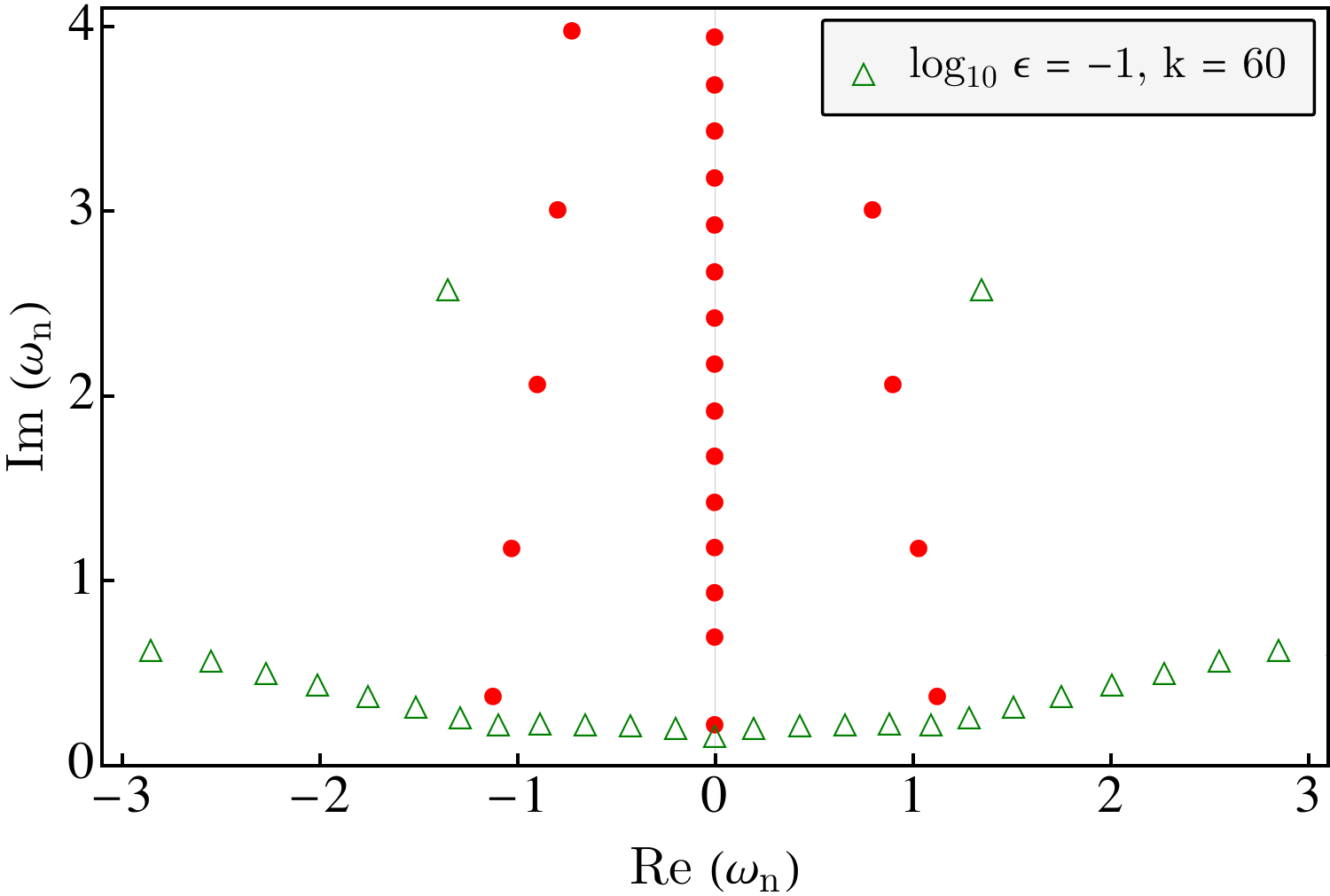}
	\endminipage
\caption{The spectra for perturbed QNM frequencies associated with deterministic perturbations with norm $||\delta V_d||={\epsilon}$ for $\ell=1$ \emph{scalar} modes ($s=0$) of a Schwarzschild de Sitter black hole with $M = 1$ and $\Lambda = 0.01$ have been presented, superimposed over the unperturbed QNM frequencies (indicated by red dots). The characteristic length scale has been set to $\lambda = 2r_+$.}
\label{fig:SdS_DET_PERT_SCALAR}
\end{figure}	

One can also consider adding random perturbations $\delta V_r(\sigma)$ to the scattering potential $V(\sigma)$: this amounts to drawing $N$ random samples from a Gaussian distribution and adding each of them to a point on the collocation grid of size $N$. This array of $N+1$ points is then converted to a diagonal matrix $\delta V_r(\sigma)$ and added to $L$ after being normalized using the energy norm. Such perturbations correspond to high-frequency perturbations by design. Random perturbations are routinely used to investigate the spectral properties of matrices \cite{Trefethen2005_pseudospectrumbook}. However, their utility is severely restricted in asymptotically flat spacetimes due to the presence of non-converging branch cut modes along the imaginary axis. The branch-cut modes manifest themselves as the ubiquitous Price's tails in the late-time decay profile of the scattered fields. But the spectra of asymptotically de Sitter black holes do not face this ``limitation", the modes which lie on the imaginary axis correspond to the so-called de Sitter modes and are convergent. Hence one can make full use of random perturbations as well to investigate the spectral stability of such spacetimes.  

It is also interesting to note that while usual investigations into the spectral stability of matrices make modifications to the whole operator, we restrict ourselves to modifications to the black hole potential (similar to \cite{Jaramillo:2020tuu,Boyanov:2022ark}). {Although the pseudospectrum indicates the presence of overall spectral instability, it does not contain any information about the nature of the perturbations that could trigger those instabilities, and it most certainly does not indicate that the spectrum will be unstable under a perturbation of the form as in \ref{e:delta_L_operator}. In this respect, the behavior of the perturbed QNMs in the presence of the perturbative probes likely corresponds to a physical effect and therefore complements the study of black hole pseudospectra.} 

\subsection{Deterministic perturbations to the scattering potential of a Schwarzschild de Sitter black hole}

We first describe the effect of deterministic oscillatory perturbations to the $\ell = 1$ scalar ($s =0$) potential of a Schwarzschild de Sitter black hole with $\Lambda = 0.01$: the results have been summarized in \ref{fig:SdS_DET_PERT_SCALAR}. We note that for a fixed $\epsilon$, increasing the wave number $k$ intensifies the instability of the QNM spectrum. For example, in the first row of \ref{fig:SdS_DET_PERT_SCALAR} corresponding to $||\delta V_d = 10^{-10}||$, we see that for a low value of $k$, the spectrum is essentially unperturbed, but as we increase $k$, a wave of instability begins to travel down the spectrum, and for $k=60$, it reaches the second set of the photon sphere modes. The new perturbed QNMs also line up in an orderly fashion along the aforementioned Nollert-Price branches and from \ref{fig:pseudospectra_scalar_l1_random_det_contour} we can confirm that these branches closely track the pseudospectral contour lines. This behavior is perfectly analogous to what is observed in asymptotically flat black holes. However, this instability is not expected to reach the fundamental mode since it lies well below the $\log_{10} \epsilon = - 10$ contour. We have observed this effect numerically to a limited degree since increasing $k$ further makes it extremely difficult to compute the perturbed QNMs with the resolution that we have used. 

However, one witnesses a rather curious behavior as one increases the strength of the perturbation. Consider the second row of \ref{fig:SdS_DET_PERT_SCALAR}, for an intermediate value of $k=20$, we progressively increase the strength of the perturbation $\epsilon$ and observe the gradual intensification of the spectral instability. It is extremely interesting to note that for the largest reasonable perturbation, that is, $\epsilon=10^{-1}$, the fundamental mode itself is destabilized! This situation is markedly different from what has been observed for asymptotically flat black holes in \cite{Jaramillo:2020tuu}, where the fundamental mode was always found to be spectrally stable and could only be dislodged by an $\mathcal{O}(1)$ modification to the scattering potential. Therefore one can conclude that the perturbed QNMs and the pseudospectra have the ability to capture the asymptotic structure of spacetime in addition to describing the near-horizon physics of black holes. However it should be noted that the spectral instability of the fundamental mode is by no means a definitive signature of asymptotically de Sitter black holes: in exotic compact objects as well, the fundamental mode is prone to destabilization against strong ($\epsilon \sim 10^{-2}$) high-frequency perturbations \cite{Boyanov:2022ark}. But the pseudospectra of an exotic compact object and a Schwarzschild de Sitter black hole are evidently different and hence the structure of the Nollert-Price branches are different as well. 

Furthermore, in asymptotically flat spacetimes, perturbations with a low wave number $ k \sim 1$ are considered to be benign in the sense that their presence leaves the spectrum unchanged \cite{Jaramillo:2020tuu}. But for asymptotically de Sitter black holes, we observe that deterministic perturbations with $k=1$ can modify the spectrum and the degree of the destabilization increases with $\epsilon$. However, such perturbations do not dramatically alter the fundamental mode. 

\begin{figure}[tbh!]
	\centering
	\minipage{0.33\textwidth}
	\includegraphics[width=\linewidth]{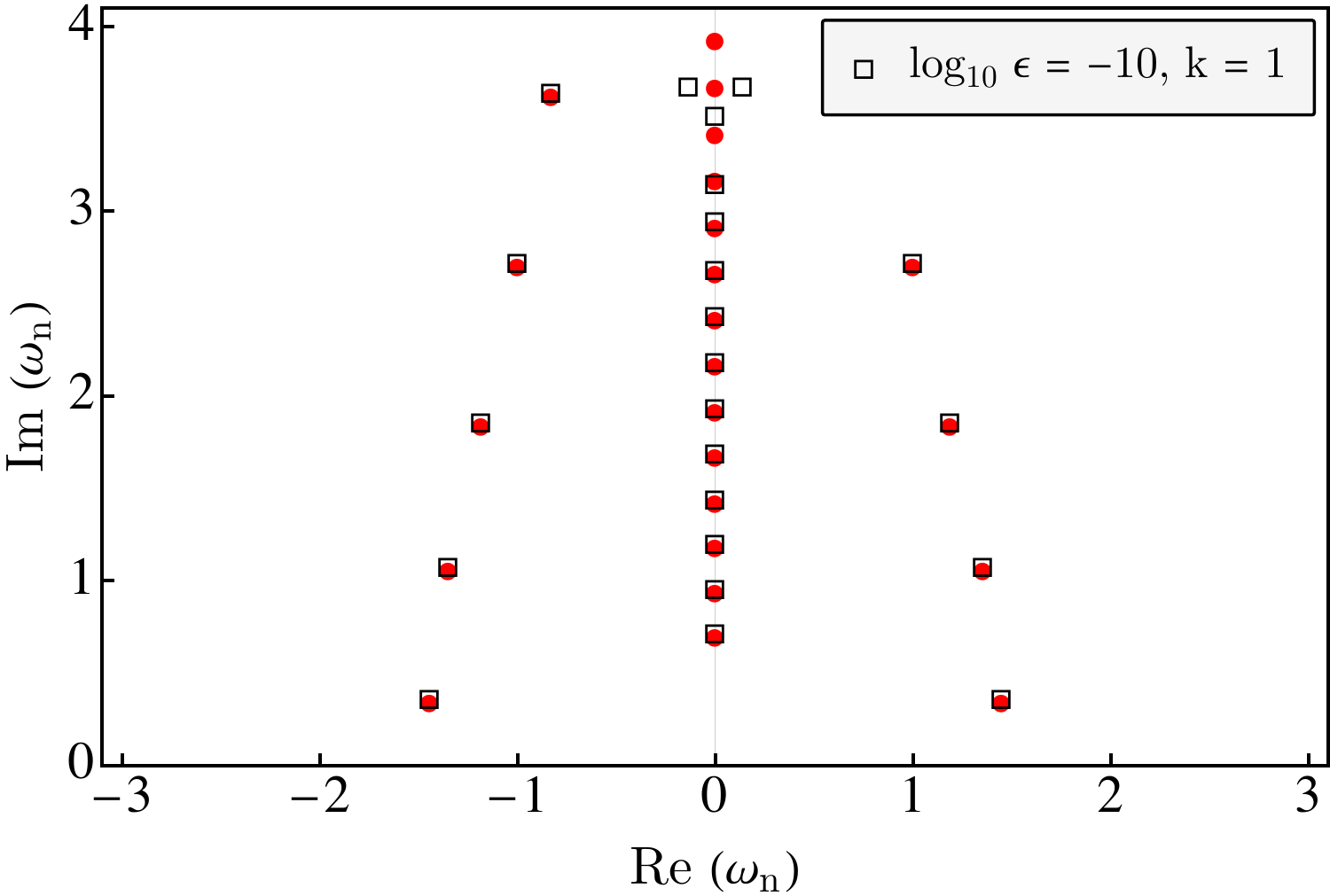}
	\endminipage\hfill
	\minipage{0.33\textwidth}
	\includegraphics[width=\linewidth]{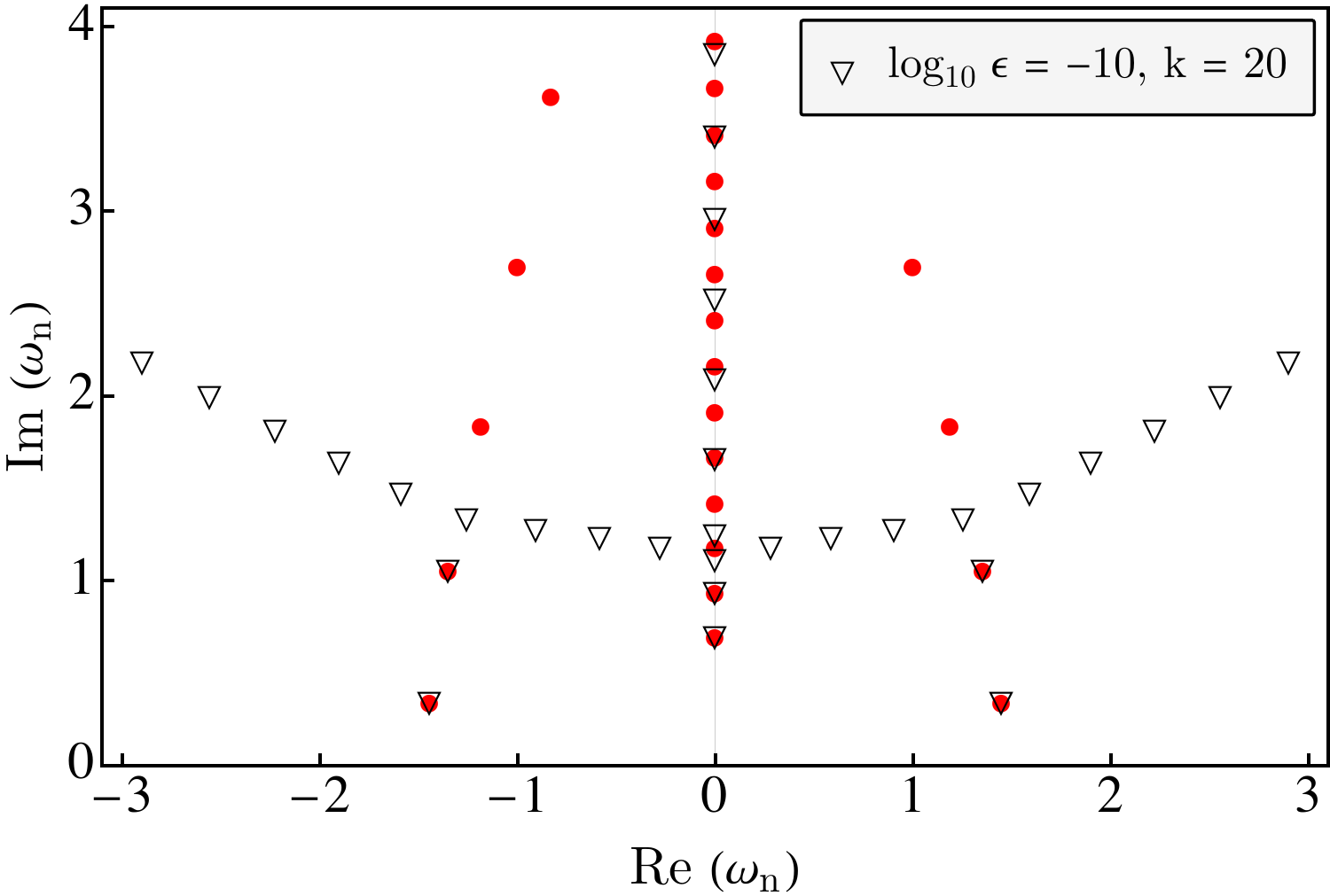}
	\endminipage\hfill
	\minipage{0.33\textwidth}
	\includegraphics[width=\linewidth]{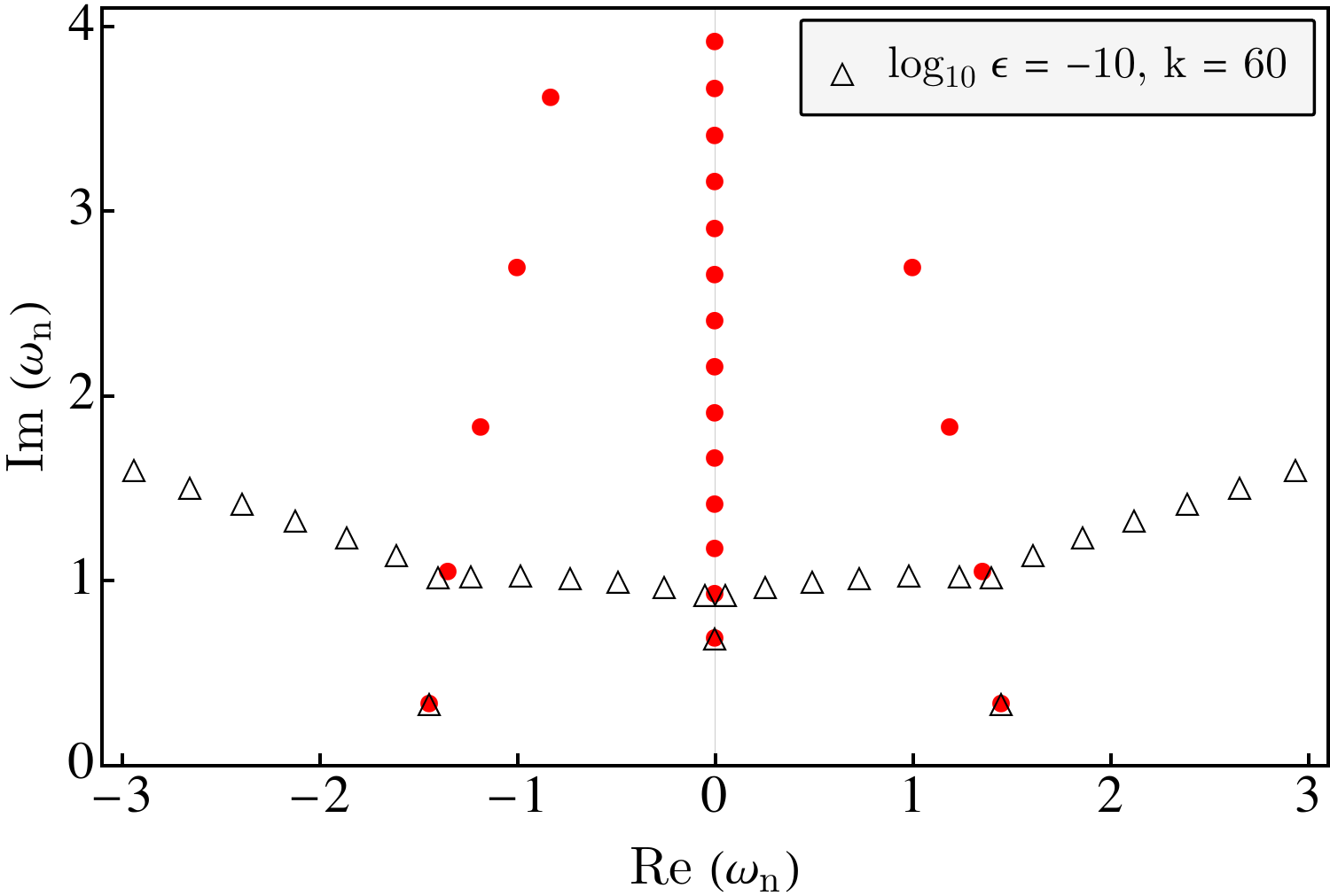}
	\endminipage\hfill
 \minipage{0.33\textwidth}
	\includegraphics[width=\linewidth]{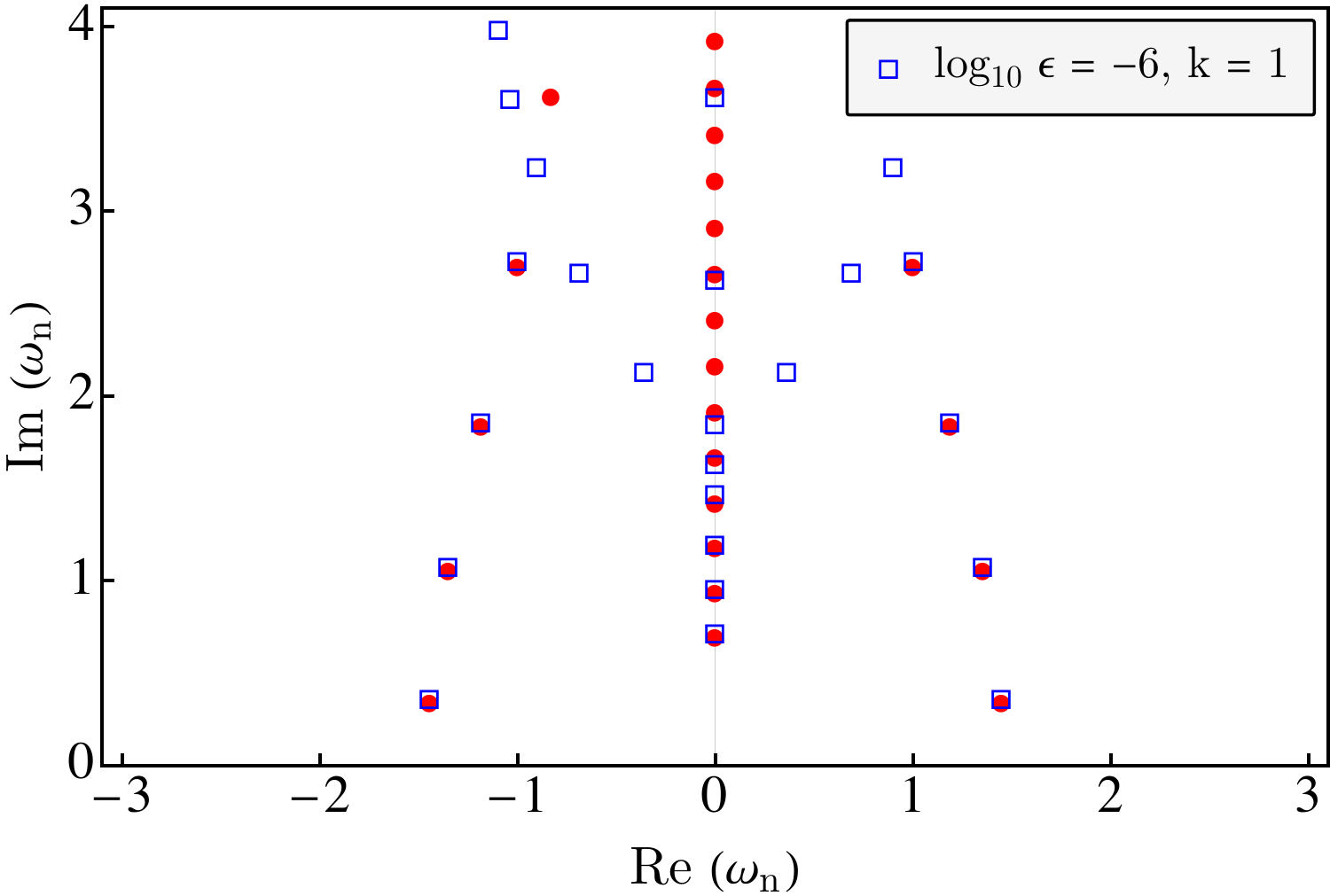}
	\endminipage\hfill
	\minipage{0.33\textwidth}
	\includegraphics[width=\linewidth]{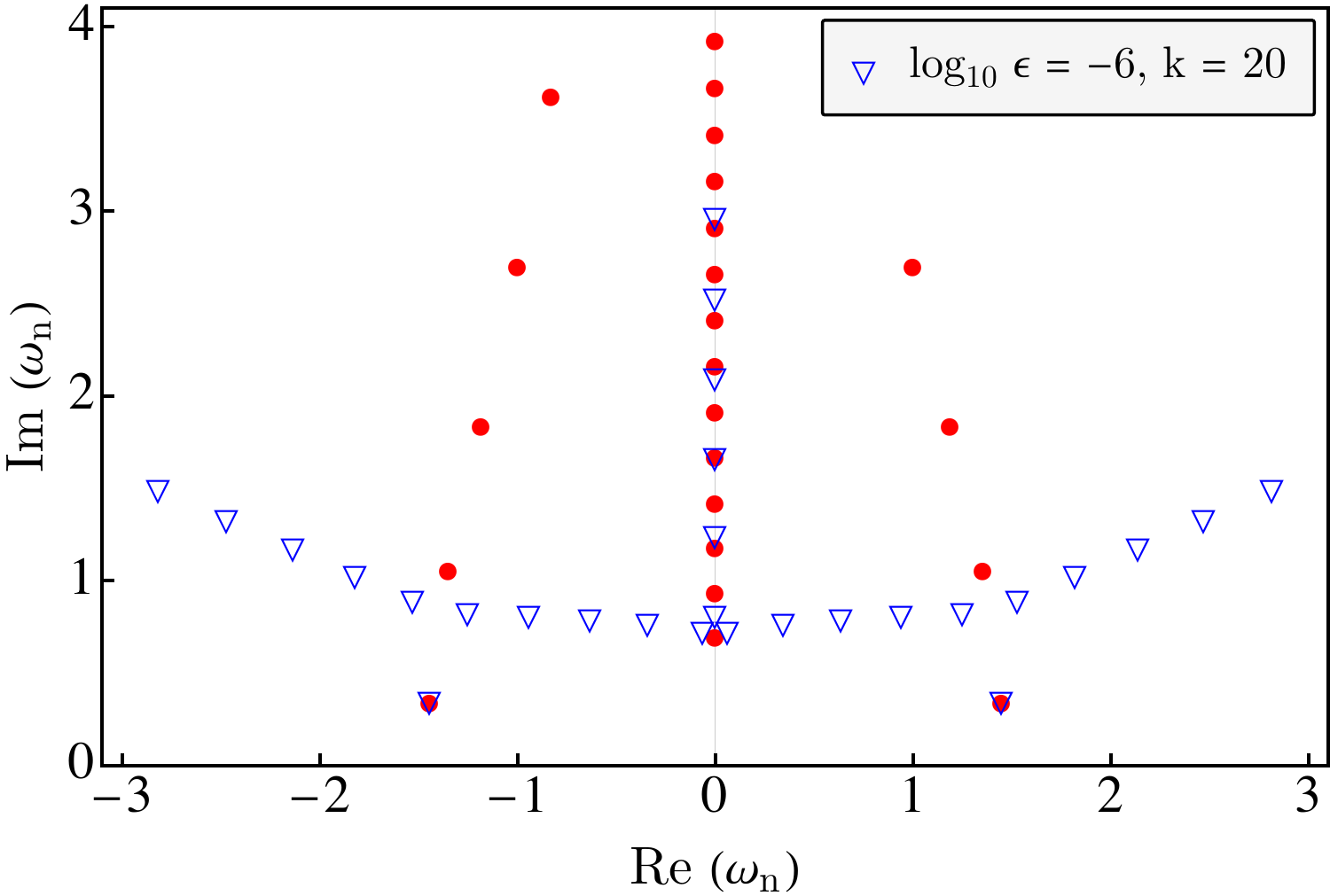}
	\endminipage\hfill
	\minipage{0.33\textwidth}
	\includegraphics[width=\linewidth]{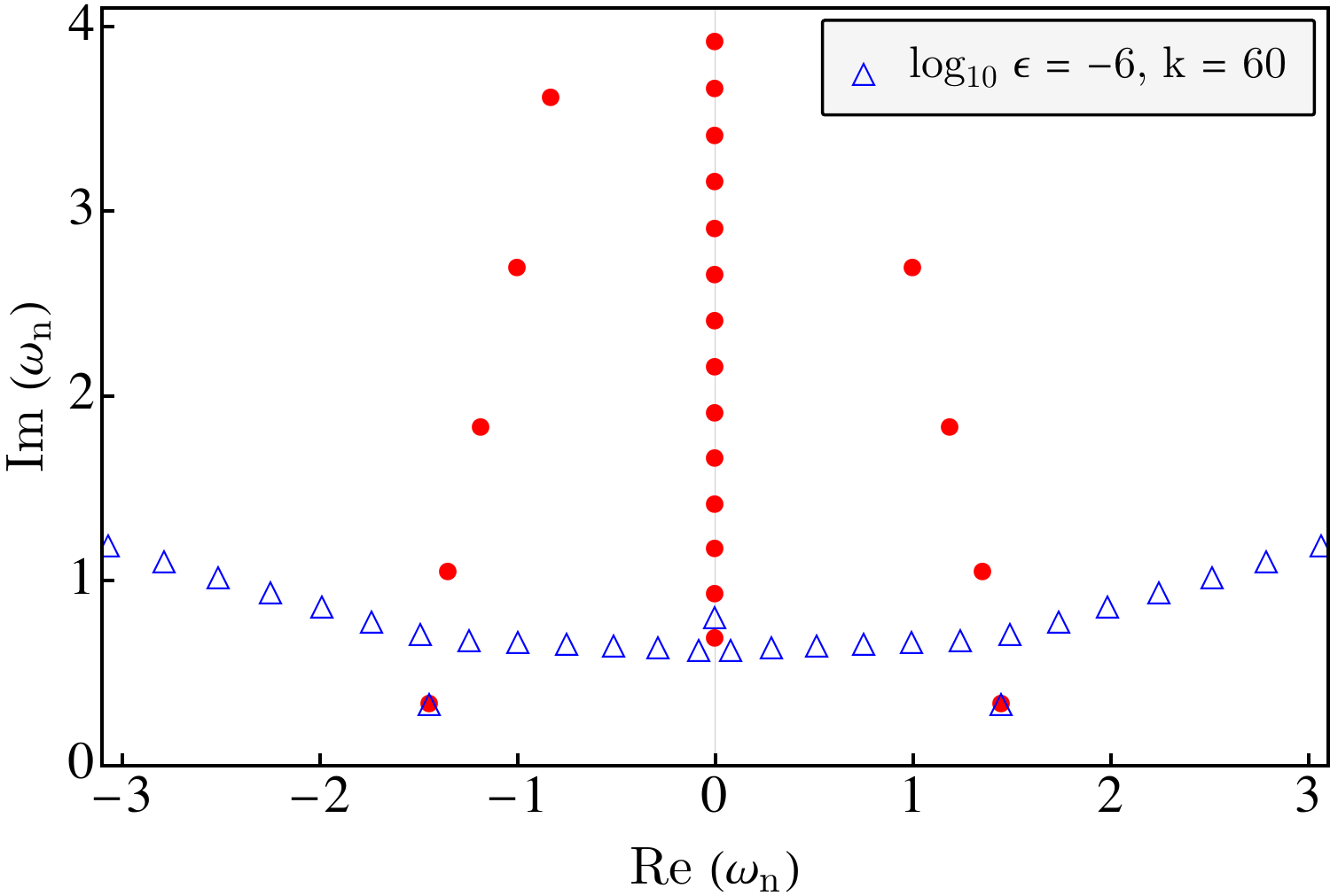}
	\endminipage\hfill
 \minipage{0.33\textwidth}
	\includegraphics[width=\linewidth]{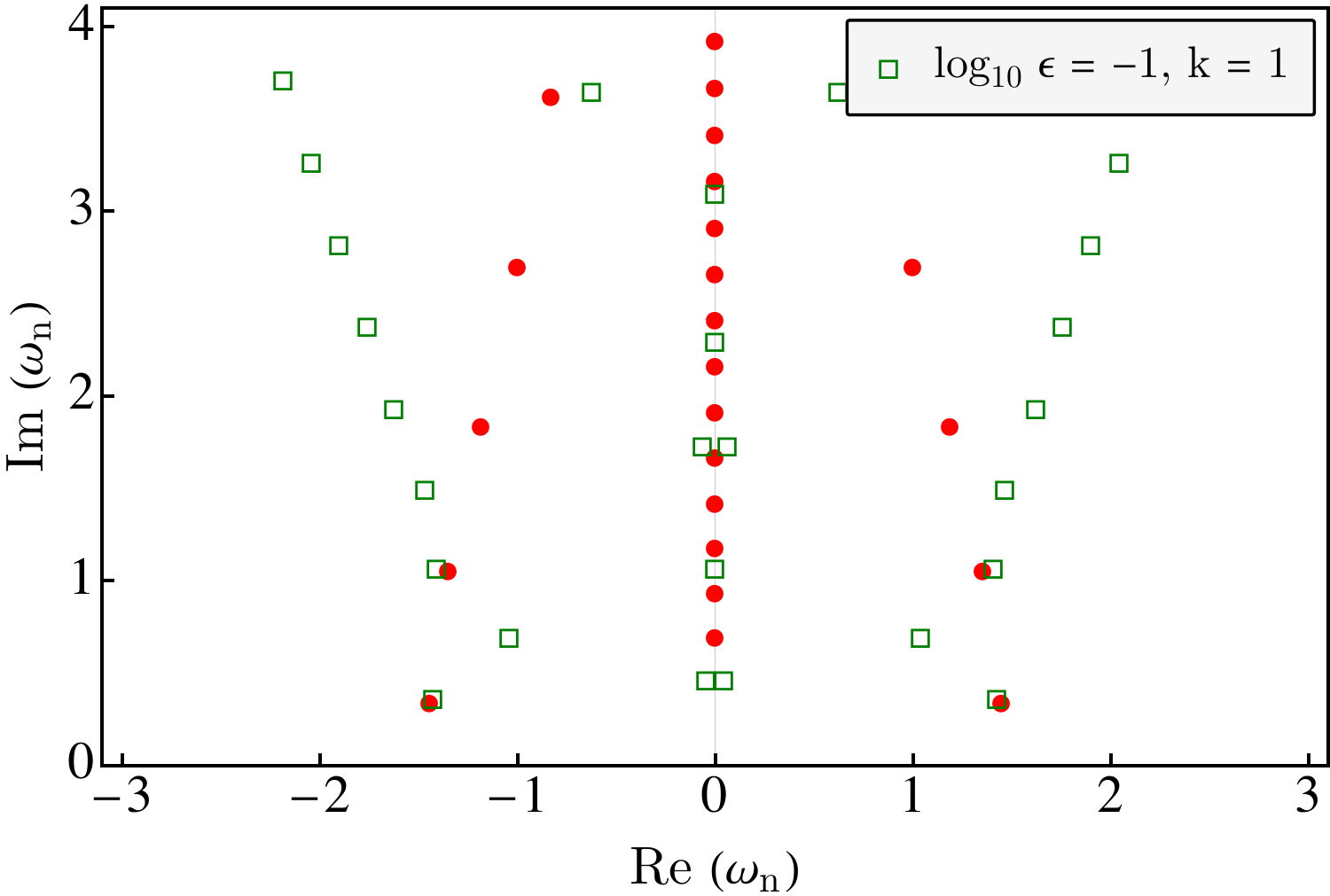}
	\endminipage\hfill
	\minipage{0.33\textwidth}
	\includegraphics[width=\linewidth]{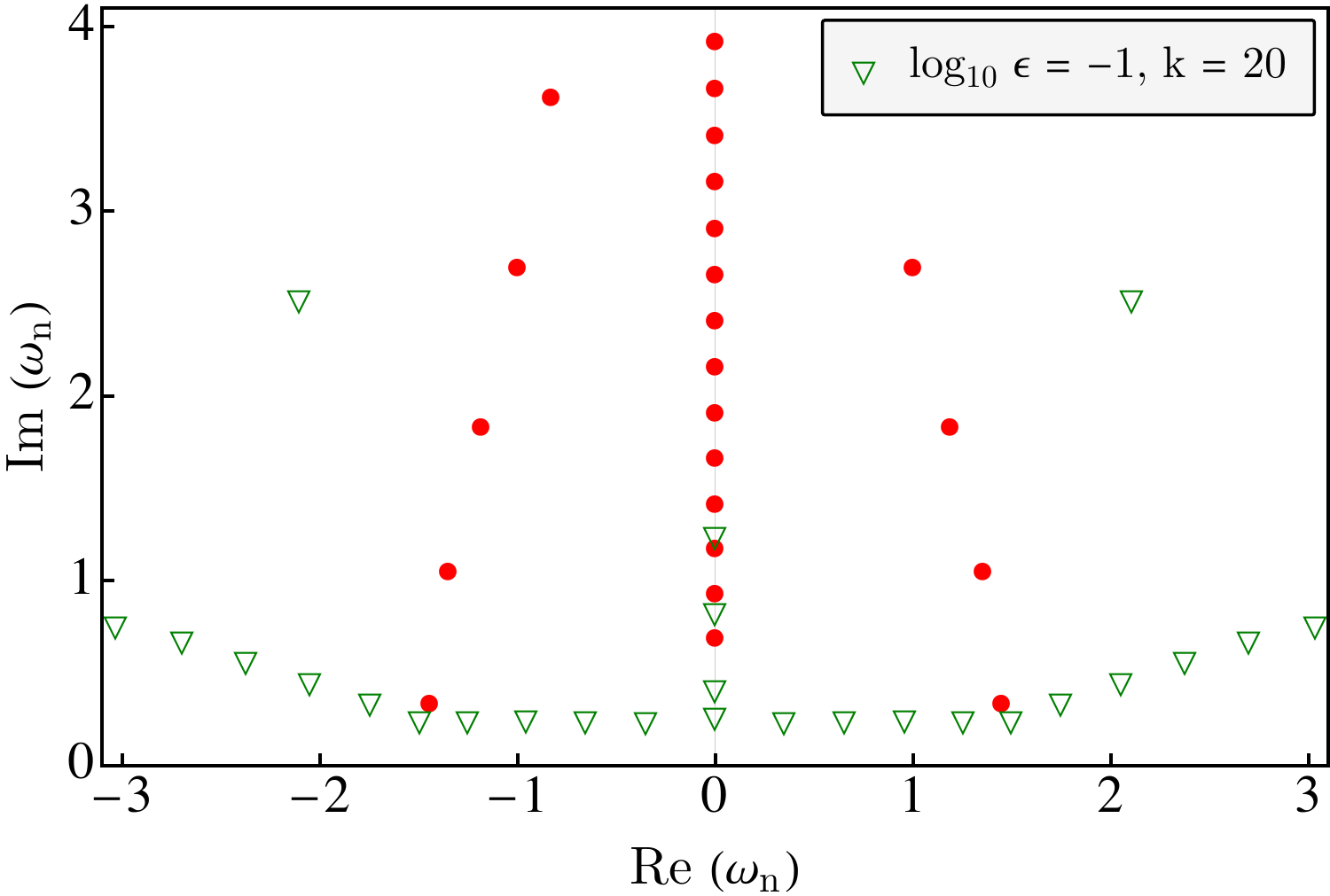}
	\endminipage\hfill
	\minipage{0.33\textwidth}
	\includegraphics[width=\linewidth]{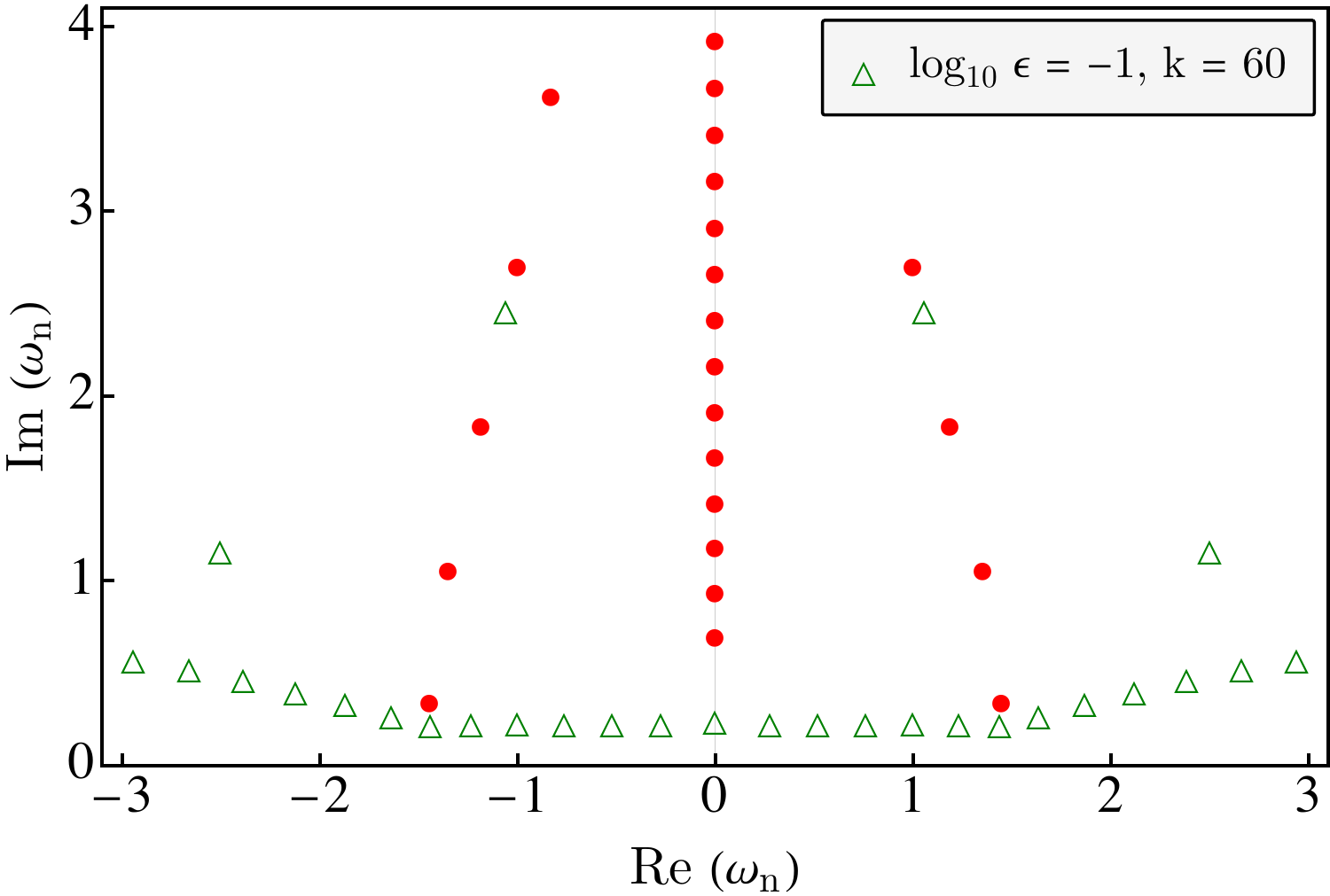}
	\endminipage
\caption{The QNM spectra for deterministic perturbations of norm $||\delta V_d||={\epsilon}$ for $\ell=2$ \emph{gravitational} perturbation ($s=2$) of a Schwarzschild de Sitter black hole with $M = 1$ and $\Lambda = 0.01$ have been superimposed over the unperturbed QNM values (indicated in red). The characteristic length scale has been set to $\lambda = 2 r_+$.}\label{fig:SdS_DET_PERT_GRAV}
\end{figure}	

The corresponding results for $\ell=2$ (axial) gravitational ($s = 2$) perturbations have been summarized in \ref{fig:SdS_DET_PERT_GRAV}. In this case as well, for large deterministic perturbation with $k=60$, the fundamental QNM frequency gets shifted from its unperturbed position. Moreover for such large a perturbation with $k=1$, the frequency overtones get significantly disturbed, though the fundamental mode remains stable. Lastly, \ref{fig:SdS_DET_PERT_GRAV} also nicely demonstrates how the instability progress to lower and lower modes with the increases in the strength of the perturbation. This trend remains for all choices of the inverse perturbation length scale $k$. Thus instability of the fundamental QNM frequency for large perturbation is a generic feature of asymptotically de Sitter spacetime as it holds for scalar as well as for gravitational perturbation. 


\begin{figure}[tbh!]
	\centering
	\minipage{0.33\textwidth}
	\includegraphics[width=\linewidth]{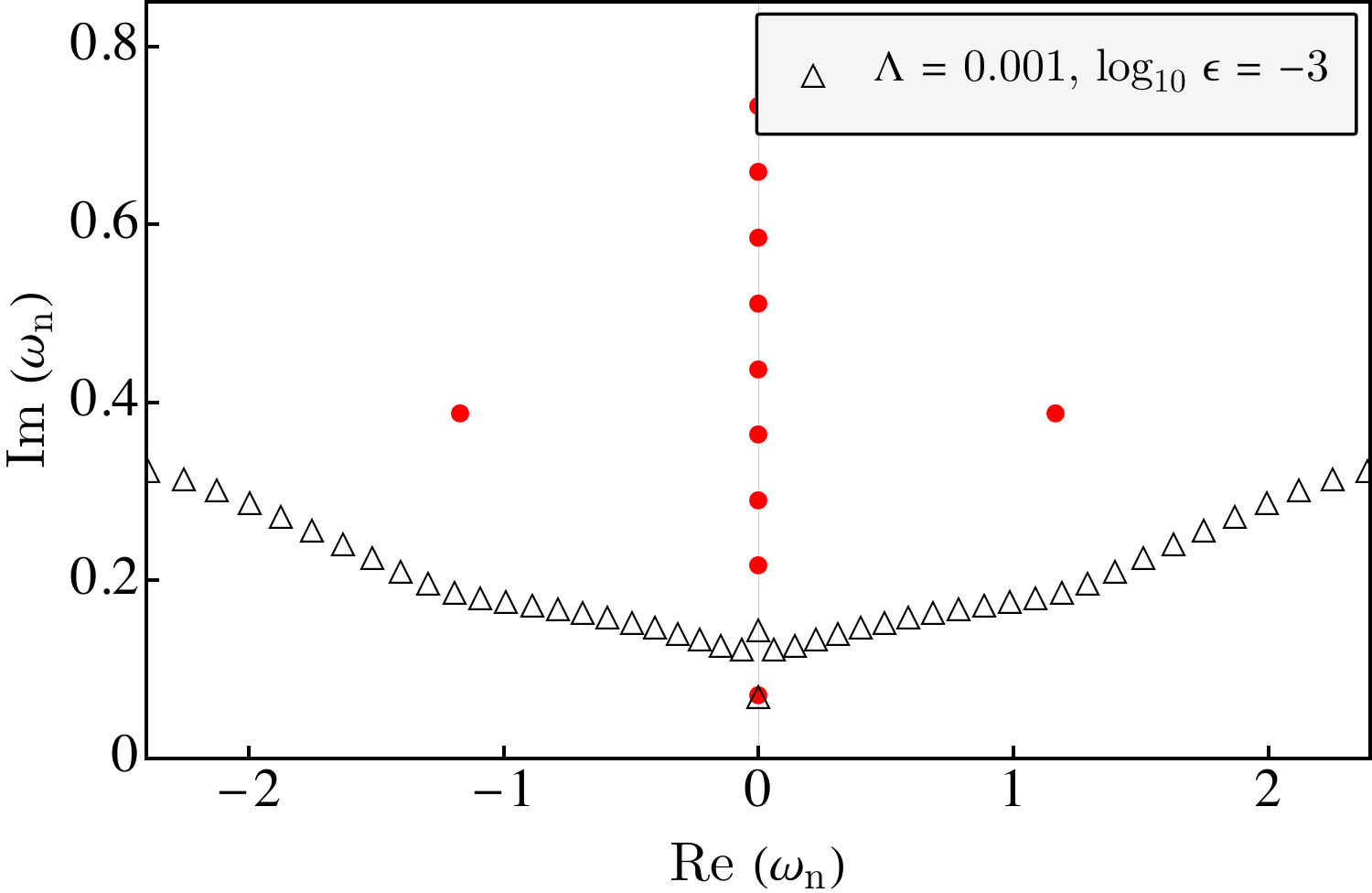}
	\endminipage\hfill
	\minipage{0.33\textwidth}
	\includegraphics[width=\linewidth]{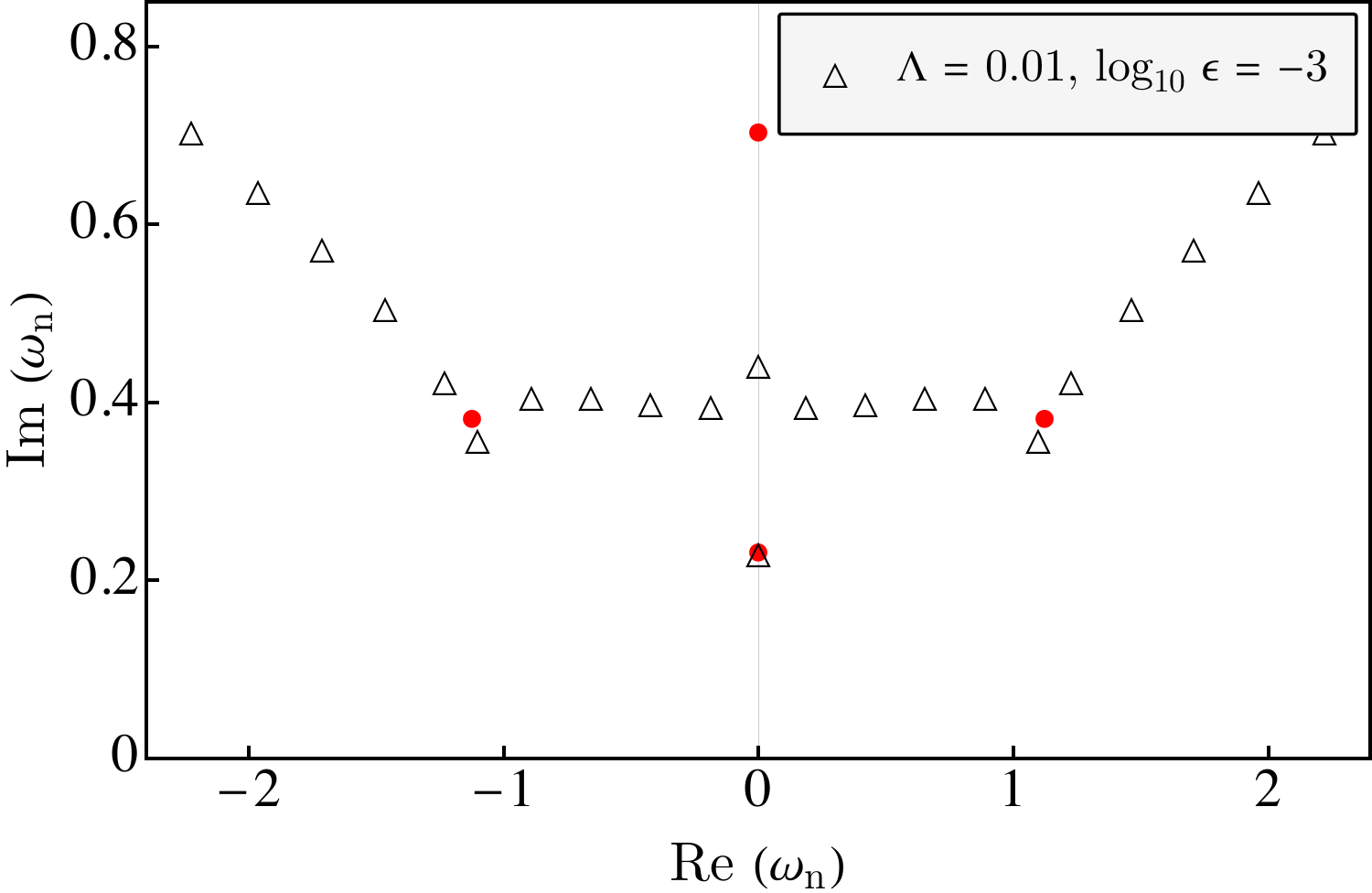}
	\endminipage\hfill
	\minipage{0.33\textwidth}
	\includegraphics[width=\linewidth]{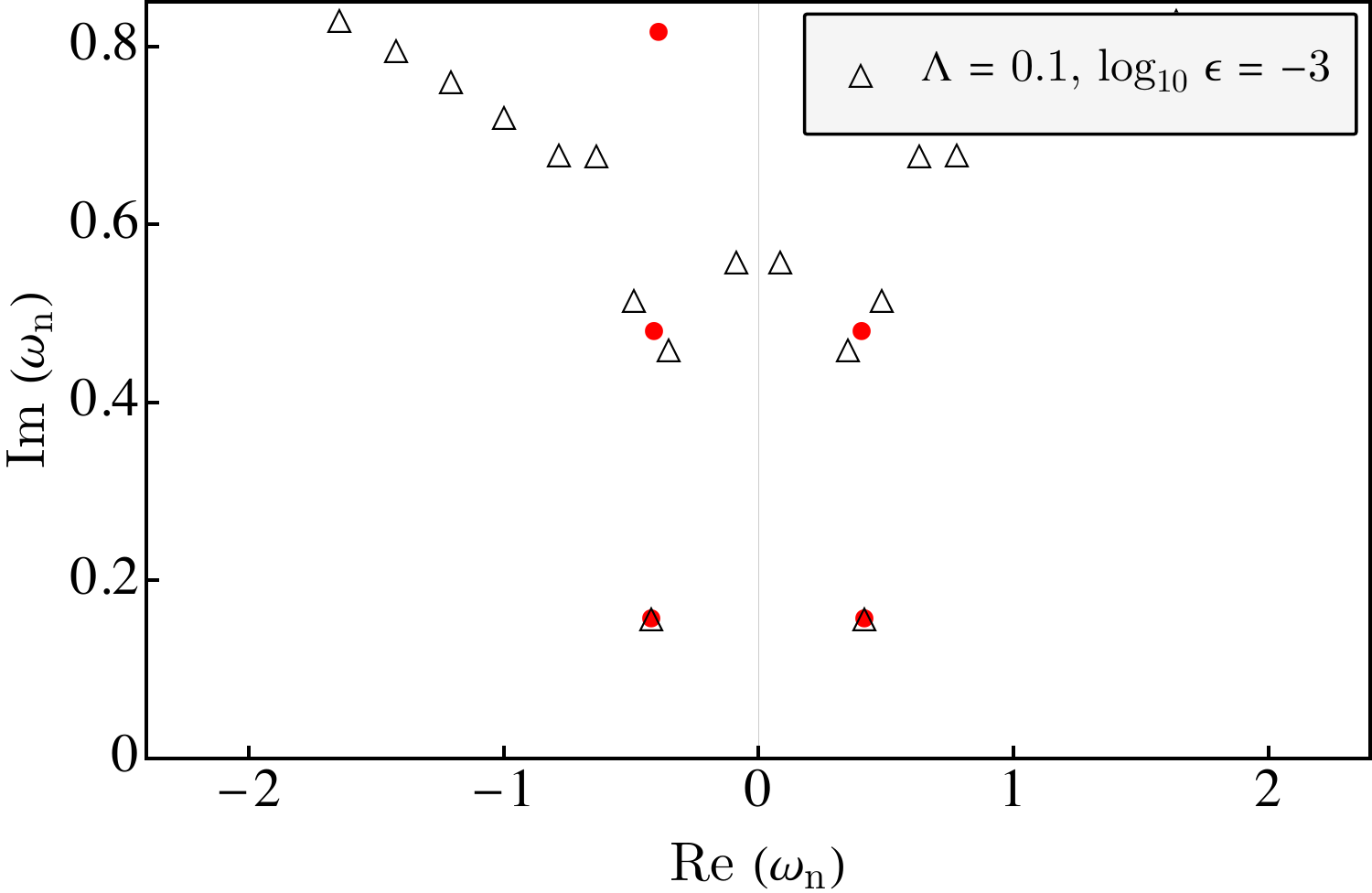}
	\endminipage\hfill
 \minipage{0.33\textwidth}
	\includegraphics[width=\linewidth]{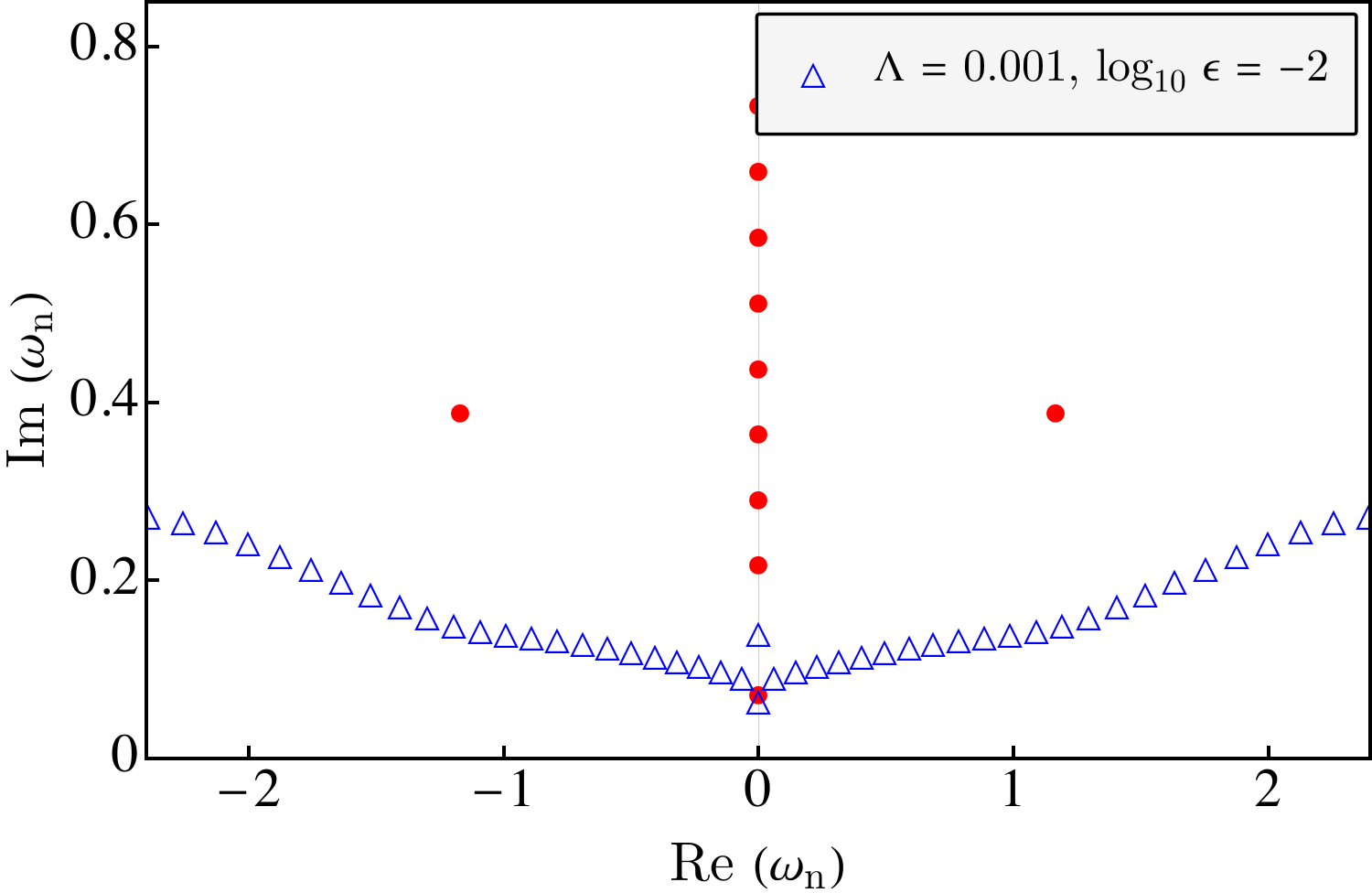}
	\endminipage\hfill
	\minipage{0.33\textwidth}
	\includegraphics[width=\linewidth]{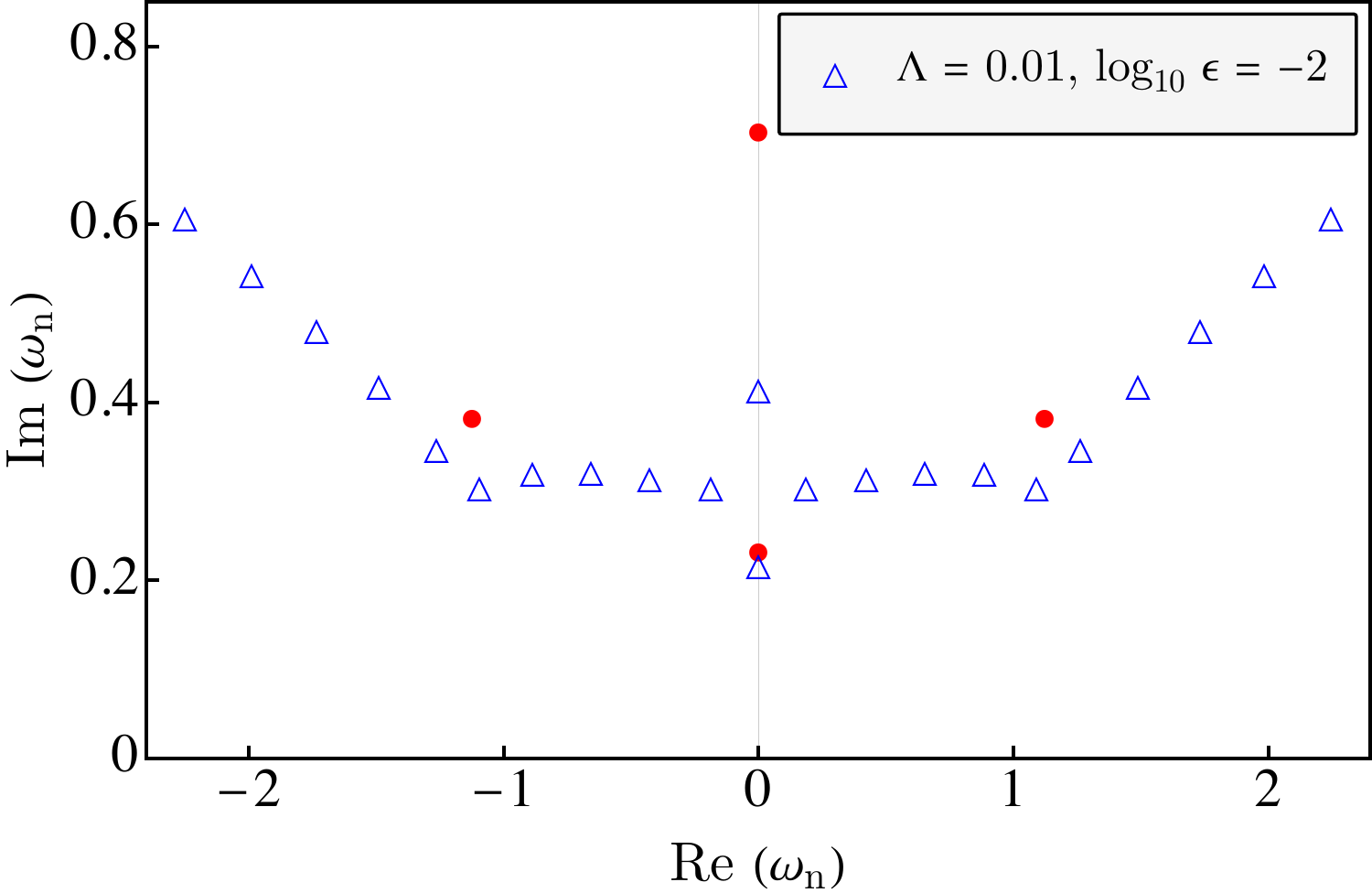}
	\endminipage\hfill
	\minipage{0.33\textwidth}
	\includegraphics[width=\linewidth]{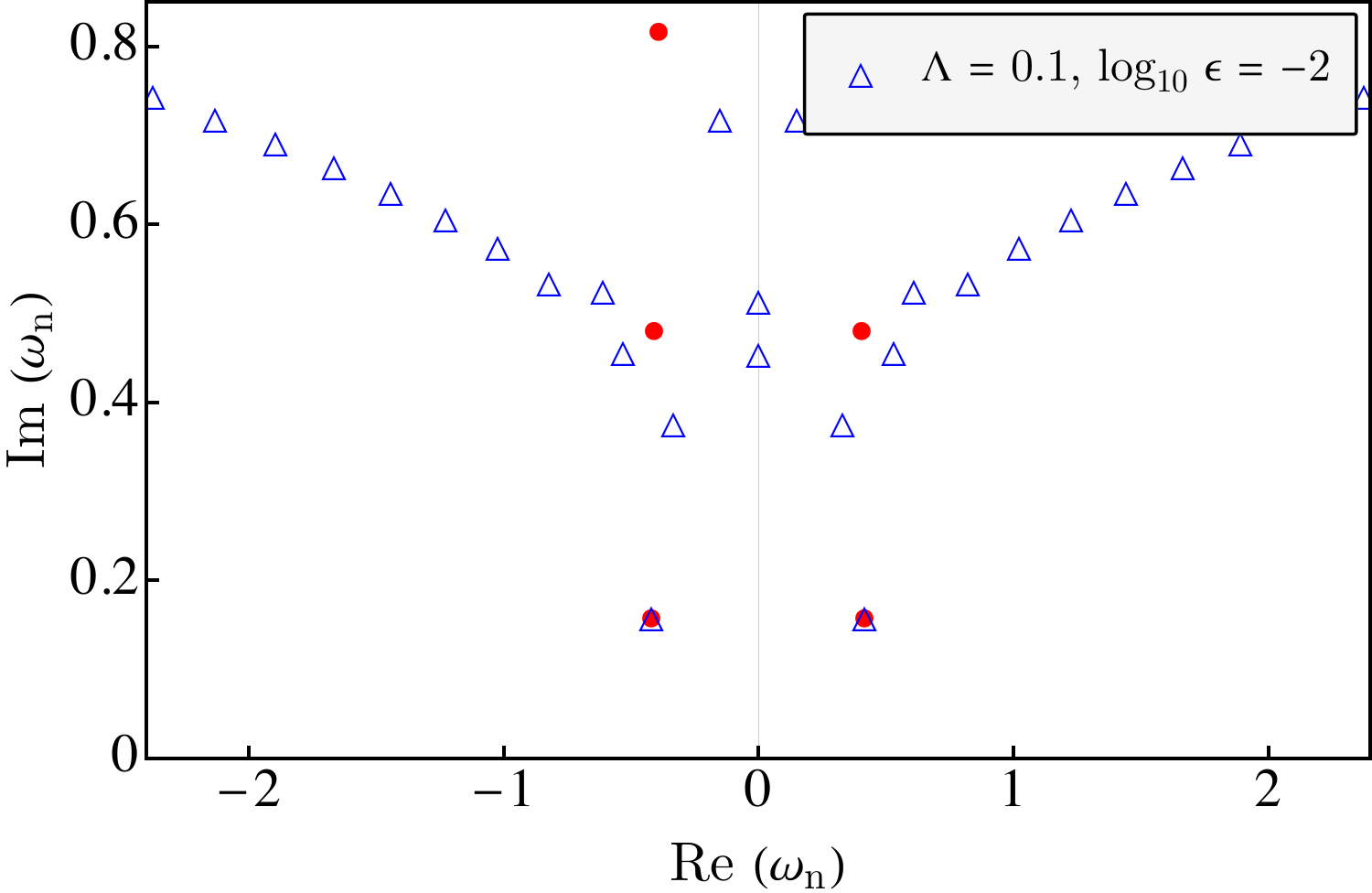}
	\endminipage\hfill
 \minipage{0.33\textwidth}
	\includegraphics[width=\linewidth]{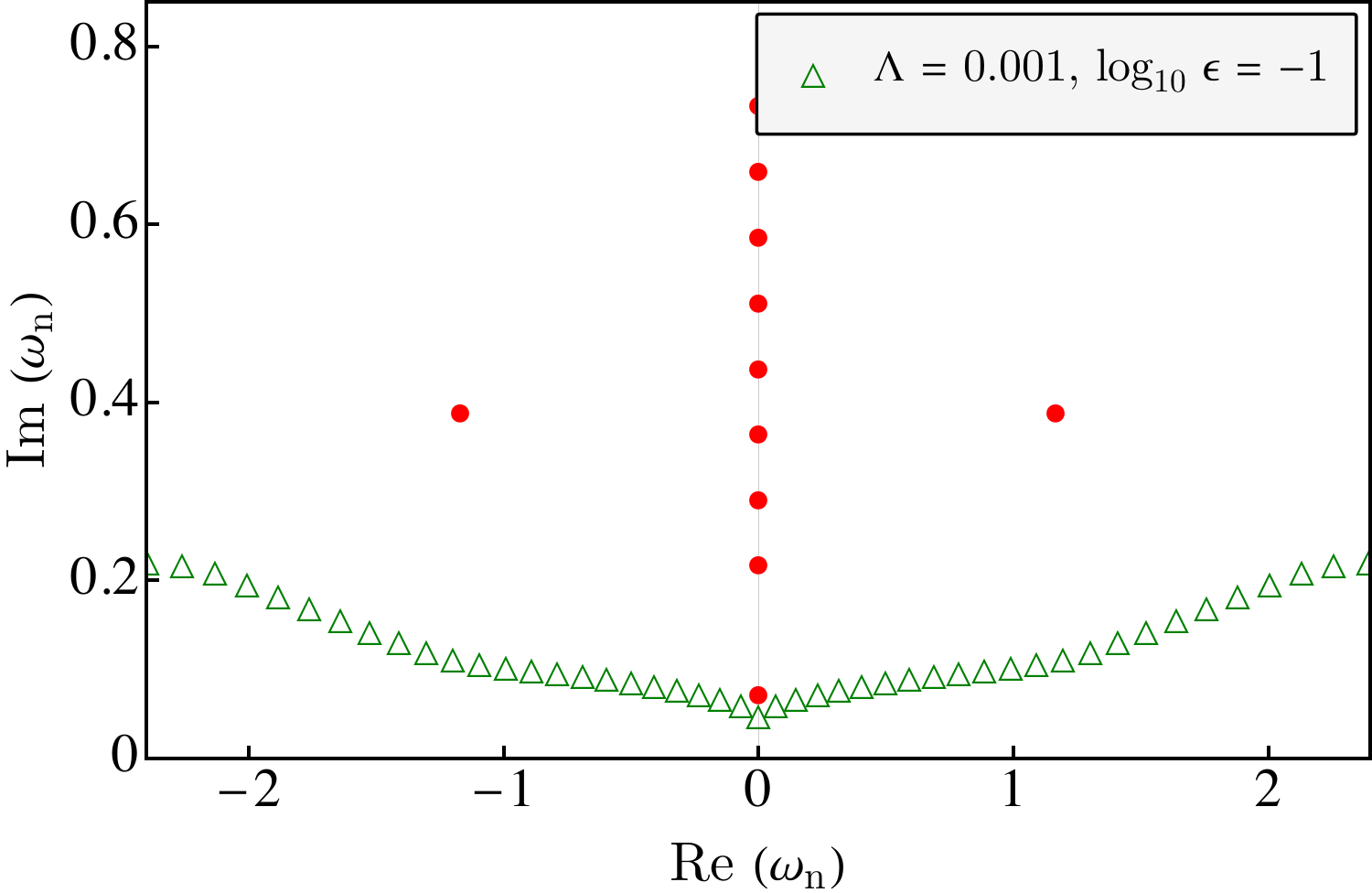}
	\endminipage\hfill
	\minipage{0.33\textwidth}
	\includegraphics[width=\linewidth]{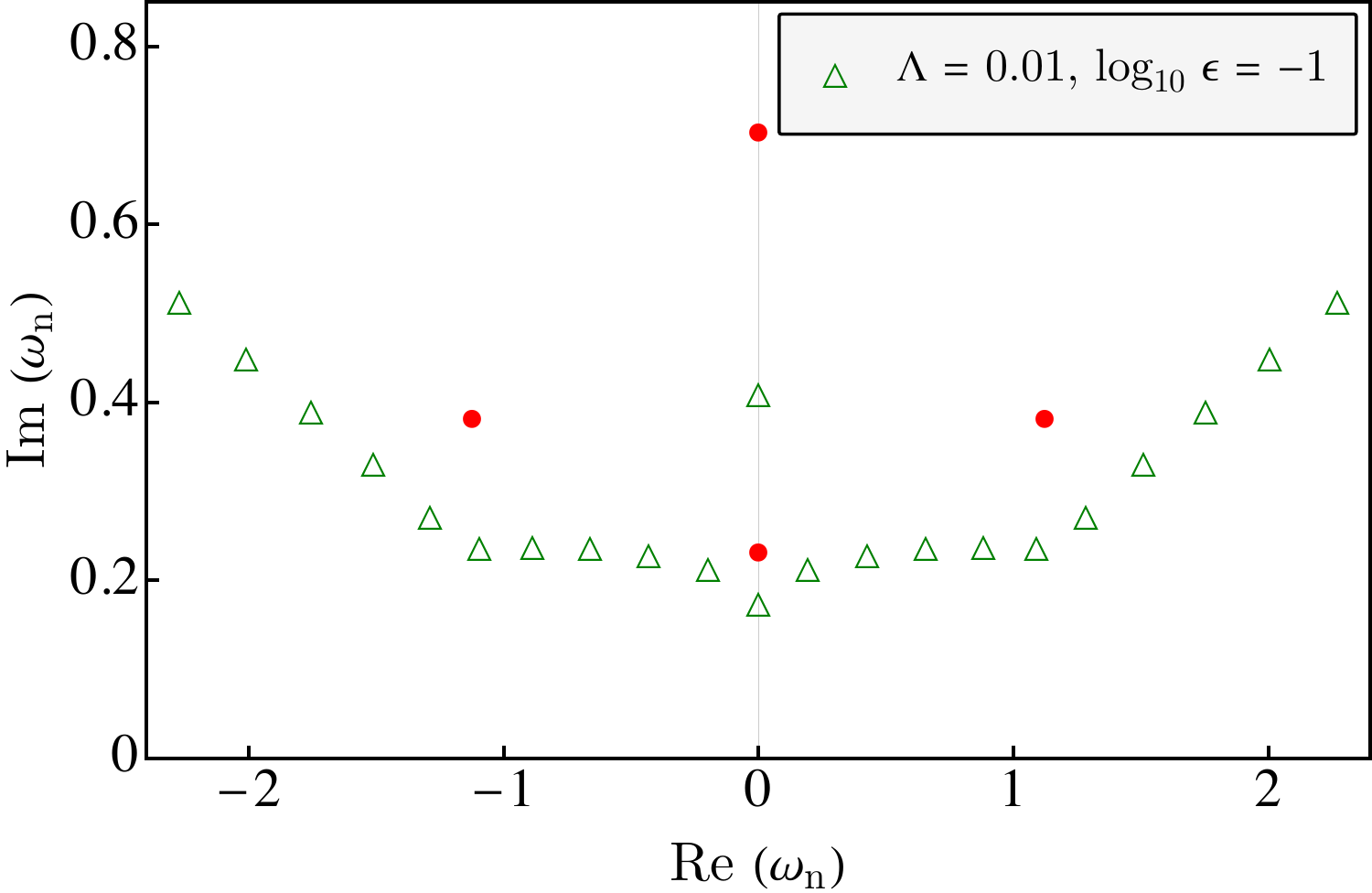}
	\endminipage\hfill
	\minipage{0.33\textwidth}
	\includegraphics[width=\linewidth]{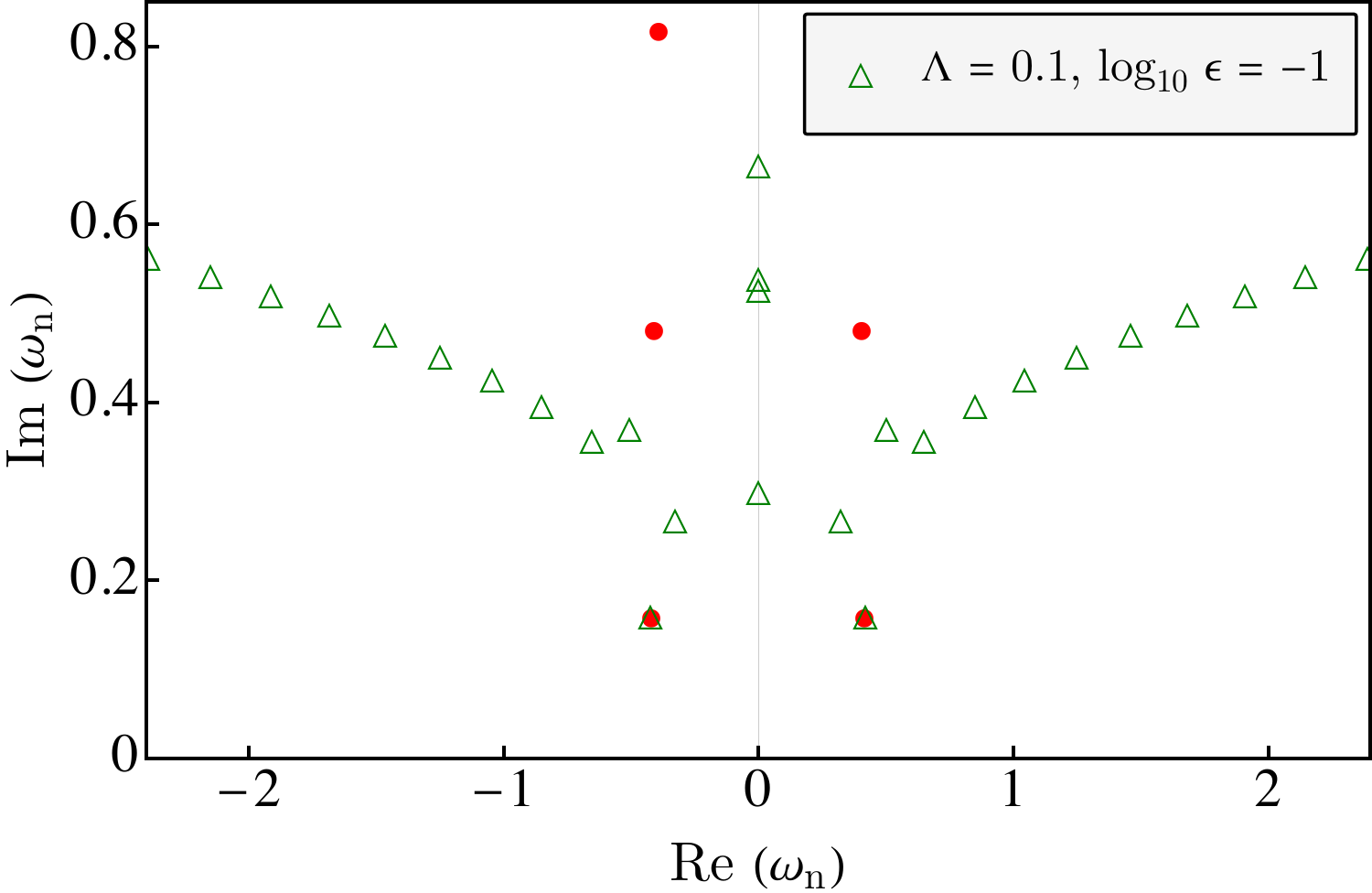}
	\endminipage
\caption{Stability/instability of the fundamental \emph{scalar} mode ($s = 0$) for $\ell=1$ of a Schwarzschild de Sitter black hole has been presented with $M = 1$ for three choices of the cosmological constant: $\Lambda = 0.001$ (left column), $\Lambda = 0.01$ (middle column) and $\Lambda = 0.1$ (right column) and the norm of the deterministic perturbations has been taken to be $||\delta V_d||={10^{-3}}$(top row), $||\delta V_d||={10^{-2}}$ (middle row), $||\delta V_d||={10^{-1}}$ (bottom row) with $k=60$. The unperturbed QNM values are indicated in red and the characteristic length scale has been set to $\lambda = 2 r_+$.}
\label{fig:SdS_scalar_det_pert_CC_variation}
\end{figure}	


\subsection{Salient features of the perturbed QNM spectrum of a Schwarzschild de Sitter black hole}

The preceding discussion provides the link between the pseudospectra of an asymptotically de Sitter black hole and the nature of the perturbations triggering spectral instabilities. Following this we present some observations solely based on studying the spectra of perturbed scattering potentials in the asymptotically de Sitter black hole spacetimes. Our aim is to explore the wide parameter space of de Sitter black holes and sniff out those regions which may prove to be interesting avenues for current and future investigations.  

We start by asking what is the effect of the magnitude of the cosmological constant on the spectral properties of the spacetime (keeping the BH mass fixed). In particular, noting that a deterministic perturbation of the order $\epsilon = 10^{-1}$ and $k = 60$ can potentially destabilize the fundamental mode, we wish to explore how this result depends on the cosmological constant. It turns out that indeed the cosmological constant $\Lambda$ has an important effect on the instability of the fundamental mode. As evident from \ref{fig:SdS_scalar_det_pert_CC_variation}, it follows that reducing the value of $\Lambda$ intensifies the spectral instability of Schwarzschild de Sitter black holes, as in these cases the lowest-lying mode corresponds to a de Sitter mode, which has no analog in the case of asymptotically flat black hole spacetimes. This effect can be clearly seen for the $\ell=1$ mode of scalar perturbation when we compare the plots in the first and second columns of \ref{fig:SdS_scalar_det_pert_CC_variation}. Since $\epsilon=10^{-1}$ is indeed a large perturbation to the perturbing potential, we have also depicted the QNM migration for perturbation with strength $\epsilon=10^{-2}$ and $\epsilon=10^{-3}$, respectively. As evident, the migration of the fundamental QNM becomes smaller as $\epsilon$ decreases. Moreover, the Nollert-Price branch comes closer to the real axis as one decreases $\Lambda$. Since we expect the Nollert-Price branch to follow the contours of the pseudospectra, we conclude that a smaller $\Lambda$ will make the perturbed QNM drift towards smaller imaginary values and hence the spectrum becomes more unstable. On the other hand, increasing the cosmological constant $\Lambda$ has the opposite effect, and in particular, we see that for $\Lambda = 1$, the fundamental mode is stable (c.f. the plots in the third column of \ref{fig:SdS_scalar_det_pert_CC_variation}) and it is a photon sphere mode. This feature can be explained as follows: in order to have two distinct horizons, the mass $M$ of the Schwarzschild de Sitter black hole and the cosmological constant $\Lambda$ must satisfy $\Lambda < (1/9M^2)$; with our chosen set of parameters, this means $\Lambda<1.111$. In the limit $\Lambda \to (1/9M^2)$, the two horizons coalesce, which is the so-called Nariai limit. In this limiting case, one can approximate the black hole potential with the P\"{o}schl-Teller potential \cite{Cardoso:2003sw}, and it has been shown that the fundamental mode of the P\"{o}schl-Teller potential is stable against such high-frequency perturbations \cite{Jaramillo:2020tuu}. Thus our findings are consistent with results previously reported in the literature.

Furthermore, the departure of the fundamental de Sitter mode from its unperturbed position, for scalar perturbation, with small $\Lambda$ and $\epsilon \sim 10^{-1}$, is small and is of the same order as $\epsilon$. Therefore, the migration of the fundamental scalar mode is consistent with the `large' perturbation and hence may not be linked to instability. Thus, as far as scalar perturbation is concerned, the fundamental mode of the Schwarzschild de Sitter black hole spacetime appears to be stable.


\begin{figure}[tbh!]
	\centering
	\minipage{0.33\textwidth}
	\includegraphics[width=\linewidth]{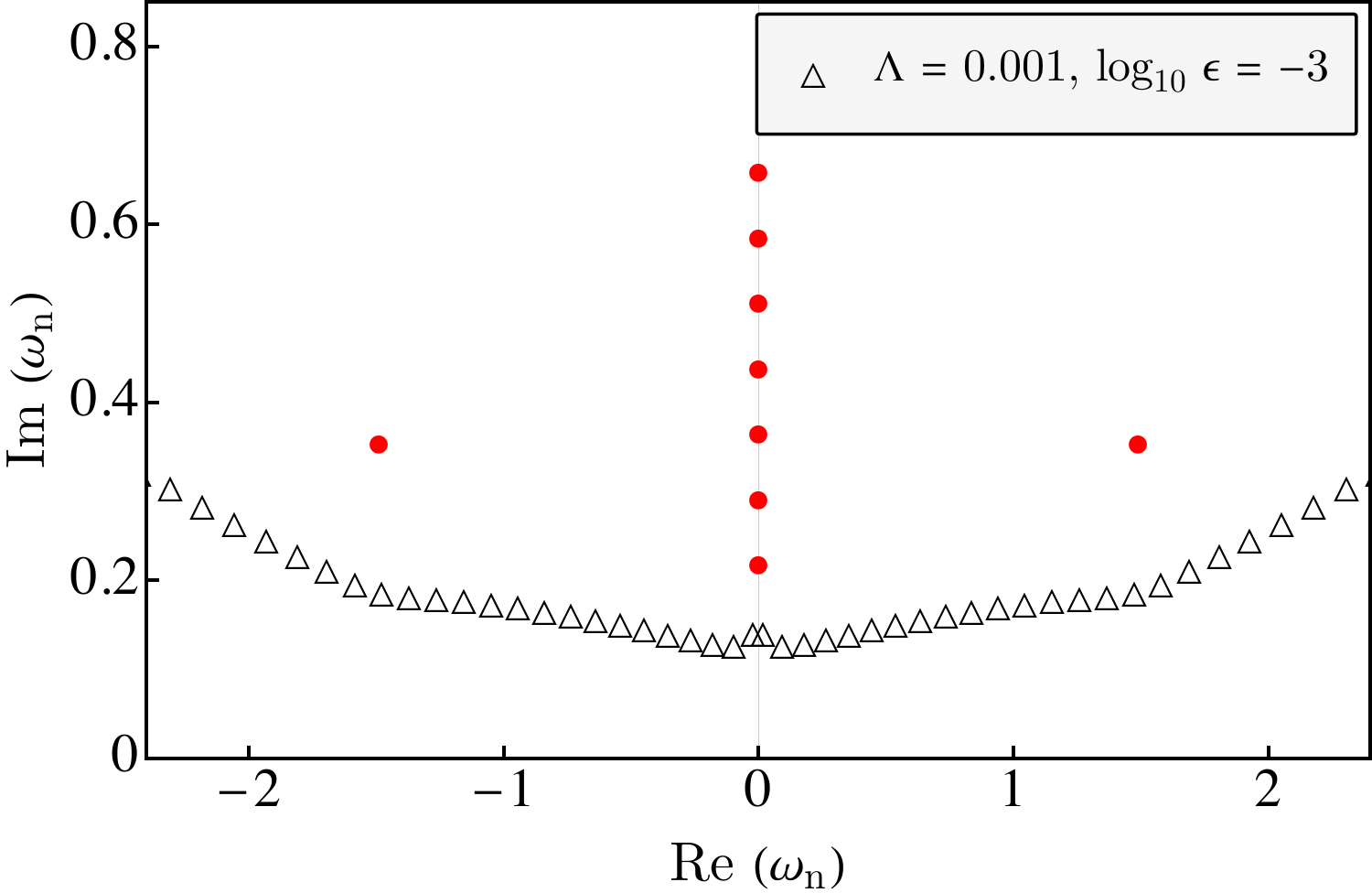}
	\endminipage\hfill
	\minipage{0.33\textwidth}
	\includegraphics[width=\linewidth]{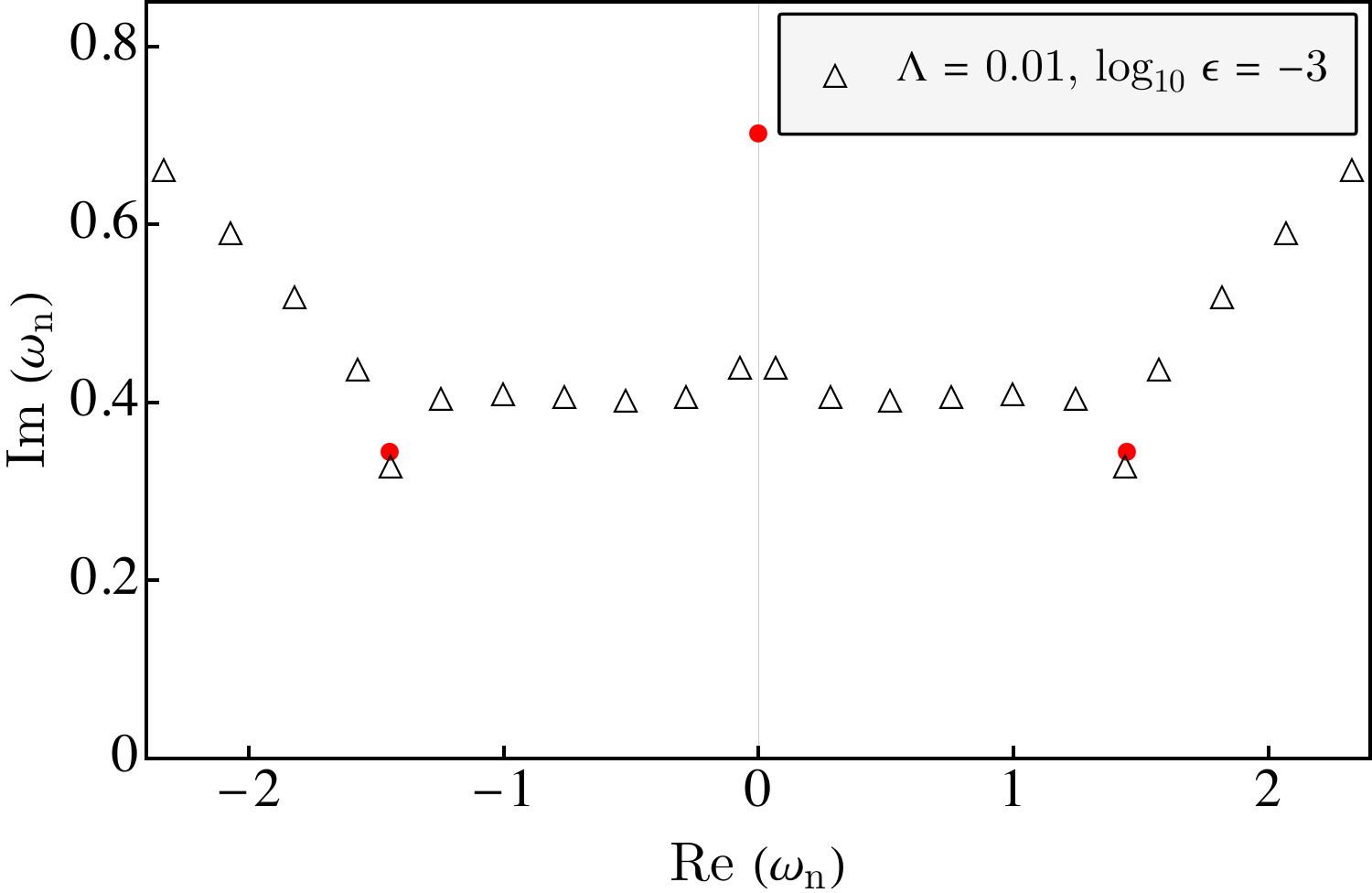}
	\endminipage\hfill
	\minipage{0.33\textwidth}
	\includegraphics[width=\linewidth]{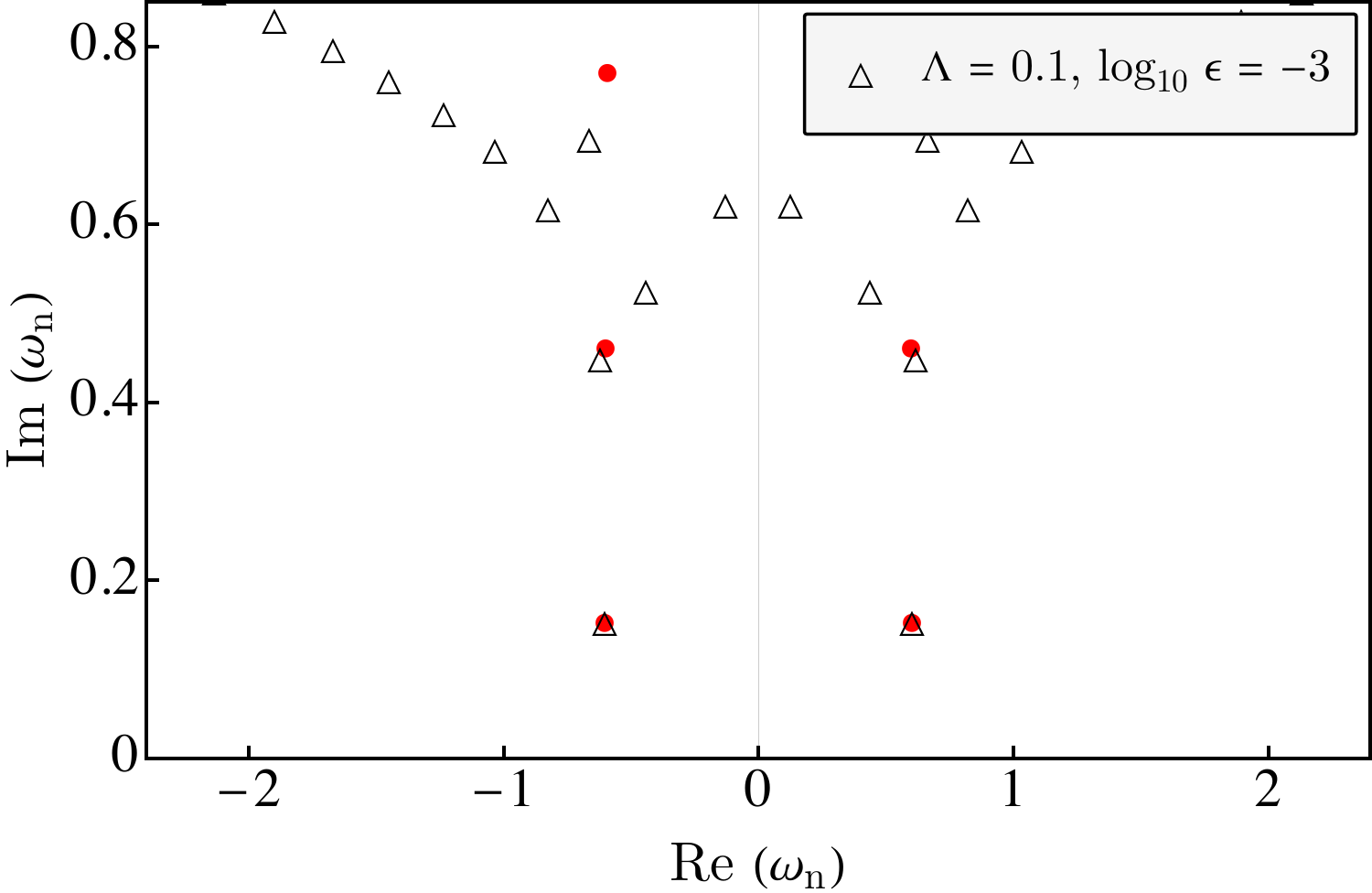}
	\endminipage\hfill
 \minipage{0.33\textwidth}
	\includegraphics[width=\linewidth]{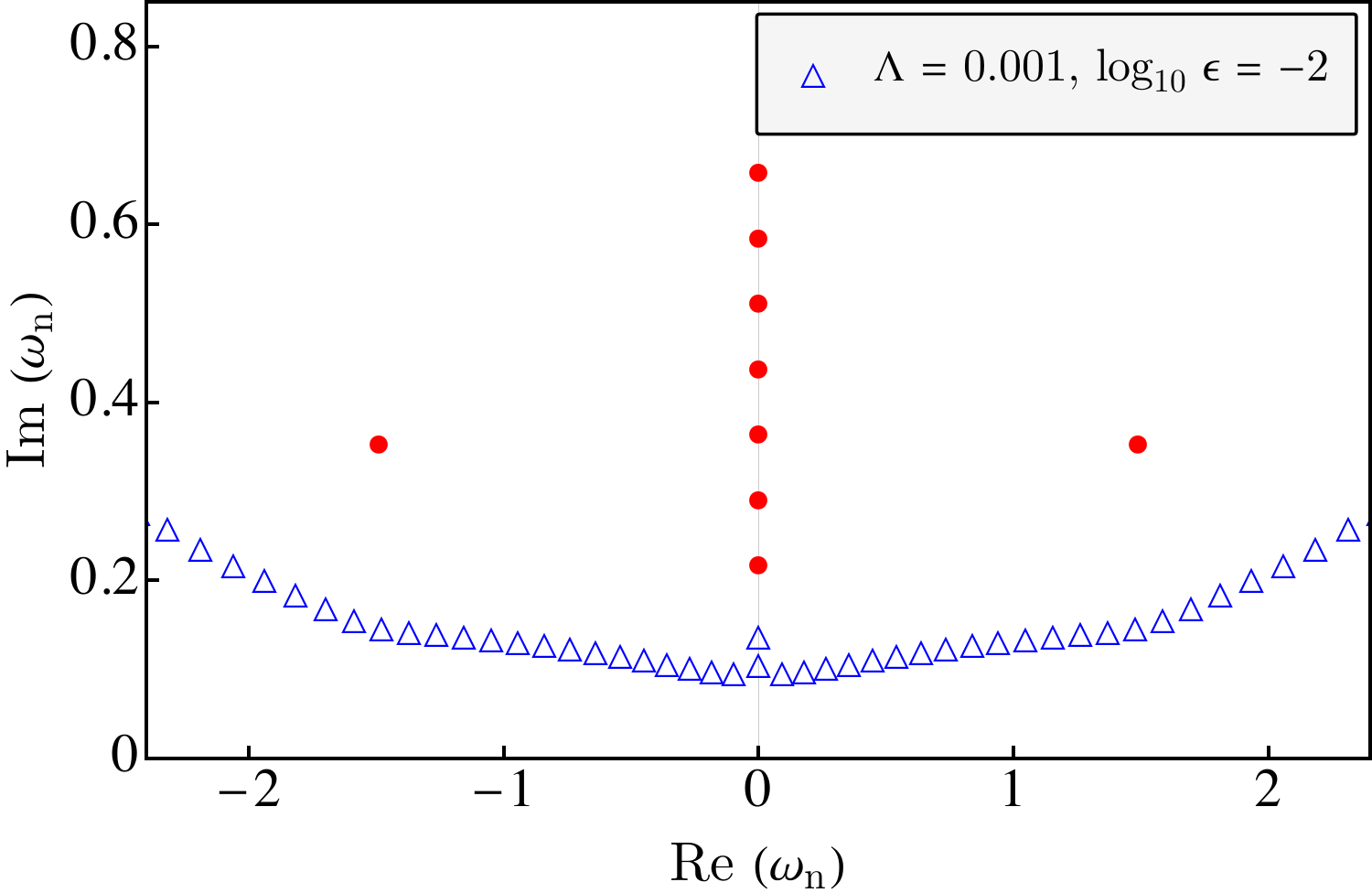}
	\endminipage\hfill
	\minipage{0.33\textwidth}
	\includegraphics[width=\linewidth]{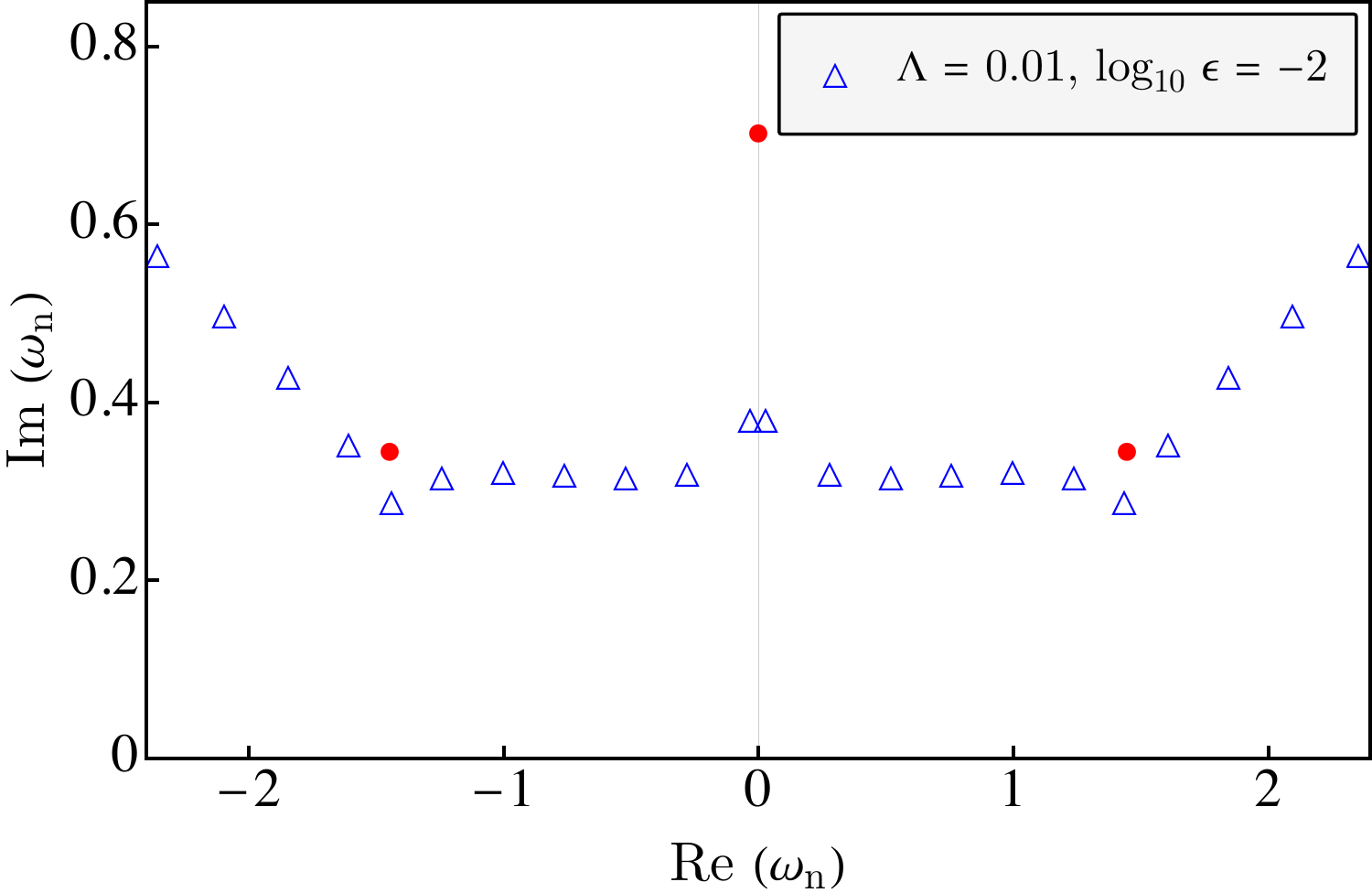}
	\endminipage\hfill
	\minipage{0.33\textwidth}
	\includegraphics[width=\linewidth]{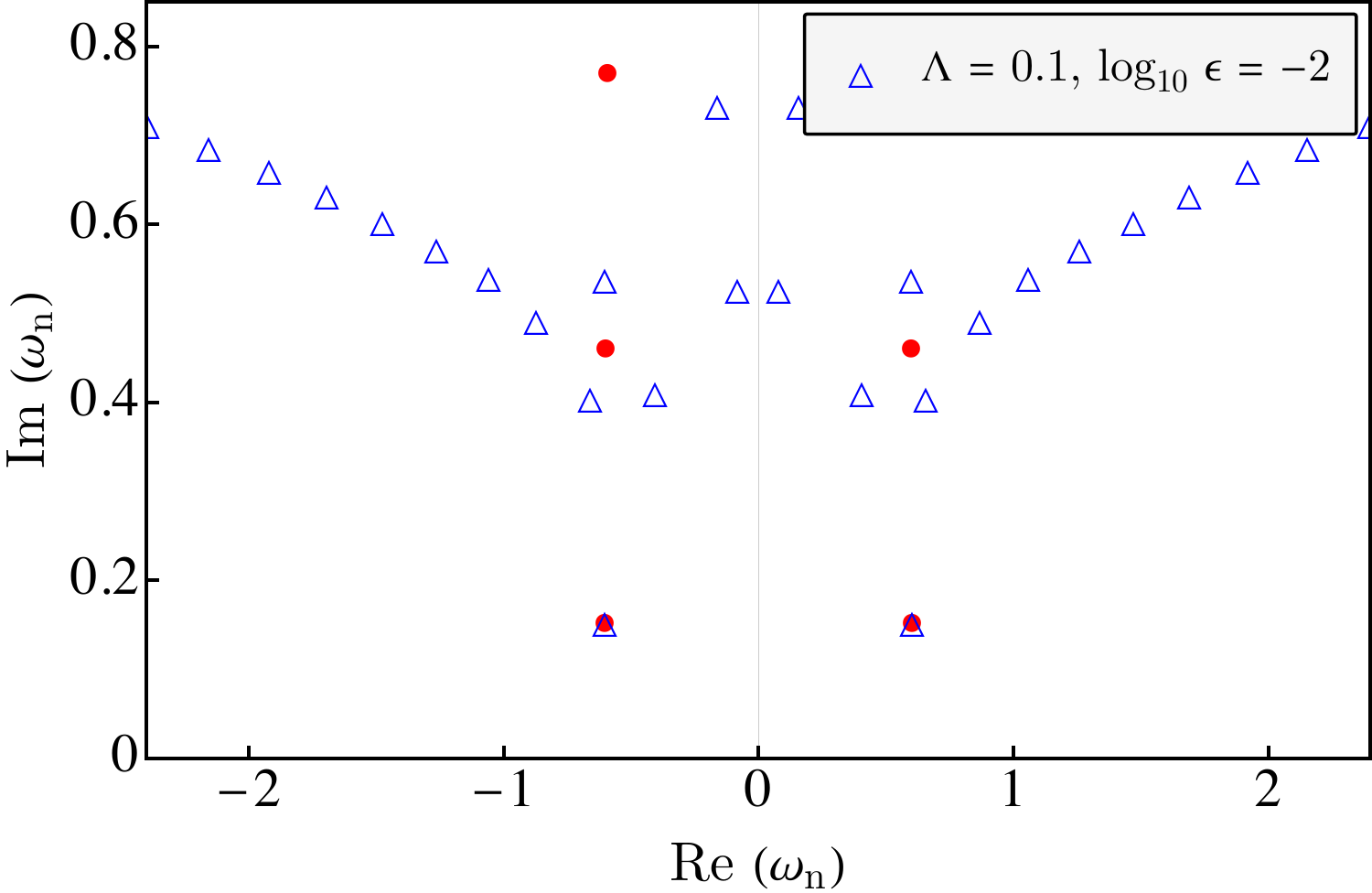}
	\endminipage\hfill
 \minipage{0.33\textwidth}
	\includegraphics[width=\linewidth]{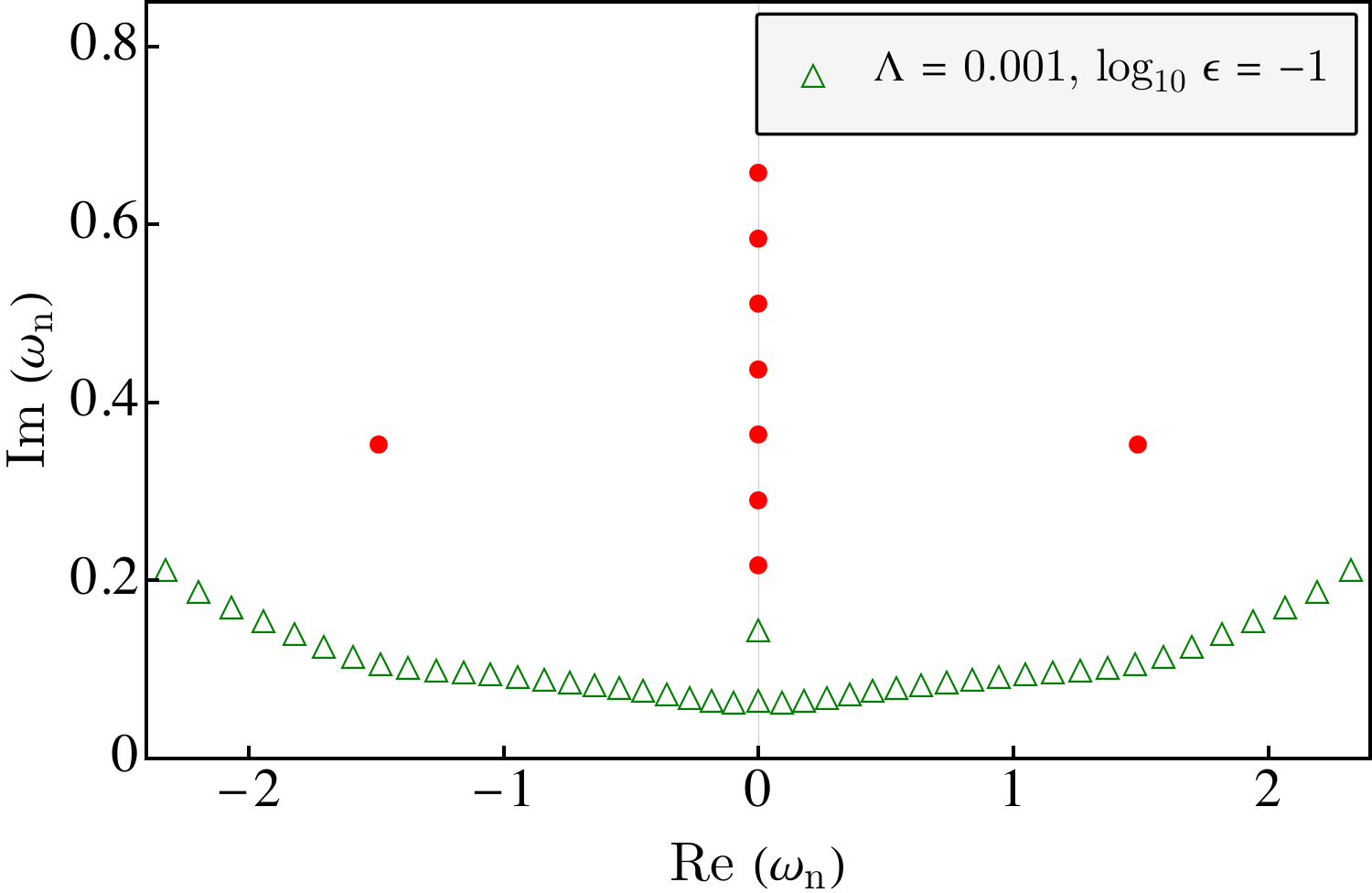}
	\endminipage\hfill
	\minipage{0.33\textwidth}
	\includegraphics[width=\linewidth]{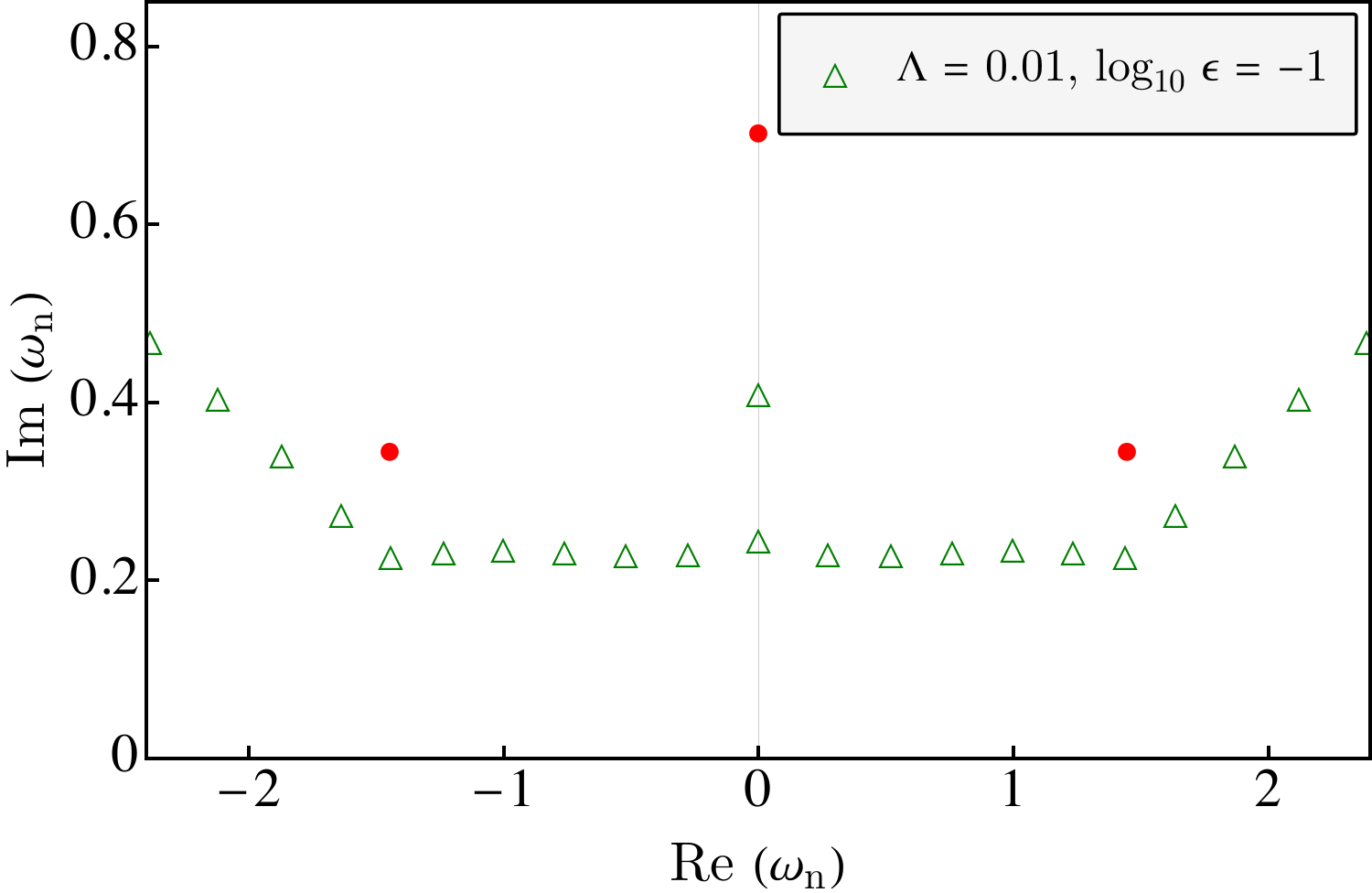}
	\endminipage\hfill
	\minipage{0.33\textwidth}
	\includegraphics[width=\linewidth]{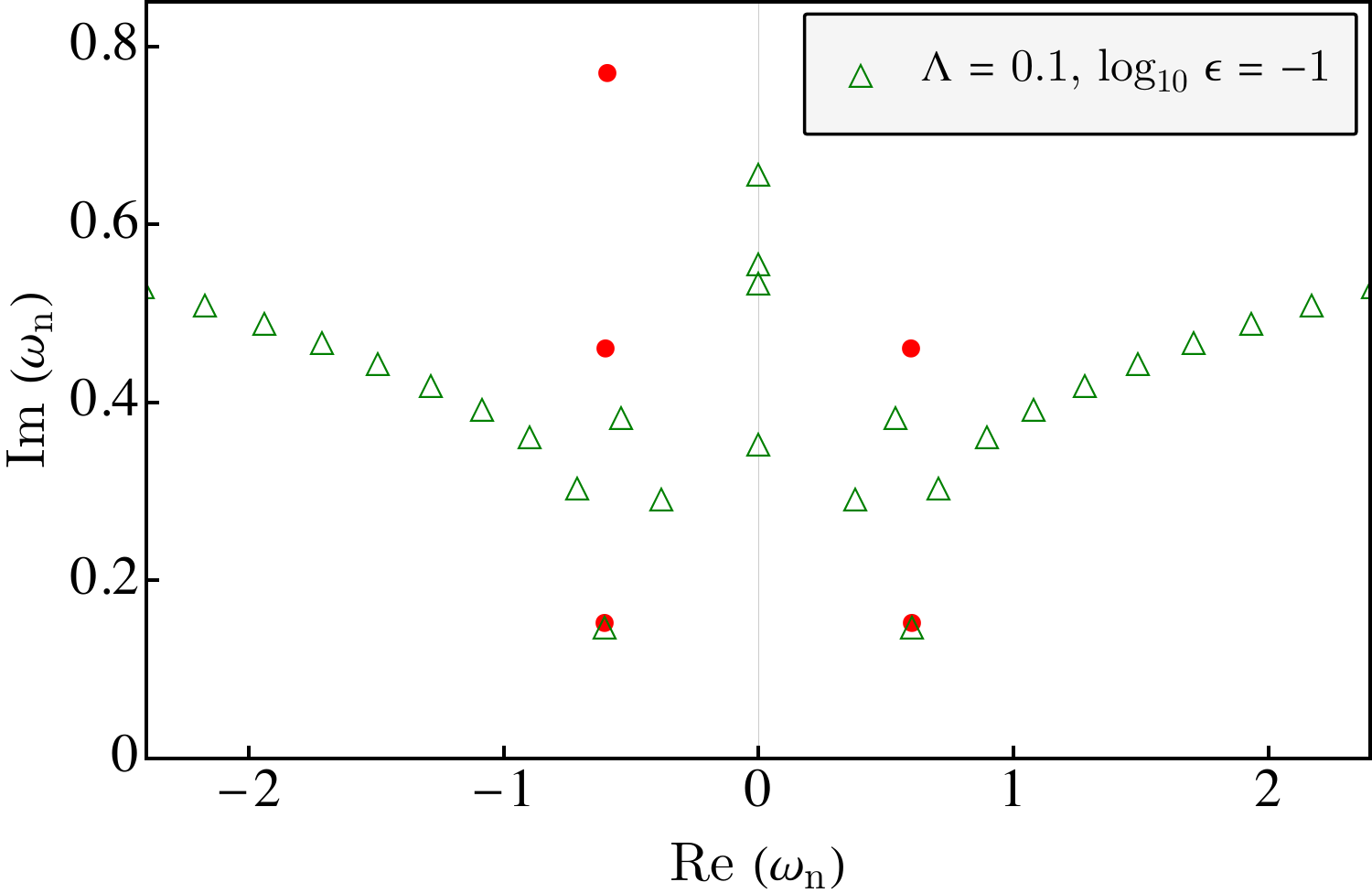}
	\endminipage
\caption{Stability/instability of the fundamental \emph{gravitational} mode ($s = 2$) for $\ell=2$ of a Schwarzschild de Sitter black hole has been presented with $M = 1$ for three choices of the cosmological constant: $\Lambda = 0.001$ (left column), $\Lambda = 0.01$ (middle column) and $\Lambda = 0.1$ (right column) and the norm of the deterministic perturbations has been taken to be $||\delta V_d||={10^{-3}}$(top row), $||\delta V_d||={10^{-2}}$ (middle row), $||\delta V_d||={10^{-1}}$ (bottom row) with $k=60$. The unperturbed QNM values are indicated in red and the characteristic length scale has been set to $\lambda = 2 r_+$.}
\label{fig:SdS_grav_det_pert_CC_variation}
\end{figure}	


However, the situation regarding stability is very different for the fundamental mode associated with gravitational perturbation. Just like the case of scalar perturbation, the cosmological constant affects the stability of the fundamental modes for gravitational perturbations as well. A decrease in the cosmological constant indeed makes the instability of the fundamental QNM frequency worse, since not only do the perturbed QNMs drift further away from the fundamental mode, but they also tend to have smaller imaginary parts. This is evident from the plots in the first two columns of \ref{fig:SdS_grav_det_pert_CC_variation}. Hence regardless of the nature of the perturbing field, we note that a smaller cosmological constant enhances the instability of the fundamental mode. For larger cosmological constant, the connection with the P\"{o}schl-Teller potential makes the fundamental mode stable, even for gravitational perturbation, as highlighted in the plots in the third column of \ref{fig:SdS_grav_det_pert_CC_variation}. But, for gravitational perturbation, in the presence of a small cosmological constant, the fundamental de Sitter mode is unstable even for perturbations with $\epsilon \sim 10^{-3}$. The migrated and the unperturbed QNMs are separated by an amount that is at least one order of magnitude larger than the strength of the perturbation. Thus, for asymptotically de Sitter spacetimes, the fundamental QNM associated with the gravitational perturbation becomes unstable for small values of the cosmological constant. This feature is exclusive to asymptotic de Sitter spacetimes and does not exist for asymptotically flat black holes. It also gives a hope to rescue the strong cosmic censorship conjecture for asymptotically de Sitter black holes, at least for gravitational perturbations.

\begin{figure}[tbh!]
	\centering
	\minipage{0.33\textwidth}
	\includegraphics[width=\linewidth]{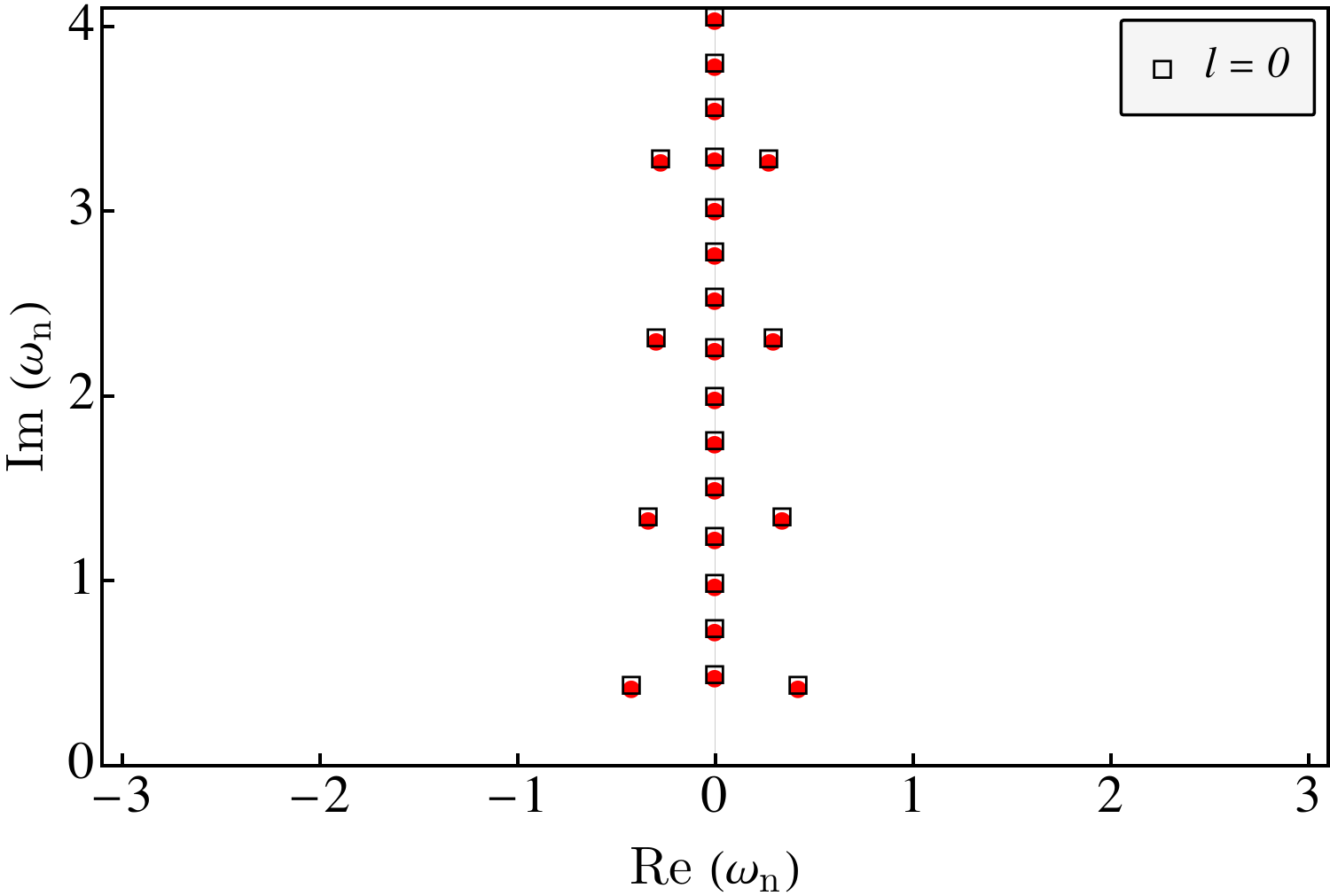}
	\endminipage\hfill
	\minipage{0.33\textwidth}
	\includegraphics[width=\linewidth]{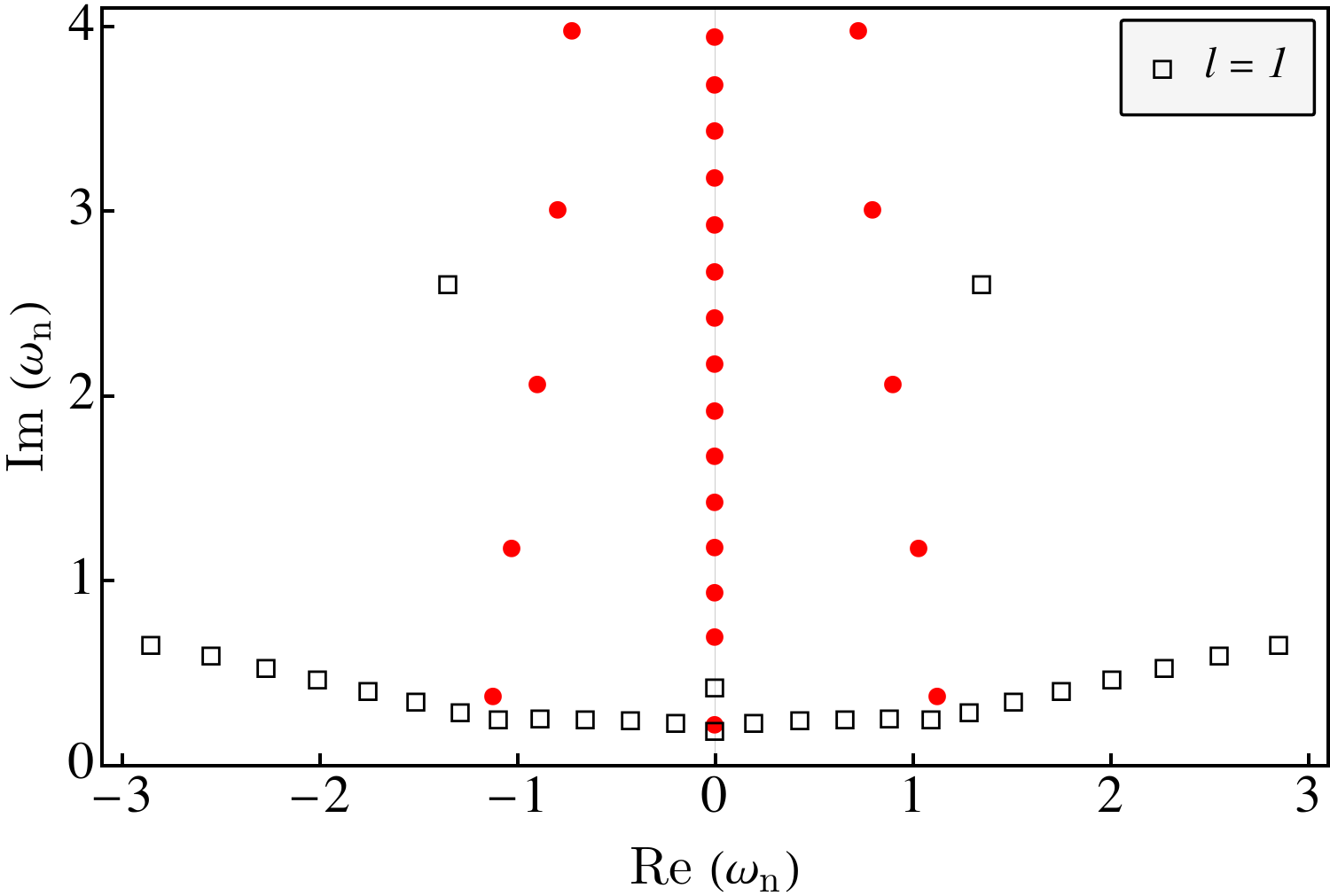}
	\endminipage\hfill
	\minipage{0.33\textwidth}
	\includegraphics[width=\linewidth]{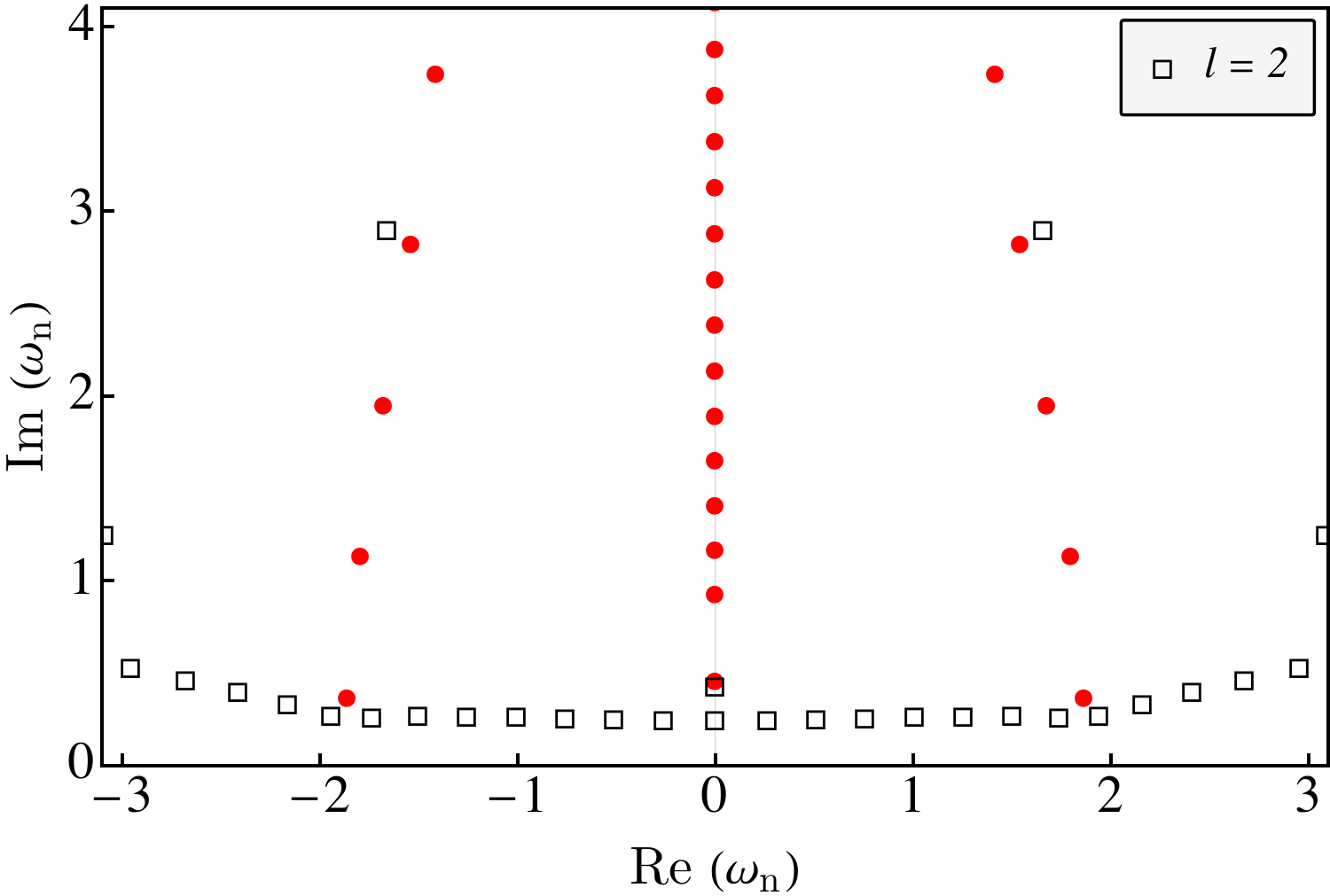}
	\endminipage
\caption{The perturbed, as well as unperturbed QNM spectra for deterministic perturbations of norm $||\delta V_d||={10^{-1}}$ with $k=60$ has been presented for different choices of the angular number --- $\ell=0$ (left panel), $\ell=1$ (middle panel) and $\ell=2$ (right panel). In these plots, the perturbation is due to \emph{scalar} modes in the background of a Schwarzschild de Sitter black hole with $M = 1$ and $\Lambda = 0.01$. The unperturbed QNM frequencies have been indicated in red and the characteristic length scale has been set to $\lambda = 2 r_+$.}\label{fig:SdS_scalar_det_pert_L_variation}
\end{figure}	

\begin{figure}[htb!]
	\centering
	\minipage{0.33\textwidth}
	\includegraphics[width=\linewidth]{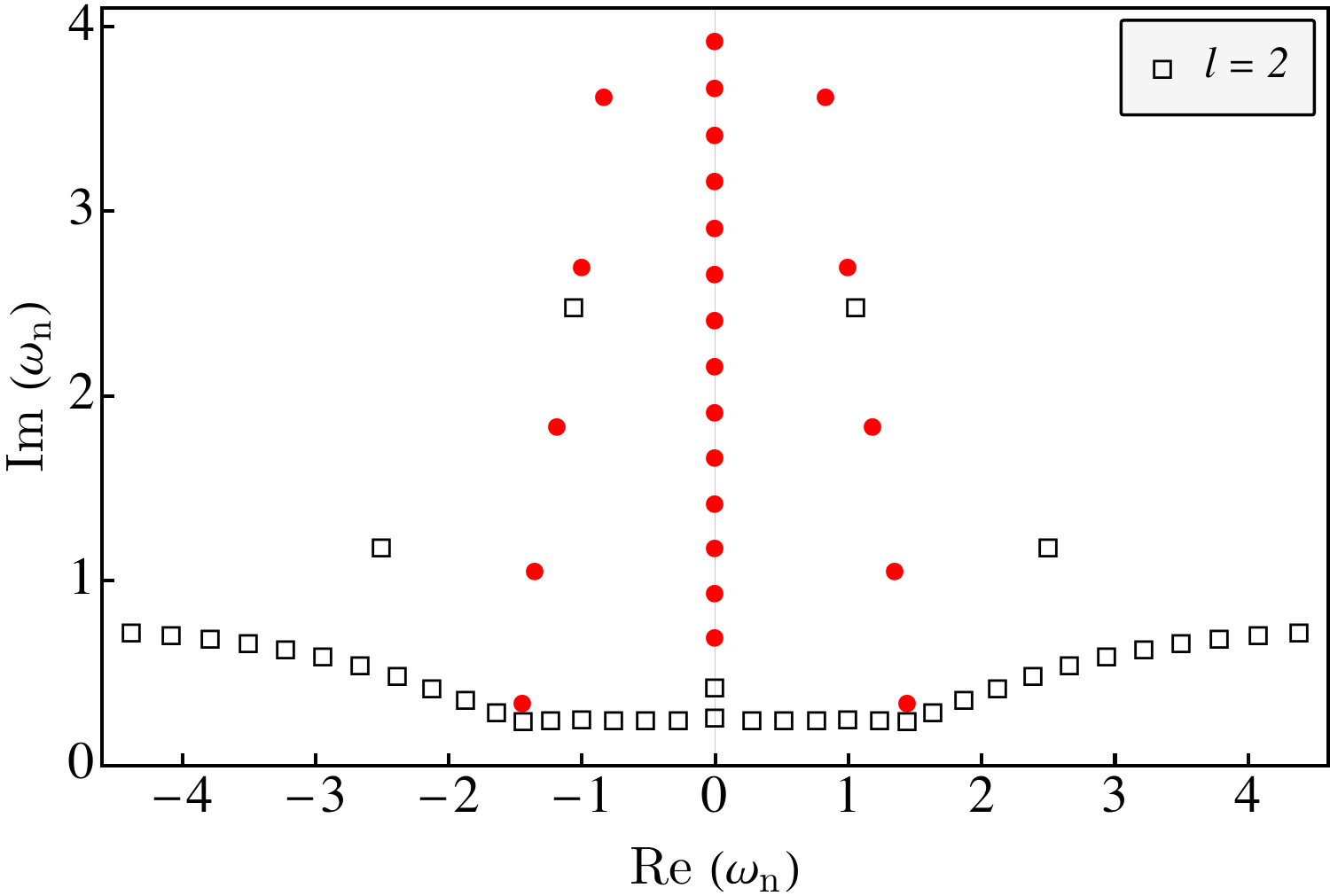}
	\endminipage\hfill
	\minipage{0.33\textwidth}
	\includegraphics[width=\linewidth]{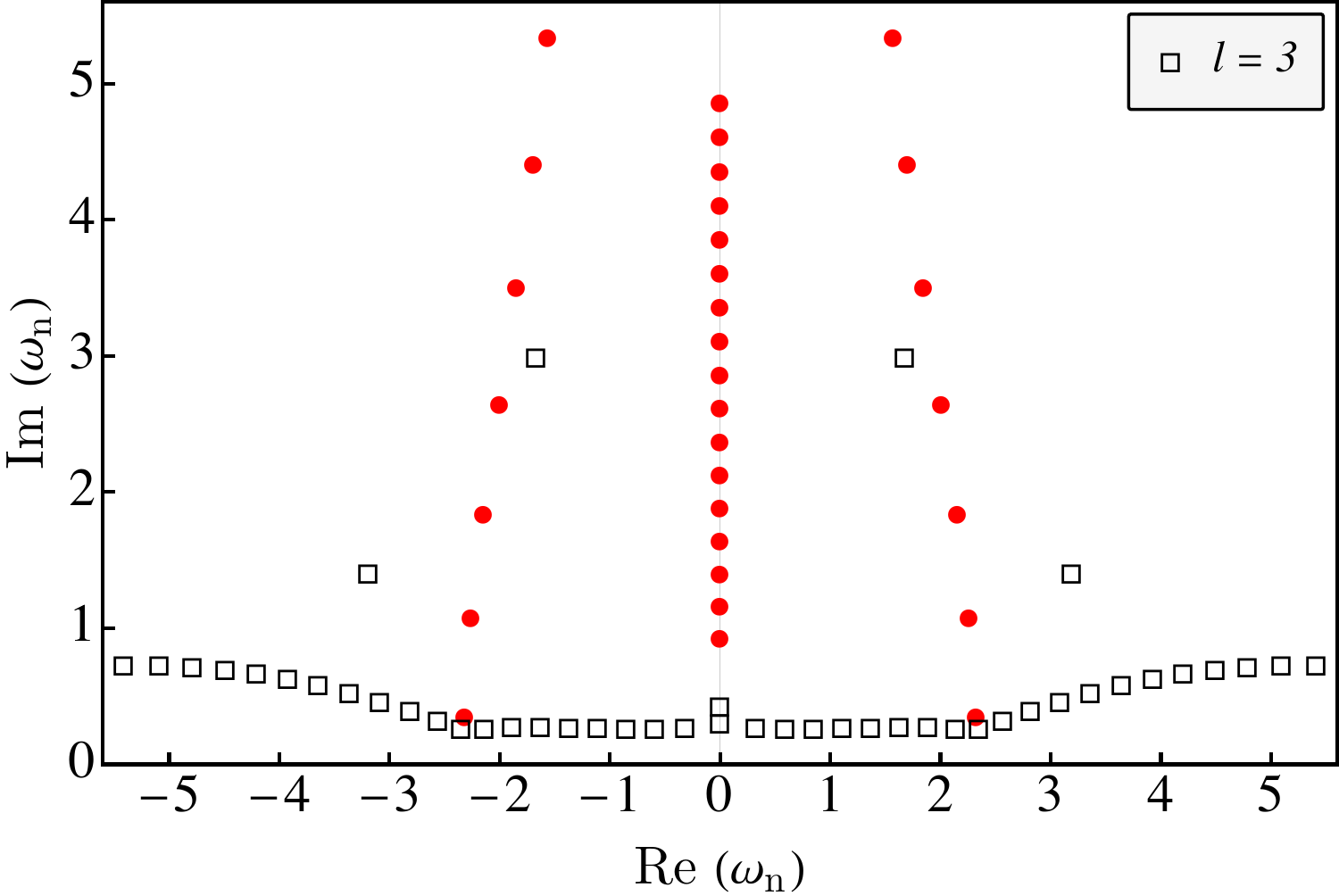}
	\endminipage\hfill
	\minipage{0.33\textwidth}
	\includegraphics[width=\linewidth]{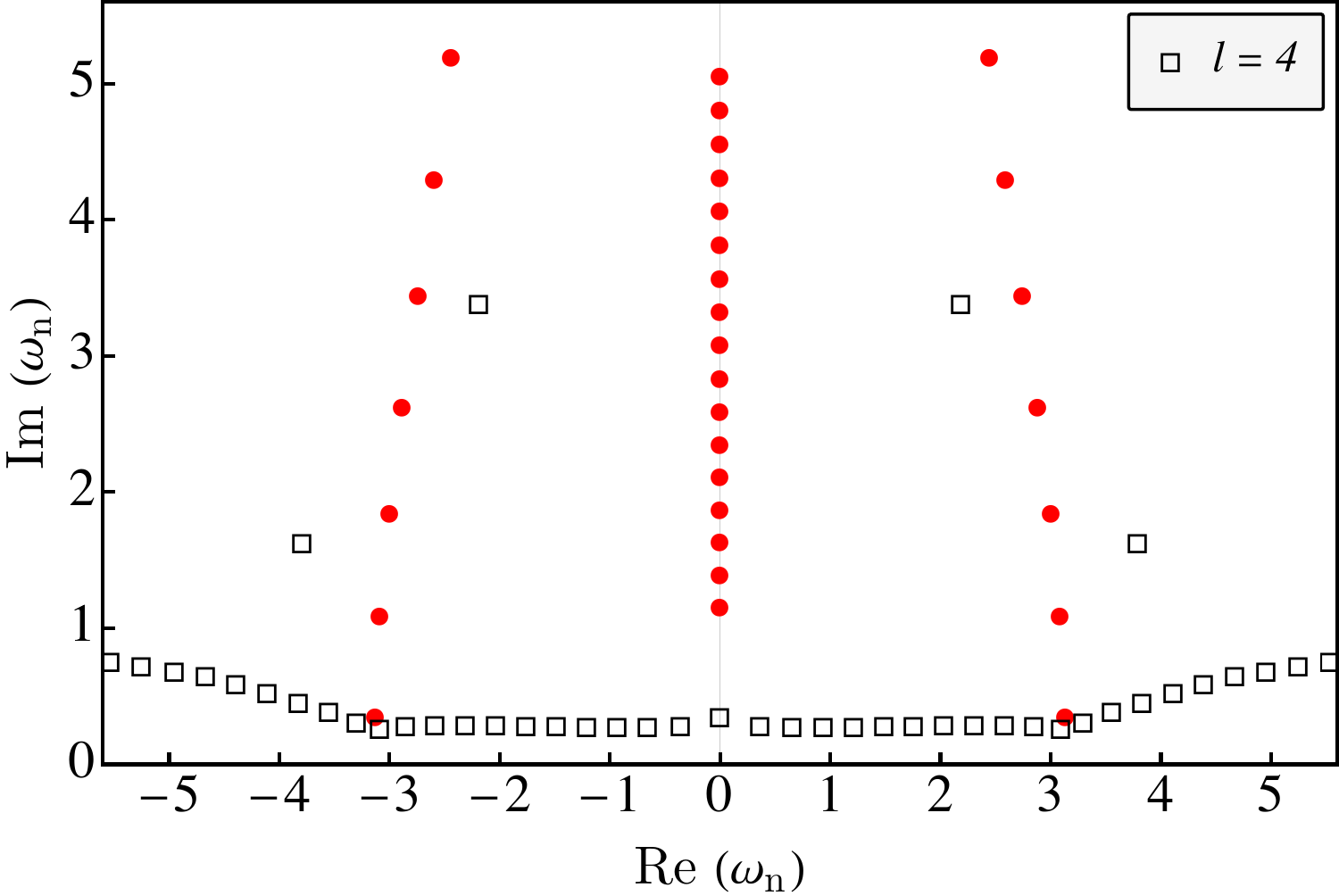}
	\endminipage
\caption{The QNM spectra for both perturbed and unperturbed modes have been presented under deterministic perturbations of norm $||\delta V_d||=10^{-1}$ with $k=60$ for different angular numbers --- $\ell=2$ (left panel), $\ell=3$ (middle panel) and $\ell=4$ (right panel). The modes are for \emph{gravitational} scattering potential in the background of a Schwarzschild de Sitter black hole with $M = 1$ and $\Lambda=0.01$. The perturbed modes have been superimposed over the unperturbed QNM frequencies, indicated in red. The characteristic length scale has been set to $\lambda = 2 r_+$.}
\label{fig:SdS_grav_det_pert_L_variation}
\end{figure}

Lastly, we consider the effect of the angular number $\ell$ (c.f. \ref{fig:SdS_scalar_det_pert_L_variation} for scalar perturbation and \ref{fig:SdS_grav_det_pert_L_variation} for gravitational perturbation) on the instability of the QNM frequencies. It turns out that increasing the angular number $\ell$ does not affect the qualitative nature of the Nollert-Price branches, and hence does not drastically alter the spectral properties of the operator governing the perturbations. This is true for both scalar and gravitational perturbations. However, there is one intriguing feature worth highlighting: for the $\ell=0$ mode, which exists \emph{only} for the scalar perturbation, the QNM spectrum is extremely stable for a high-frequency perturbation of significant strength, e.g., it is stable even for deterministic perturbation with norm $\epsilon \sim 10^{-1}$. This intriguing aspect merits further investigation, but we have found that the computation of the pseudospectrum for the $\ell=0$ mode to be exceptionally difficult, since the $(1/r^2)$ term in the potential completely disappears, rendering the resultant matrix $L$, governing the scattering potential experienced by the perturbations, highly ill-conditioned. This feature appears to impose several numerical difficulties and we shall attempt the computation again elsewhere. 

\subsection{Random perturbation and the QNM frequencies in Schwarzschild de Sitter spacetime}

\begin{figure}[tbh!]
 \centering
  \begin{subfigure}[t]{0.48\textwidth}
    \centering
    \includegraphics[width=\textwidth]
    {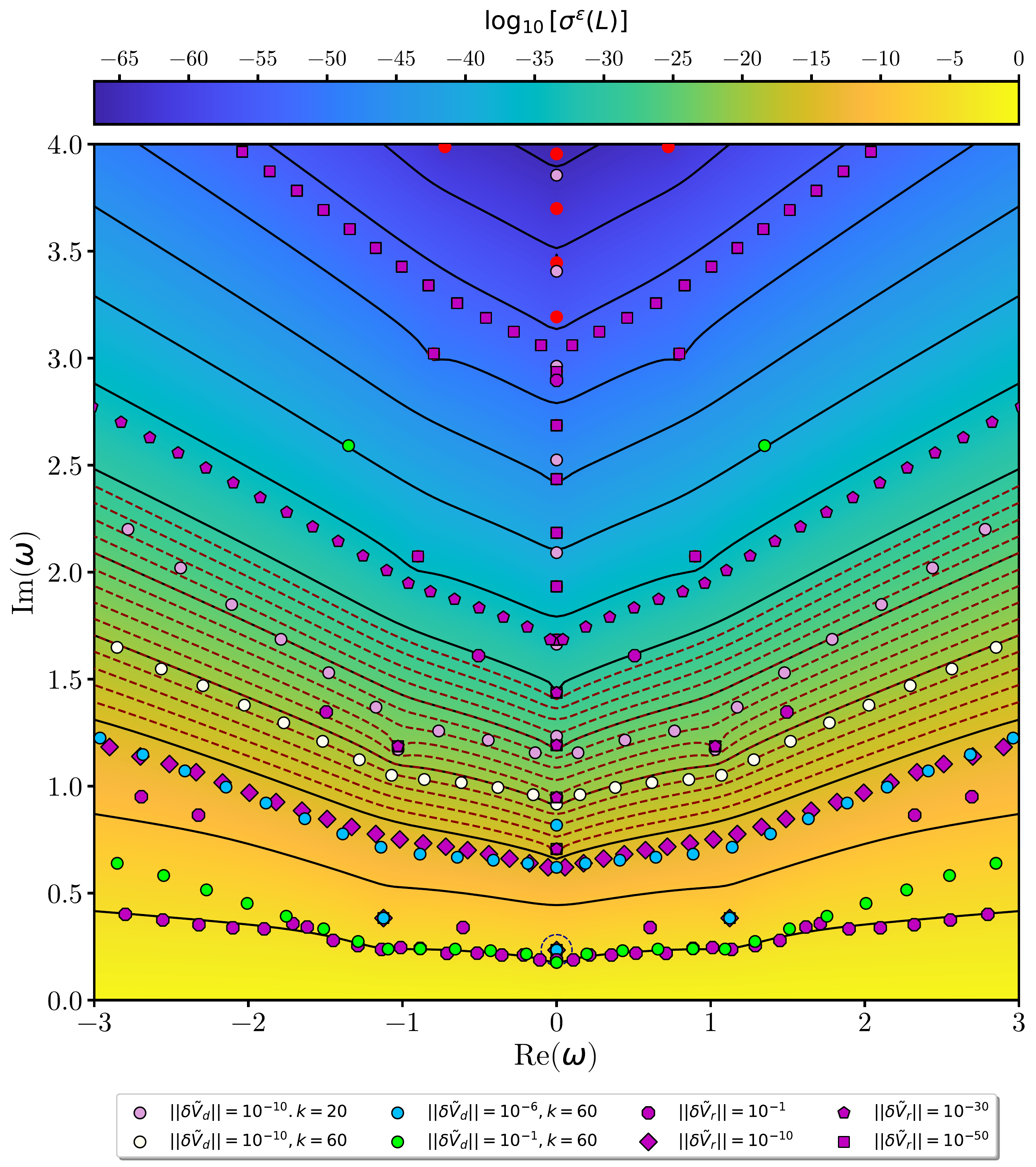}
\caption{The $\ell=1$ \emph{scalar} $\epsilon$ pseudospectrum and Nollert-Price branches arising from various perturbations to the scattering potential of Schwarzschild de Sitter black hole. Observe that the lowest-lying de Sitter mode (indicated by a dashed blue circle) is displaced by both high frequency random and deterministic oscillatory perturbations of the order $||\delta V|| = 10^{-1}$. This is in sharp contrast to the corresponding asymptotically flat scenario.}
    \label{fig:pseudospectra_scalar_l1_random_det_contour}
  \end{subfigure}\hspace{1em} 
  \begin{subfigure}[t]{0.48\textwidth}
  \centering
    \includegraphics[width=\textwidth]{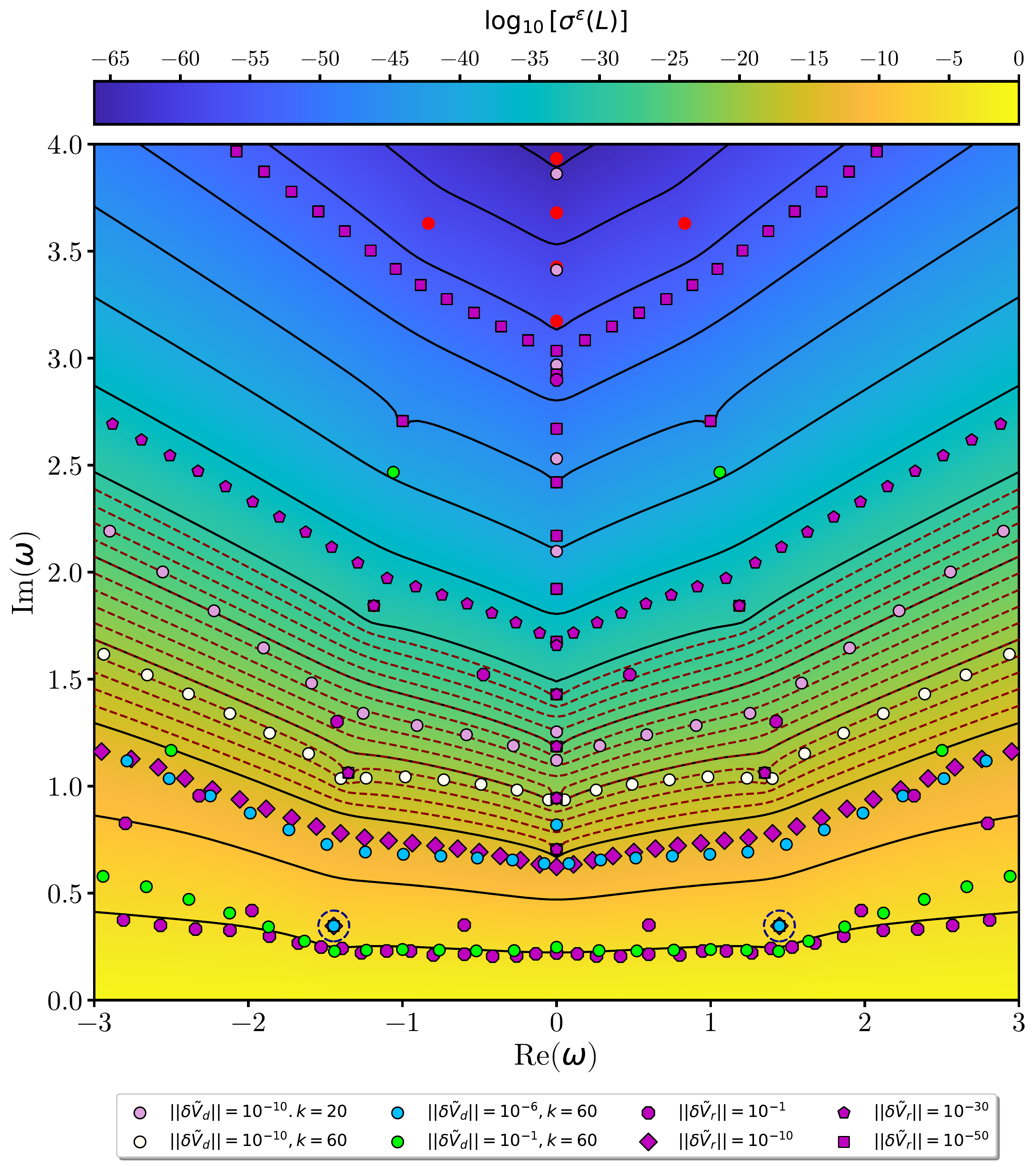}
\caption{The $\ell=2$ \emph{gravitational} $\epsilon$ pseudospectrum and Nollert-Price branches arising from both deterministic and random perturbations to the scattering potential of Schwarzschild de Sitter black hole. Observe that the two lowest-lying photon sphere modes (indicated by dashed blue circles) are displaced by both high frequency random and deterministic oscillatory perturbations of the order $||\delta V|| = 10^{-1}$.}
    \label{fig:pseudospectra_grav_l2_random_det_contour}
  \end{subfigure}
\caption{Nollert-Price branches and destabilization of the QNM spectra versus the $\epsilon$ pseudospectra have been depicted for a Schwarzschild de Sitter black hole with $M=1$ and $\Lambda=0.01$. In both the figures, the solid black contour lines correspond to $\log_{10}\epsilon$, which range from $-65$ (top level) to $-5$ (bottom level) in steps of $5$, whereas the dashed red contour lines range from $-29$ (top level) to $-16$ (bottom level) in steps of $1$. The QNMs (filled red circles) have also been indicated for reference.}
  \label{fig:fig:pseudospectra_CC_0.01_rand_det_pert_contour}
\end{figure}

The overall behaviour of the spectral instability is further corroborated by the inclusion of random perturbations as well. The results have been compiled together in \ref{fig:fig:pseudospectra_CC_0.01_rand_det_pert_contour} for the benefit of the reader. On an operational level, the inclusion of a random perturbation translates to adding a set of random numbers to the scattering potential. It is designed to mimic a high-frequency sinusoidal perturbation and its strength is controlled purely by the energy norm of the probe field. With reference to \ref{fig:fig:pseudospectra_CC_0.01_rand_det_pert_contour}, we observe that high-frequency random  perturbations of the order $||\delta V_r|| = \epsilon$ to the black hole potential destabilizes the QNM spectra lying above the $\log_{10}{\epsilon}$ contour level of the pseudospectrum, but the spectra below the said contour level remain unaffected, as expected. In particular, a large high-frequency perturbation of the order $\epsilon \sim 10^{-1}$ is able to destabilize the fundamental mode(s) (indicated by dashed blue circles in \ref{fig:fig:pseudospectra_CC_0.01_rand_det_pert_contour}). Lastly, the perturbed QNM spectra follow an orderly arrangement in the complex plane that approximately mimics the pseudospectral contour lines. This migration of the perturbed QNMs (arising from explicit perturbations to the BH potential) along the so-called Nollert-Price branches establishes how the pseudospectra (a property of the unperturbed potential) is able to capture the spectral instabilities underlying scattering phenomena in black hole spacetimes. In \cite{Jaramillo:2020tuu}, it was noted that the pseudospectrum of the P\"{o}schl-Tellers potential under random perturbations has a high degree of spectral stability and this paradoxically regularizing effect of random perturbations improved with increasing the strength of the perturbation. The behaviour was the consequence of certain theorems, which states that this behaviour is connected to a Weyl law for the QNMs (see \cite{Jaramillo:2020tuu} and the references there in for further details). Even though random perturbations proved to be limited in probing the pseudospectrum of asymptotically flat black holes, nevertheless a Weyl law for black hole QNMs has been formulated, based on several mathematical results, to probe the geometry of black holes in gravitational wave signals \cite{Jaramillo:2021tmt, Jaramillo:2022zvf}. These interesting questions have not been explored in the context of asymptotically de Sitter spacetimes. Therefore developing an equivalent Weyl law for asymptotically de Sitter black holes and forging a connection with the pseudospectrum of the perturbed operator will be discussed elsewhere.


\subsection{Convergence of the perturbed and unperturbed QNMs for Schwarzschild de Sitter black hole}

Let us now focus on the issues related to the convergence of the QNMs associated with the unperturbed scattering potential, as well as the convergence of the perturbed QNMs related to the perturbed scattering potentials. The accurate computation of quasi-normal modes, especially the overtones, has been a challenging enterprise for the community. The need to use extended precision in intermediate steps of the computation is well known \cite{Jansen:2017oag,Lin_2019,Fortuna:2020obg,Jaramillo:2020tuu,Mamani:2022akq} and this numerical difficulty is a manifestation of the spectral instability of the underlying eigenvalue problem \cite{Jaramillo:2020tuu}. There are also issues related to the appearance of spurious eigenvalues when one approximates an operator with a matrix \cite{Boyd2001-kt,Dias:2015nua}. These problems also arise in the computation of the perturbed QNMs, which are essentially the eigenvalues of the perturbed potential. However, there is an added layer of subtlety in their computation, an issue prominently emphasized in \cite{Jaramillo:2020tuu}. If one does not use sufficient precision while computing the eigenvalues, then the internal rounding-off errors will accumulate and give rise to a spectrum contaminated with numerical artifacts that will surprisingly be arranged along the Nollert-Price branches, that is, along the contours of the pseudospectra. Therefore it is crucial to determine that the spectra of perturbed QNMs reported here are indeed (physical) eigenvalues of the perturbed potential through rigorous convergence tests. 

\begin{figure}[tbh!]
	\centering
	\minipage{0.33\textwidth}
	\includegraphics[width=\linewidth]{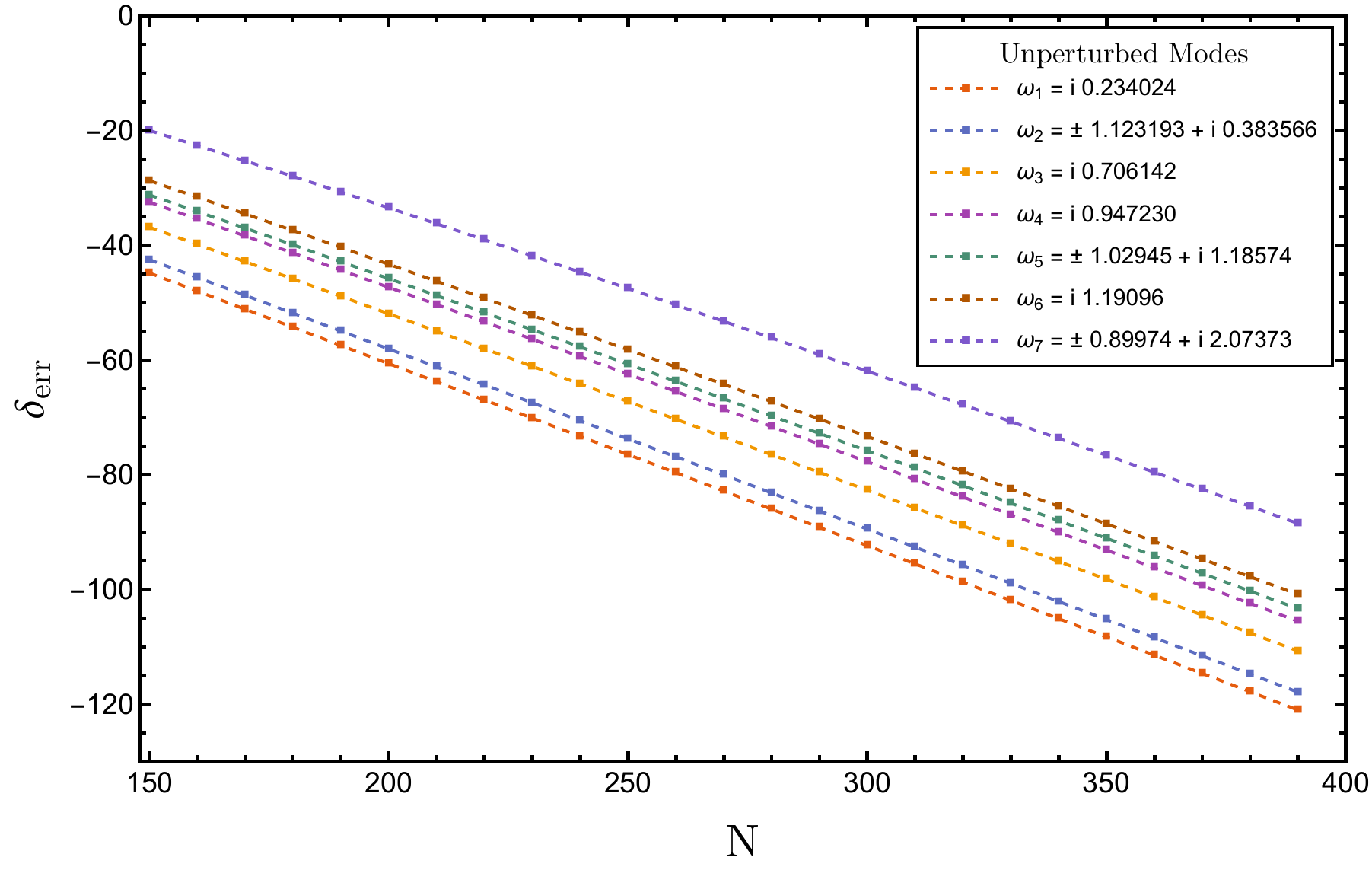}
	\endminipage\hfill
	\minipage{0.33\textwidth}
	\includegraphics[width=\linewidth]{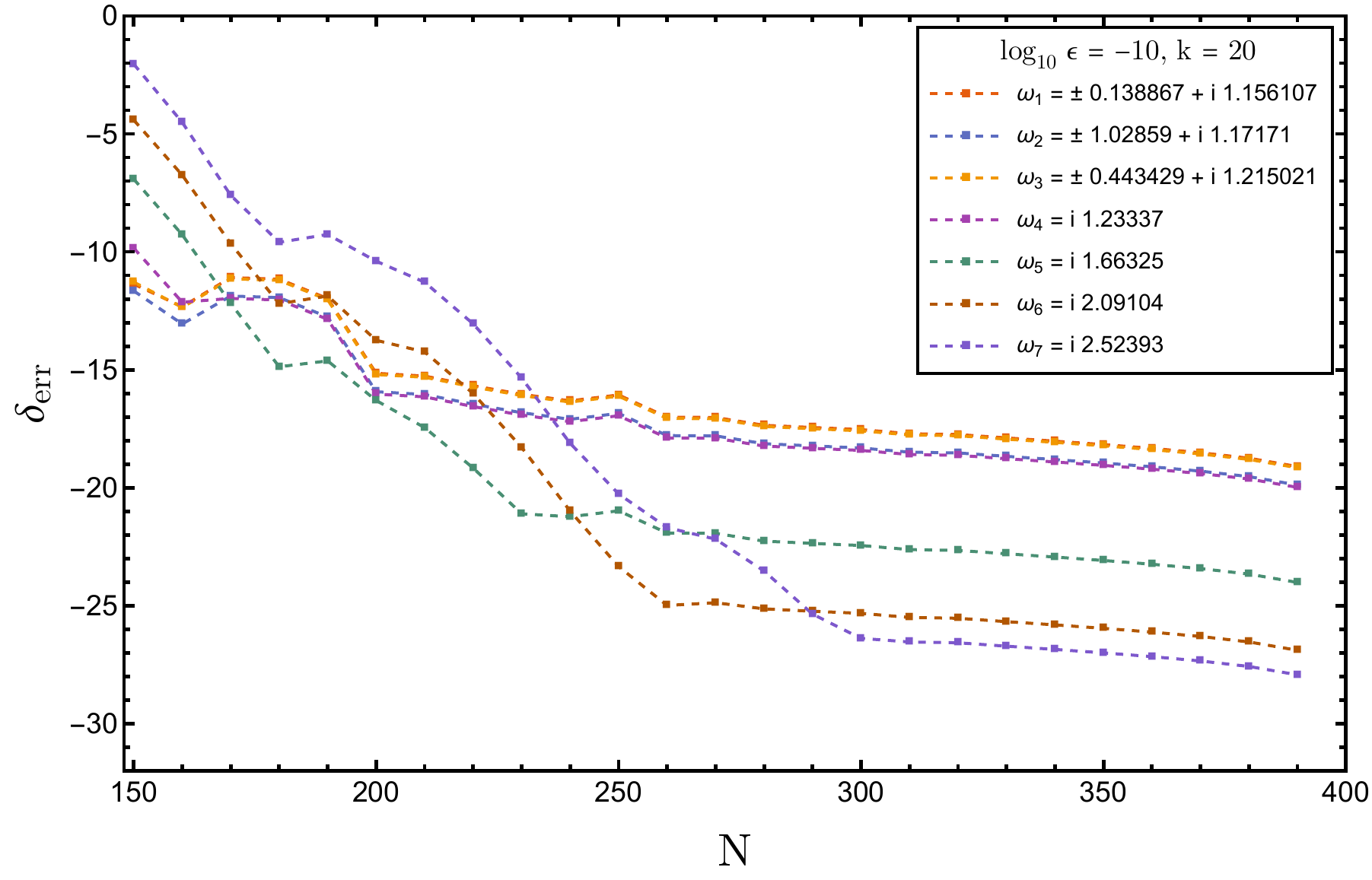}
	\endminipage\hfill
	\minipage{0.33\textwidth}
	\includegraphics[width=\linewidth]{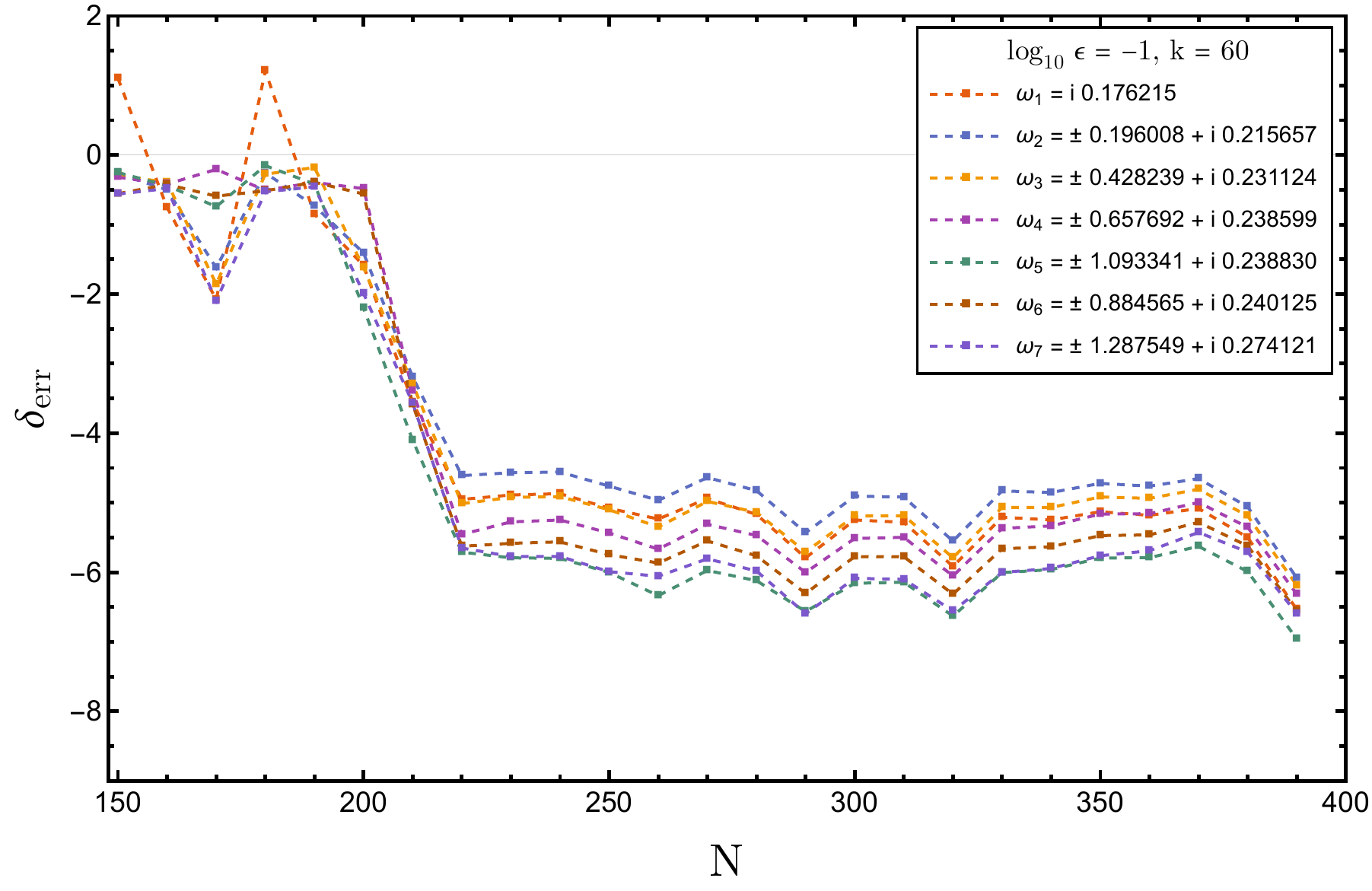}
	\endminipage
\caption{Convergence of the unperturbed (left panel), and perturbed QNM spectra for deterministic perturbations of norm $||\delta V_d||= 10^{-10}$ with $k= 20 $ (middle panel) and  $||\delta V_d||= 10^{-1}$ with $k= 60 $ (right panel) for $\ell=1$ \emph{scalar} modes of a Schwarzschild de Sitter black hole with $M = 1$ and $\Lambda = 0.01$ has been presented. Here $\delta_\mathrm{err}$ corresponds to the relative error between the complex frequencies of the modes at particular value of CGL grid size $N$ with respect to a reference value corresponding to $N=400$.}\label{fig:SdS_scalar_convergence}
\end{figure}	

\begin{figure}[tbh!]
	\centering
	\minipage{0.33\textwidth}
	\includegraphics[width=\linewidth]{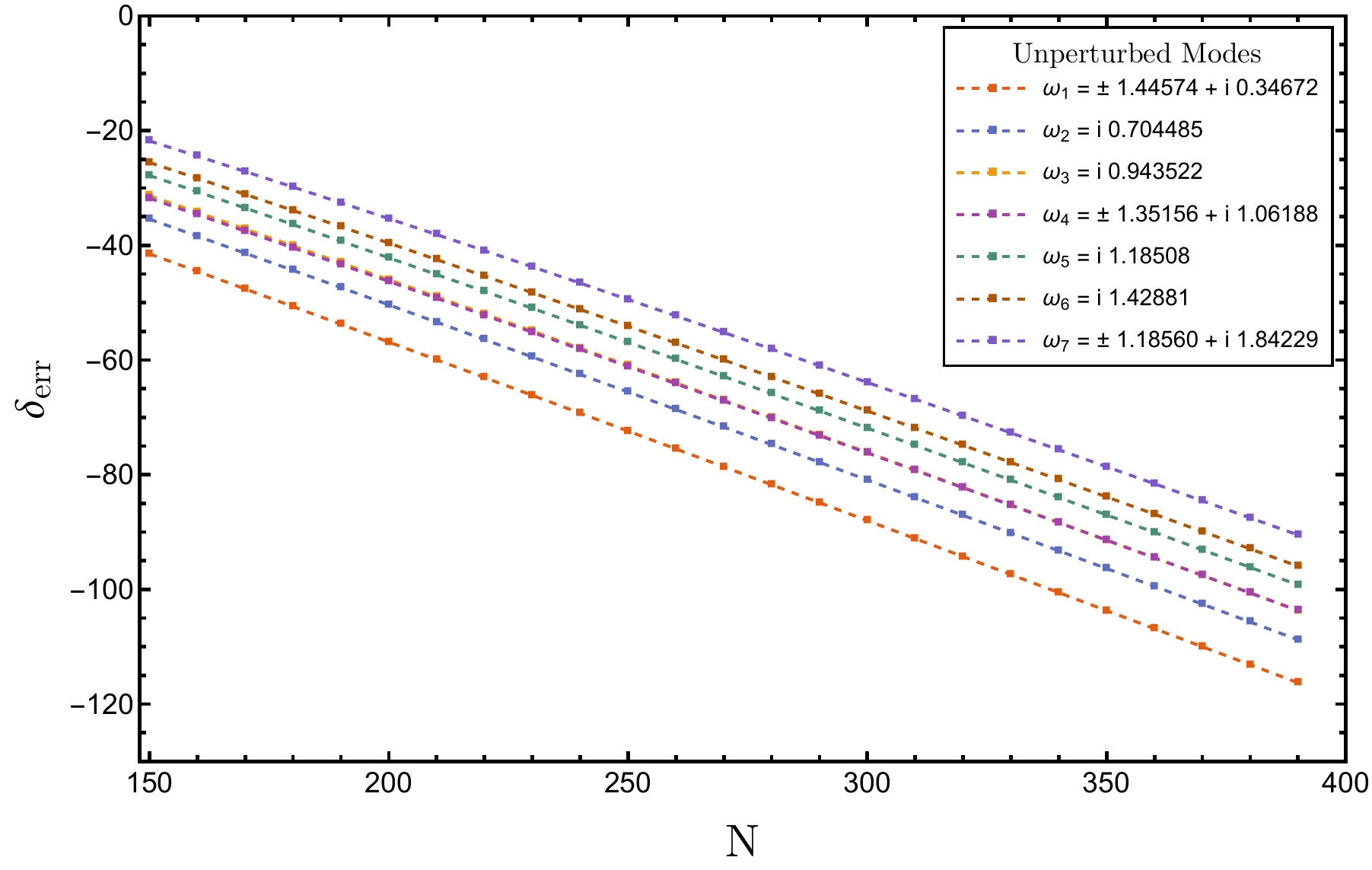}
	\endminipage\hfill
	\minipage{0.33\textwidth}
	\includegraphics[width=\linewidth]{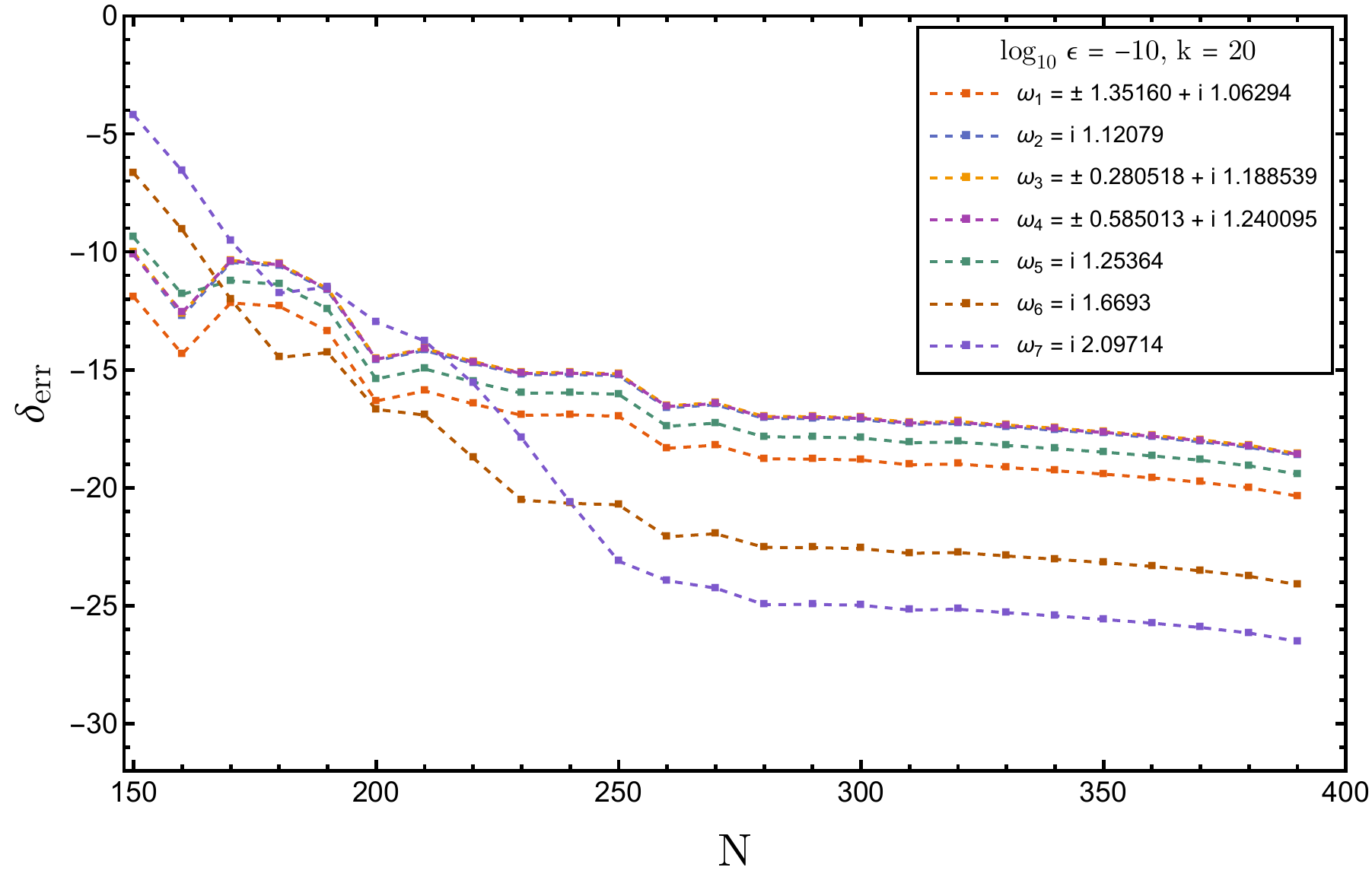}
	\endminipage\hfill
	\minipage{0.33\textwidth}
	\includegraphics[width=\linewidth]{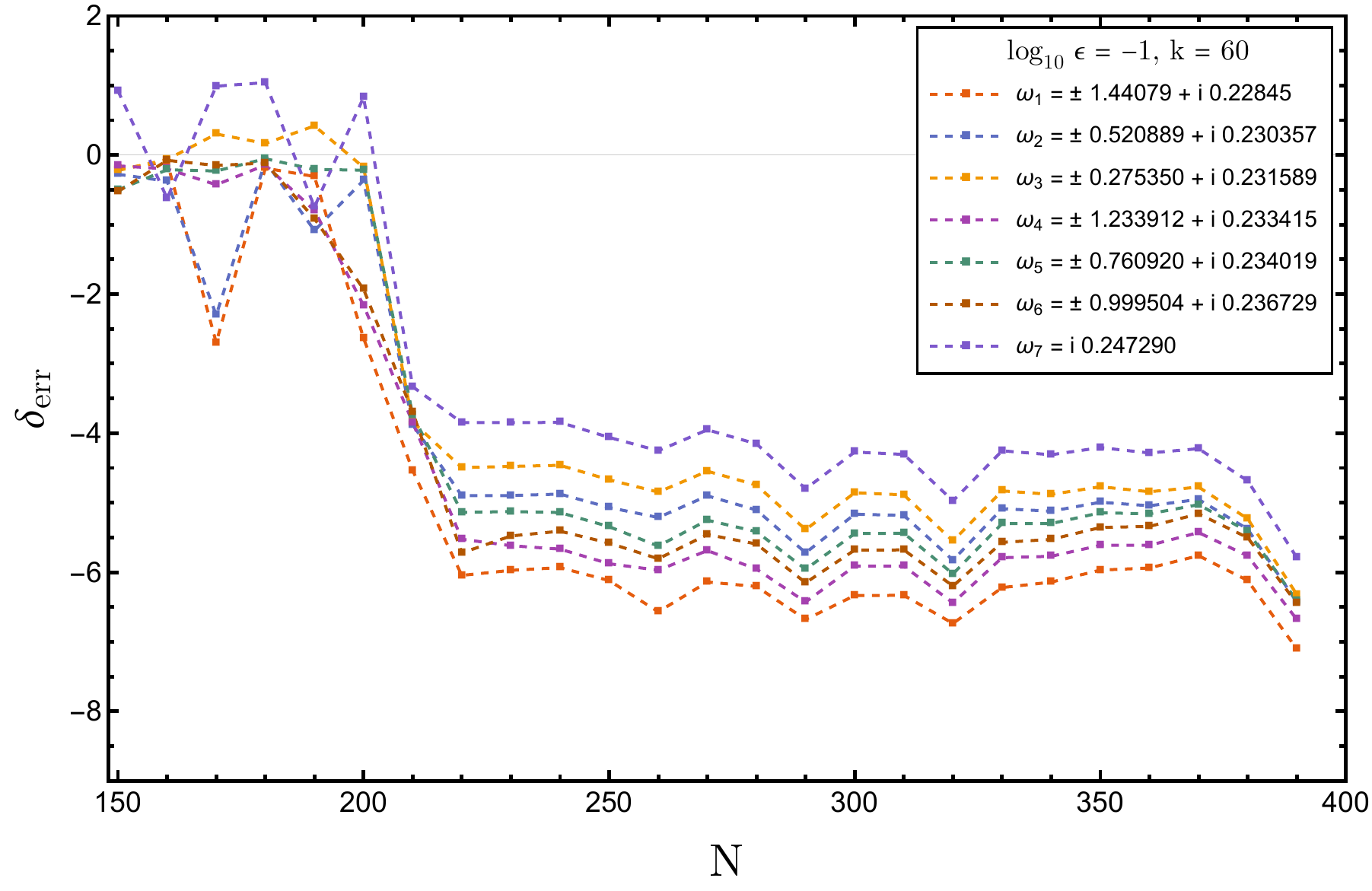}
	\endminipage
\caption{Convergence of the unperturbed (left panel), and perturbed QNM spectra for deterministic perturbations of norm $||\delta V_d||= 10^{-10}$ with $k= 20 $ (middle panel) and  $||\delta V_d||= 10^{-1}$ with $k= 60 $ (right panel) have been presented for the $\ell=2$ mode of \emph{gravitational} perturbation of a Schwarzschild de Sitter black hole with $M = 1$ and $\Lambda = 0.01$. The convergence is assured as the relative error $\delta_\mathrm{err}$ goes over to smaller vales as the grid size $N$ becomes larger. The characteristic length scale has been set to $\lambda = 2 r_+$.}
\label{fig:SdSgrav_convergence}
\end{figure}

In the present work, we have validated the convergence of our results in the following ways: At the very outset, we set the internal precision to a very high value ($\sim 10$ times the machine precision). We then compute the spectrum of the unperturbed potential for $N= 150, 160, \cdots, 390, 400$ where $N$ is the size of the CGL grid (c.f. \ref{cgl_grid}). Taking the eigenvalues obtained for $N = 400$ as our reference, we then compute the relative error between the corresponding eigenvalues for all the other values of $N$, viz.,
\begin{equation}
\mathscr{E}^N_n =\Bigr|1-\dfrac{\omega^N_n}{\omega^{N=400}_n}\Bigr|~,
\end{equation}
where $N= 150, 160, \cdots, 380, 390$ are the sizes of the CGL grid and $n =1, 2, 3, \cdots $ are the eigenfrequencies. While studying the eigenvalues $\omega_n$ of the unperturbed potential, that is, the QNM frequencies, we have restricted ourselves to the range $0\leq \mathrm{Im}(\omega_n)\leq 4$ since we are able to recover a sufficient number of overtones in that range with our choice of characteristic length scale $\lambda = 2 r_+$. We have then plotted the logarithm of the relative errors, $\delta_\mathrm{err}$ for each $n$ against $N$ in the first panel of \ref{fig:SdS_scalar_convergence} for $\ell = 1$ mode of scalar perturbation, and \ref{fig:SdSgrav_convergence} for $\ell = 2$ (axial) gravitational QNMs for a  Schwarzschild de Sitter black hole with $M = 1$ and $\Lambda = 0.01$. We observe a nice exponential convergence, where the relative error for the unperturbed QNM frequencies drops by at least $\sim 80$ orders of magnitude as one increases the grid size from $N=150$ to $N =400$. After taking into account our choice of scaling and sign convention for the Fourier modes, we have also confirmed that our results match those that have been reported previously \cite{Zhidenko:2003wq, Cho:2009cj, Jansen:2017oag}.

We have performed the same convergence test for various deterministic oscillatory perturbations to the black hole potential. In the second and third panel of \ref{fig:SdS_scalar_convergence}  we have plotted $\delta_\mathrm{err}$ for the pseduoQNMs resulting from a deterministic perturbation of  norm $||\delta V_d||= 10^{-10}$ with $k= 20 $ and  $||\delta V_d||= 10^{-1}$ with $k= 60 $, respectively, for $\ell = 1$ {scalar} modes. The same for $\ell=2$ (axial) gravitational modes are shown in the second and third panels of \ref{fig:SdSgrav_convergence}. In all of these cases, we have focused our attention on the first few modes lying on the respective Nollert-Price branches near the imaginary axis and a couple of purely imaginary modes that appear just above the respective Nollert-Price branches. We note that on changing the grid size from $N=150$ to $N=400$, the drop in relative error is close to at least $10$ orders of magnitude for small low-frequency perturbations (second panel of \ref{fig:SdS_scalar_convergence} and \ref{fig:SdSgrav_convergence}), and close to at least $6$ orders of magnitude for large high-frequency perturbations (third panel of \ref{fig:SdS_scalar_convergence} and \ref{fig:SdSgrav_convergence}). The convergence is poorer when we compare it to the unperturbed case, and a possible reason for this behavior might lie in the fact that the eigenfunctions of the perturbed operators are less smooth than those of the unperturbed QNMs (recall that the convergence of Chebyshev's spectral method is exponential only if the functions are smooth \cite{trefethenMATLAB10.5555/357801, Dias:2015nua}). Lastly, the methodology that we have described above is a computationally expensive affair. Moreover, the convergence of the modes lying far away from the imaginary axis but on the Nollert-Price is rather poor. So, after establishing the convergence of a few perturbed QNMs arising from deterministic perturbations to the BH potential, we have performed the following test for the entire spectrum of all deterministic perturbations reported here: we have chosen two values of $N$, viz., $N=230$ and $N=300$, and calculated the spectra of the perturbed potential for these two values of $N$. We have then compared all the corresponding modes for two values of $N$ lying in the region of interest ($-3\leq \mathrm{Re}(\omega_n)\leq 3, 0 \leq \mathrm{Im}(\omega_n)\leq 4$) and reported only those modes whose relative difference is less than $10^{-2}$ in all the plots presented in this work.

\subsection{Deterministic perturbation to the scattering potential of a Reissner Nordstr\"{o}m de Sitter black hole} 

We conclude with some exploratory analysis of the instabilities in the scalar QNM spectrum of a Reissner-Nordstr\"{o}m de Sitter black hole. In this case, there are three horizons --- (a) the cosmological horizon, where a purely outgoing boundary condition has been imposed, (b) the outer event horizon, where a purely ingoing boundary condition has been imposed, and finally (c) the Cauchy horizon. In the asymptotically flat case, there are gauge degrees of freedom to fix certain properties of the Cauchy horizon, which is not available in the case of asymptotically de Sitter spacetime. Also in the case of Reissner-Nordstr\"{o}m de Sitter black hole, there are two possible limits --- (a) the Nariai limit, where the cosmological and the outer event horizon come closer to each other, and (b) the near-extremal limit, in which the outer event horizon and the Cauchy horizon comes on top of each other. As we will observe the behavior of the perturbed QNMs in the Nariai limit remains identical to that of the Schwarzschild de Sitter black hole, but the near-extremal limit of Reissner-Nordstr\"{o}m de Sitter black hole depicts unique features. Like before, we consider deterministic perturbations, with $\delta V_{d}=\cos(2\pi k\sigma)$, and consider various possible choices of its norm $||\delta V_{d}||$ and the wave vector $k$ which corresponds to inverse length scale associated with the perturbation. 

\begin{figure}[tbh!]
	\centering
	\minipage{0.33\textwidth}
	\includegraphics[width=\linewidth]{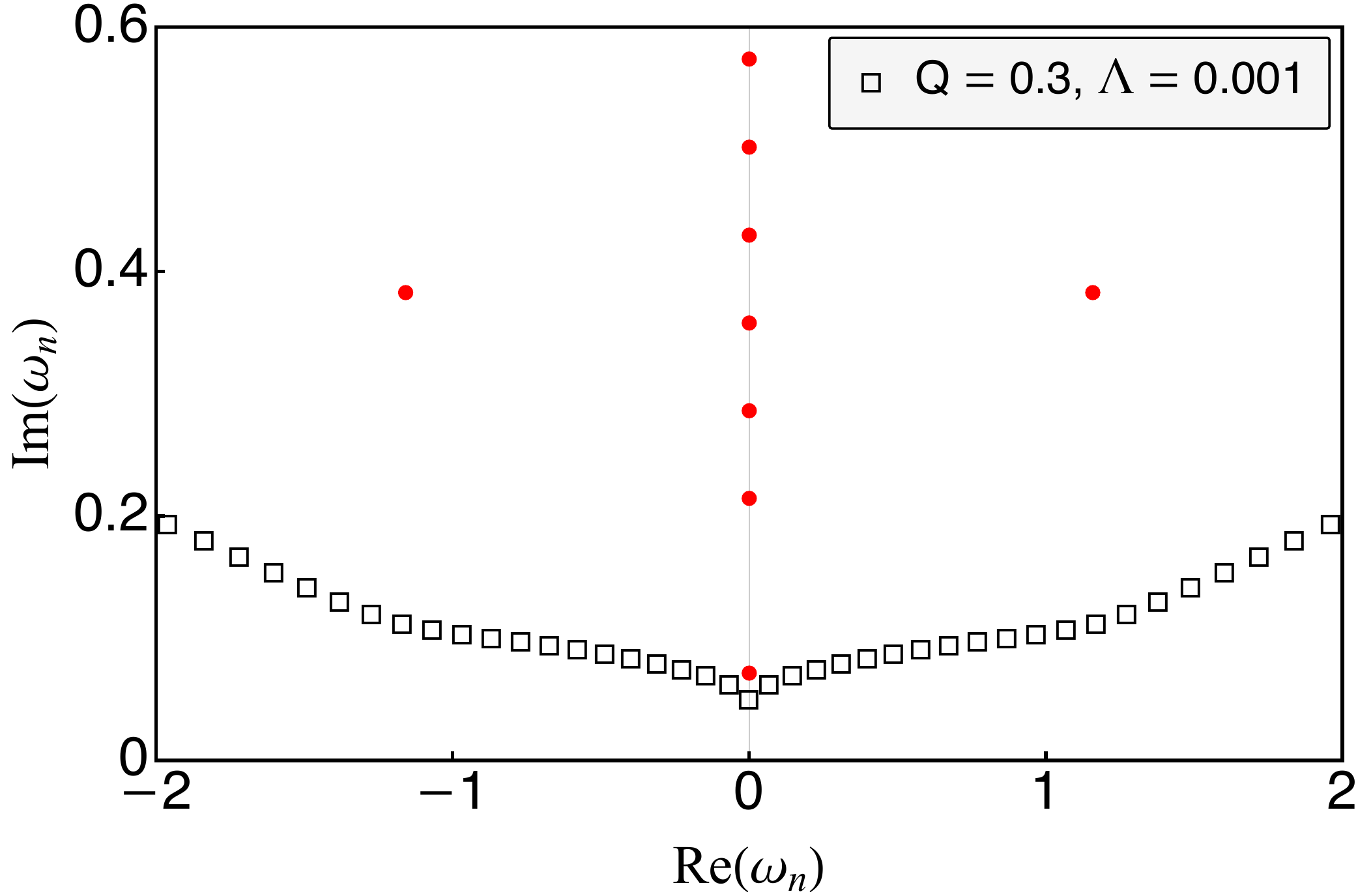}
	\endminipage\hfill
	\minipage{0.33\textwidth}
	\includegraphics[width=\linewidth]{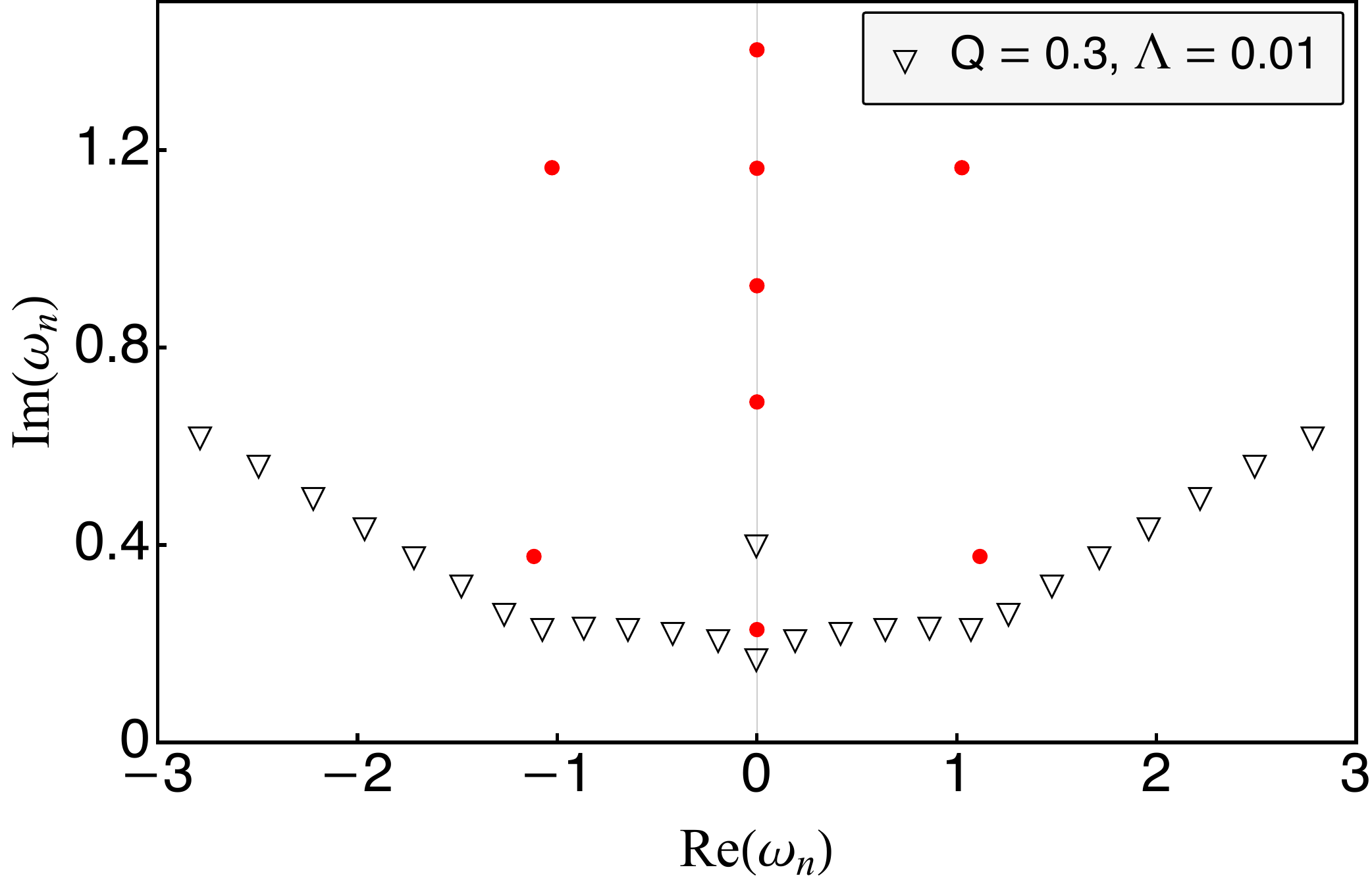}
	\endminipage\hfill
	\minipage{0.33\textwidth}
	\includegraphics[width=\linewidth]{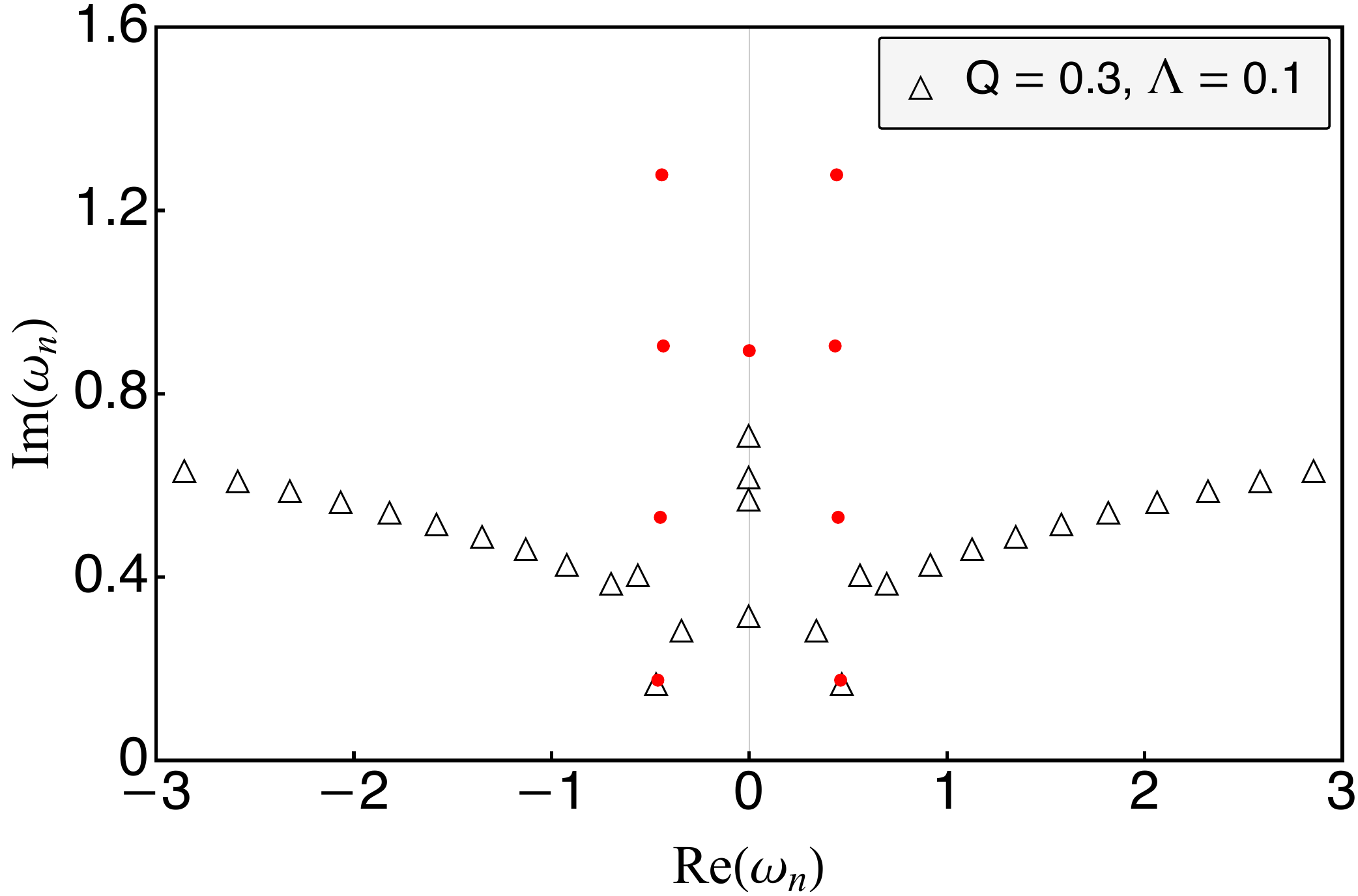}
	\endminipage\hfill
 \minipage{0.33\textwidth}
	\includegraphics[width=\linewidth]{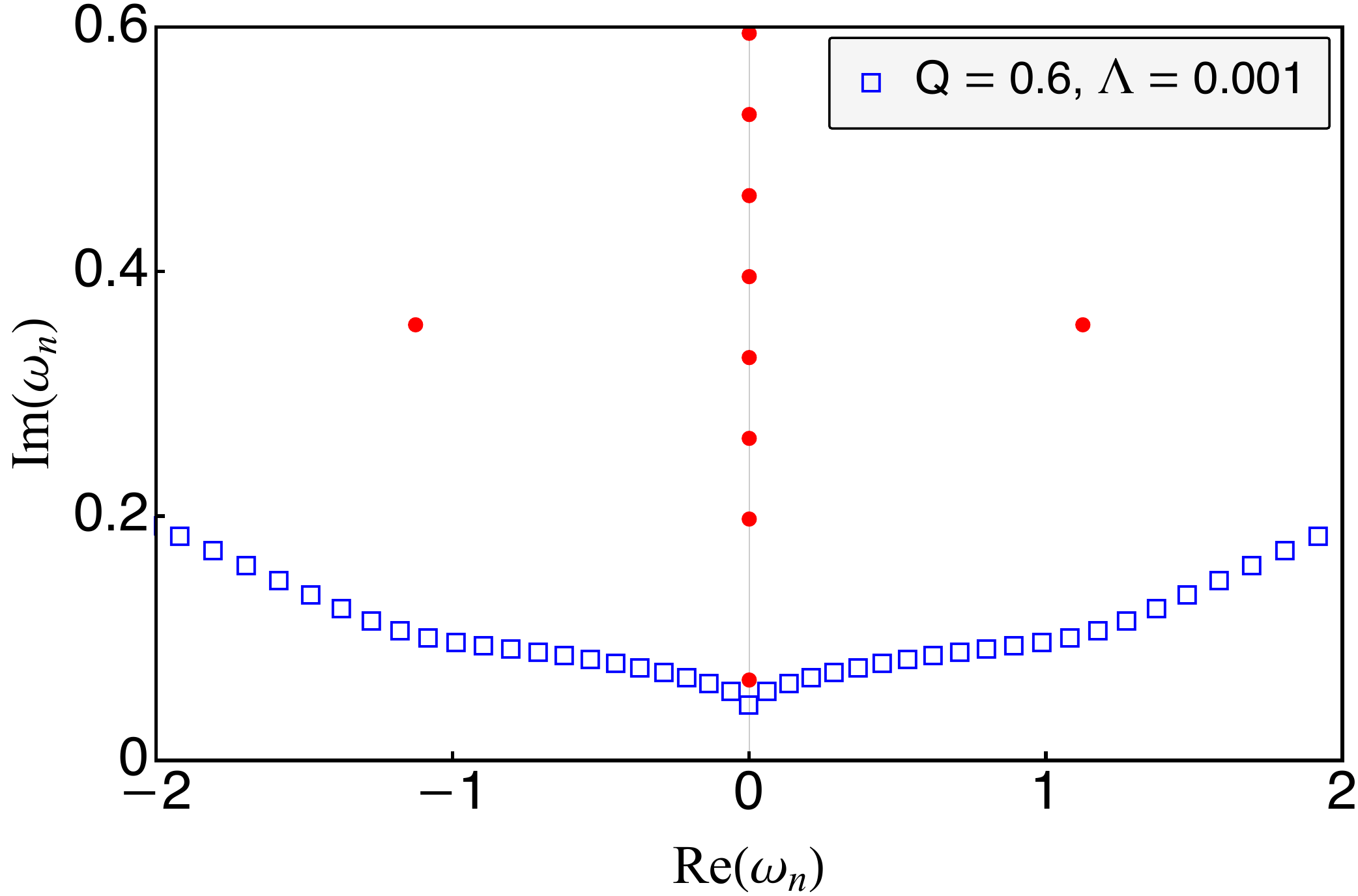}
	\endminipage\hfill
	\minipage{0.33\textwidth}
	\includegraphics[width=\linewidth]{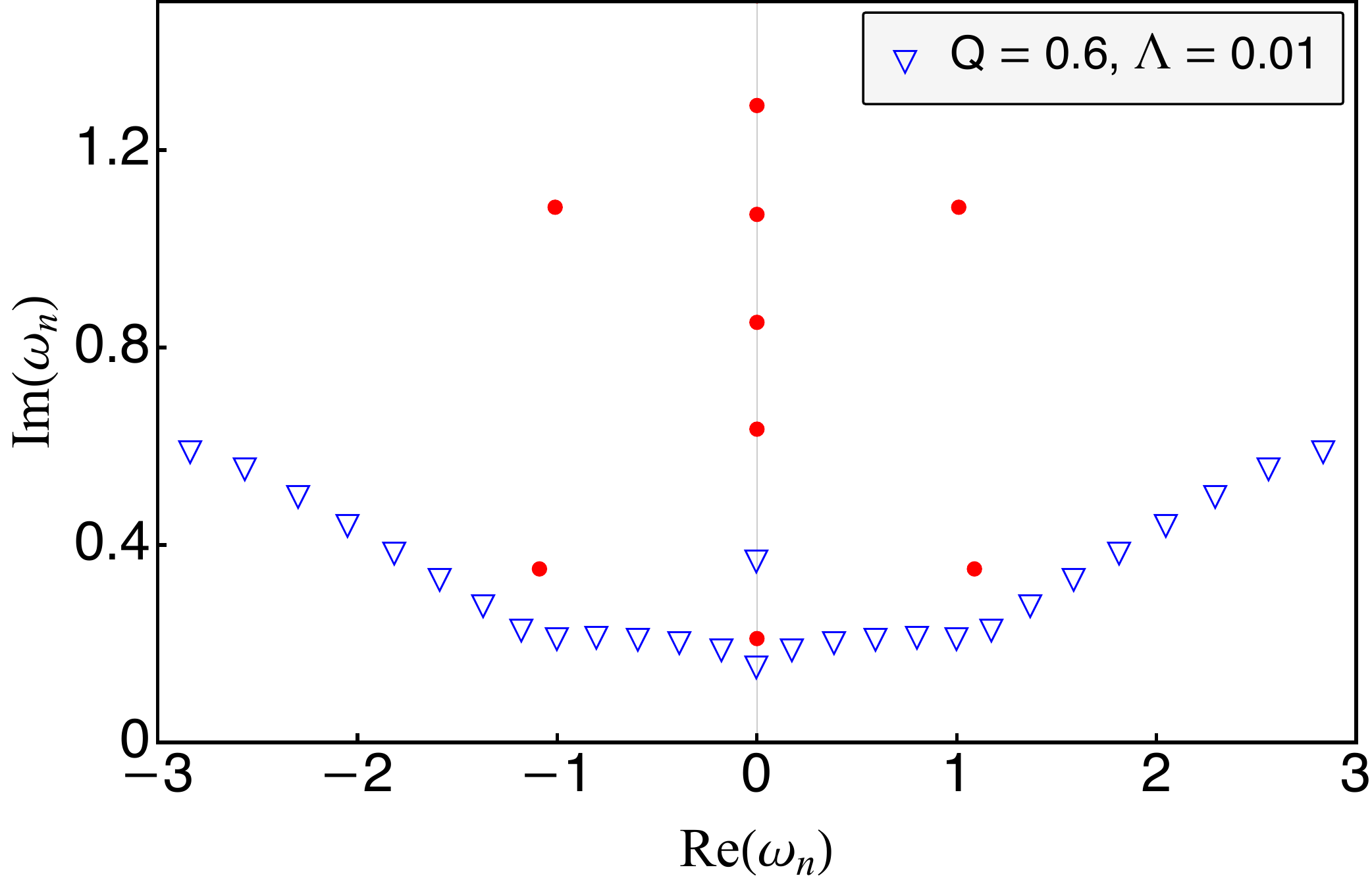}
	\endminipage\hfill
	\minipage{0.33\textwidth}
	\includegraphics[width=\linewidth]{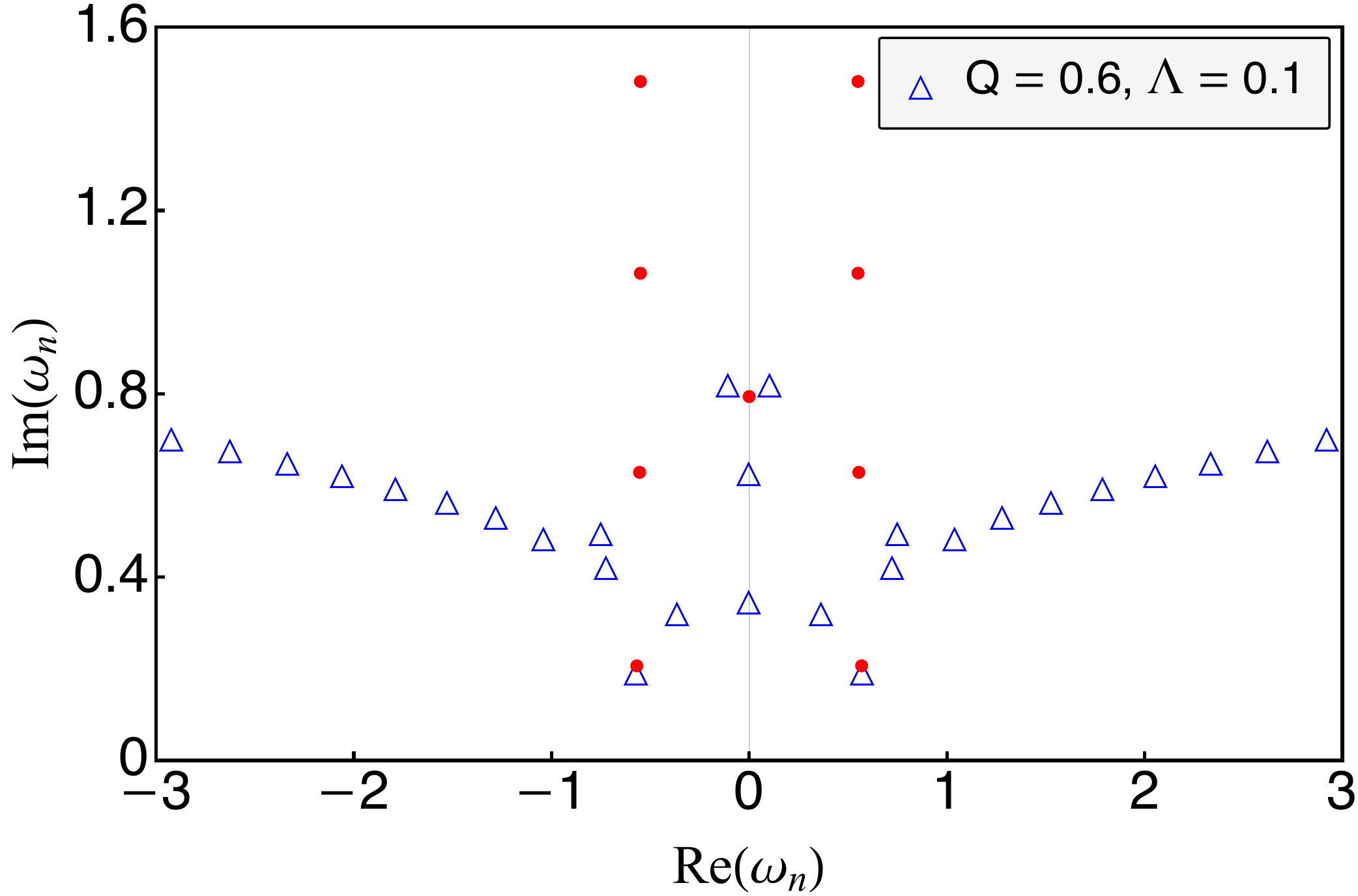}
	\endminipage\hfill
 \minipage{0.33\textwidth}
	\includegraphics[width=\linewidth]{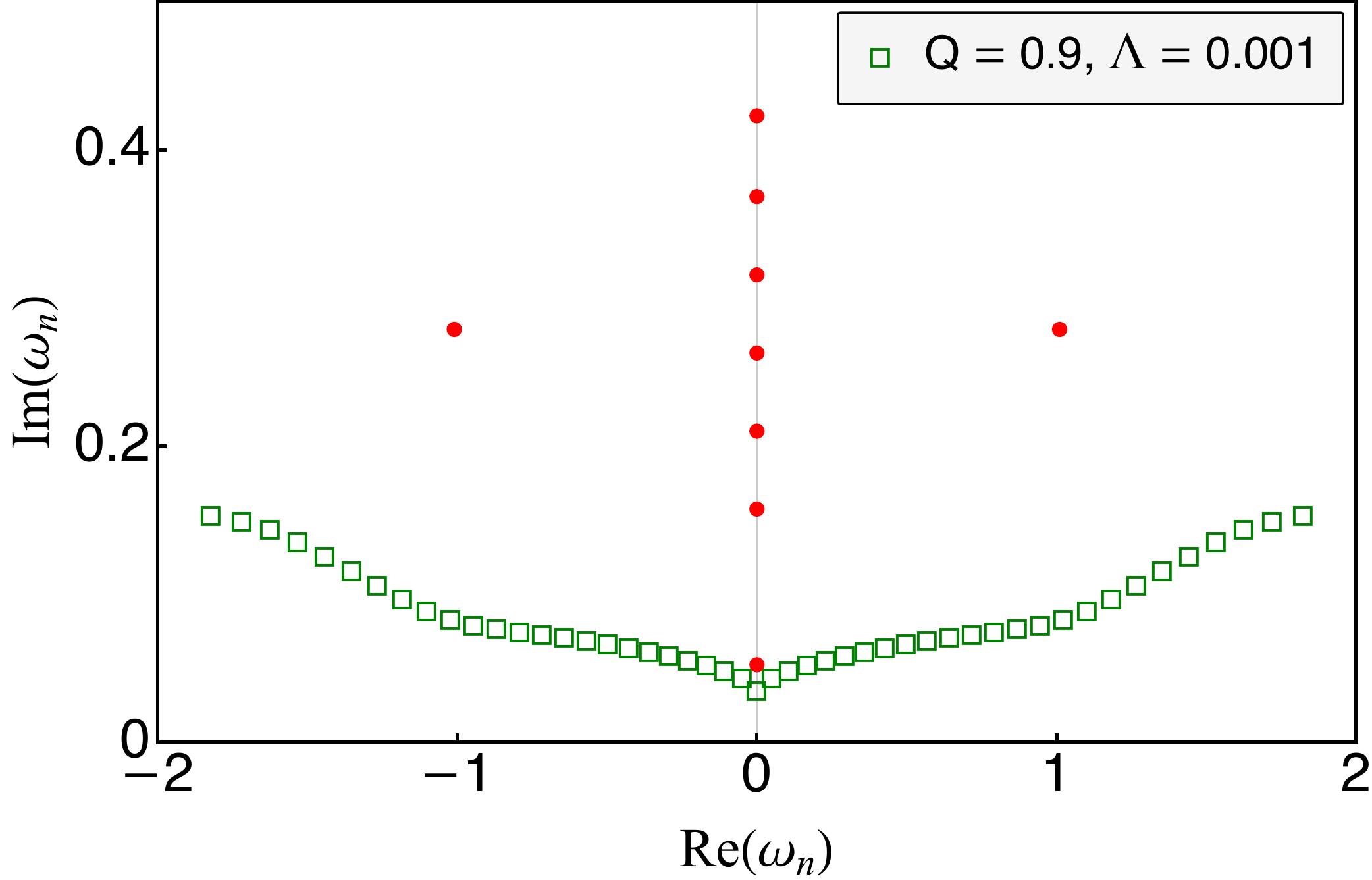}
	\endminipage\hfill
	\minipage{0.33\textwidth}
	\includegraphics[width=\linewidth]{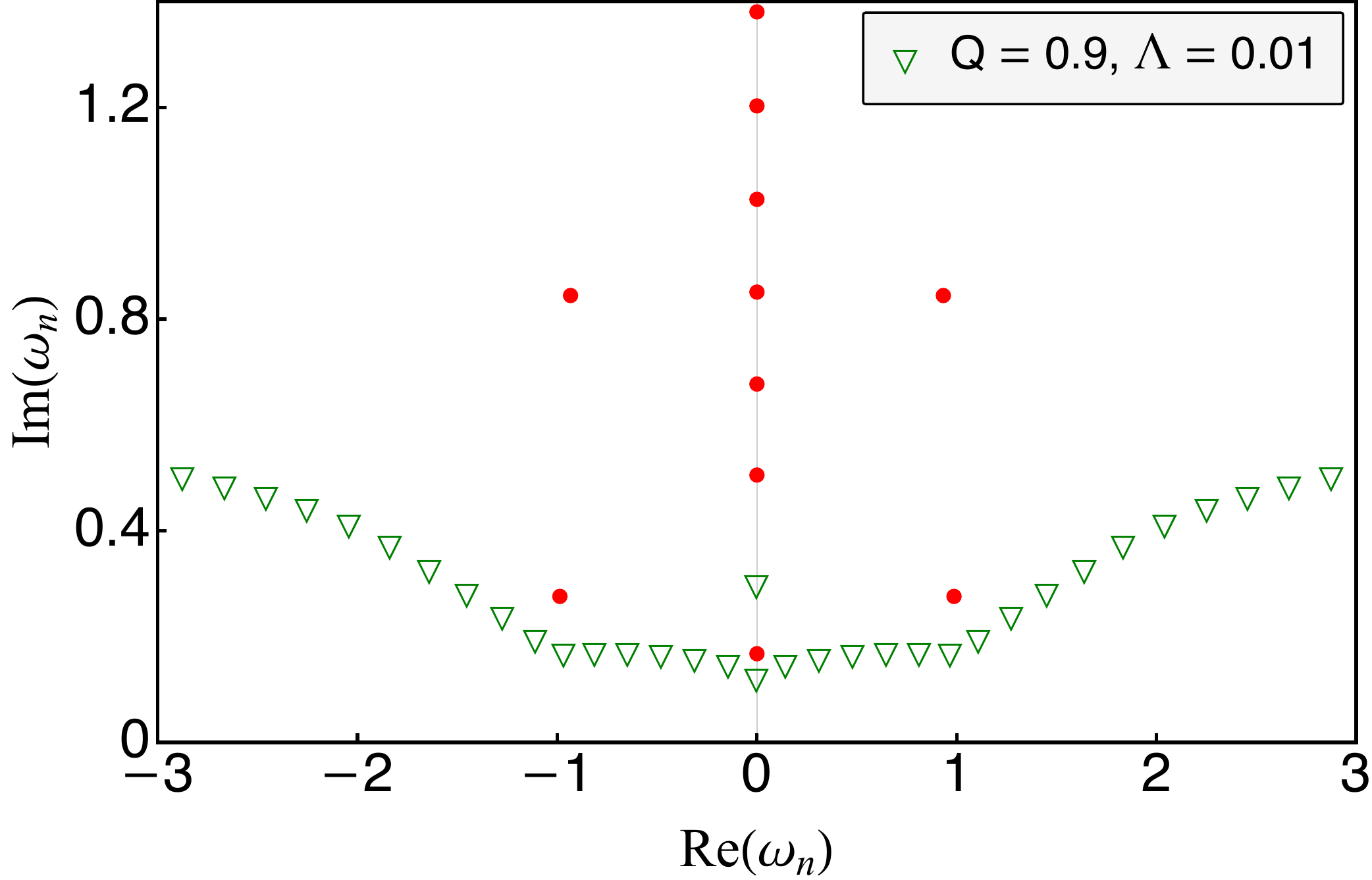}
	\endminipage\hfill
	\minipage{0.33\textwidth}
	\includegraphics[width=\linewidth]{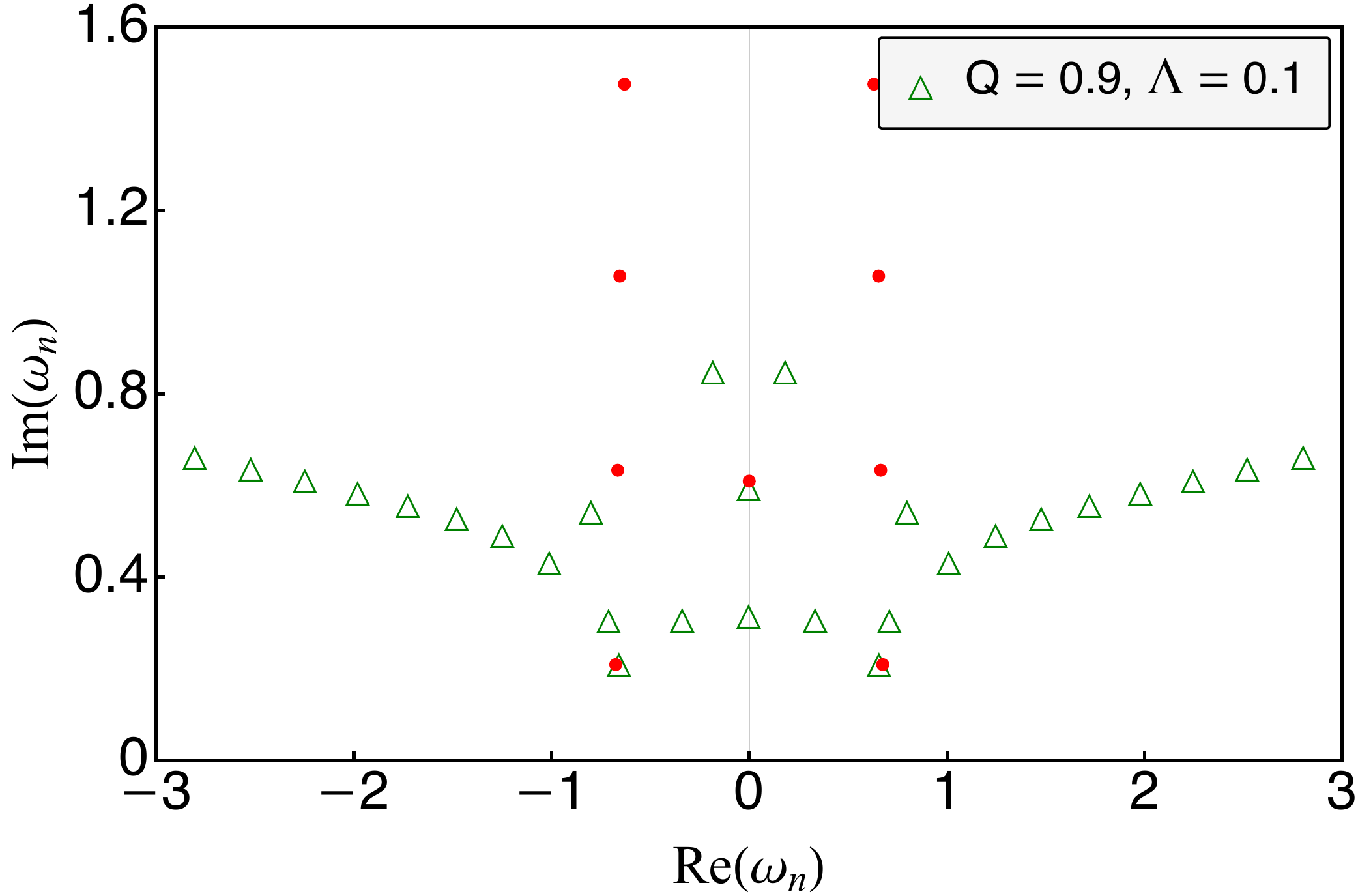}
	\endminipage
	
	\caption{We have presented the QNM spectra associated with deterministic perturbations to the scattering potential of norm $||\delta V_d||={10^{-1}}$ and $k=60$ for $\ell=1$ \emph{scalar} modes of a \RN\-de Sitter black hole with $M = 1$ and for various different choices of the cosmological constant $\Lambda$ and electric charge $Q$. The unperturbed QNM values (indicated in red) have also been superimposed over the perturbed ones. The characteristic length scale \eqref{rsigma} has been set to $\lambda = 2 r_+$.}\label{fig:RNdS_DET_PERT_SCALAR}
\end{figure}	

Taking a cue from the analysis done in the context of Schwarzschild de Sitter black hole, here we present the perturbed QNM frequencies due to a deterministic perturbation of a fixed norm and also with a definite $k$, for various choices of the charge, and the cosmological constant. The result of such an analysis has been presented in \ref{fig:RNdS_DET_PERT_SCALAR} for the $\ell=1$ modes of scalar perturbation. Since our main interest is in assessing the stability of the fundamental QNM, we consider the largest reasonable perturbation, i.e., we take the norm of the deterministic perturbation to be $||\delta V_{d}||=10^{-1}$ and the wave vector as $k=60$. As expected, for smaller values of the cosmological constant and a smaller value of the electric charge $Q$, the fundamental QNM indeed migrates to smaller imaginary values of the frequency. However, the amount of migration is of the same order as the perturbing potential and hence is most likely not an instability. Moreover, the migration of the fundamental QNM remains even when the electric charge is increased. On the other hand, the fundamental QNM becomes more stable when $\Lambda$ becomes large, i.e., in the Nariai limit. This behavior is qualitatively similar to what has been observed in the context of Schwarzschild de Sitter black holes. It is worth reiterating the surprising fact that the charge of the black hole has no bearing on the instability of the fundamental QNM frequency. This is because, for a fixed cosmological constant, we observe the same general behavior for the perturbed QNMs for diverse choices of the electric charge, viz., $Q = 0.3$ to $Q=0.9$., and it can be seen in \ref{fig:RNdS_DET_PERT_SCALAR} by moving vertically along any of the columns. While the variation with the cosmological constant for a fixed electric charge can be observed in \ref{fig:RNdS_DET_PERT_SCALAR} by moving horizontally across any of the rows. One could speculate that the large perturbation has washed out any effect of varying the charge of the black hole, but this issue needs to be investigated further. Also, implications of such deterministic perturbation of gravitational origin on the fundamental QNM of Reissner-Nordstr\"{o}m-de Sitter black hole spacetime will be discussed in future work.

\begin{figure}[tbh!]
	\centering
	\minipage{0.33\textwidth}
	\includegraphics[width=\linewidth]{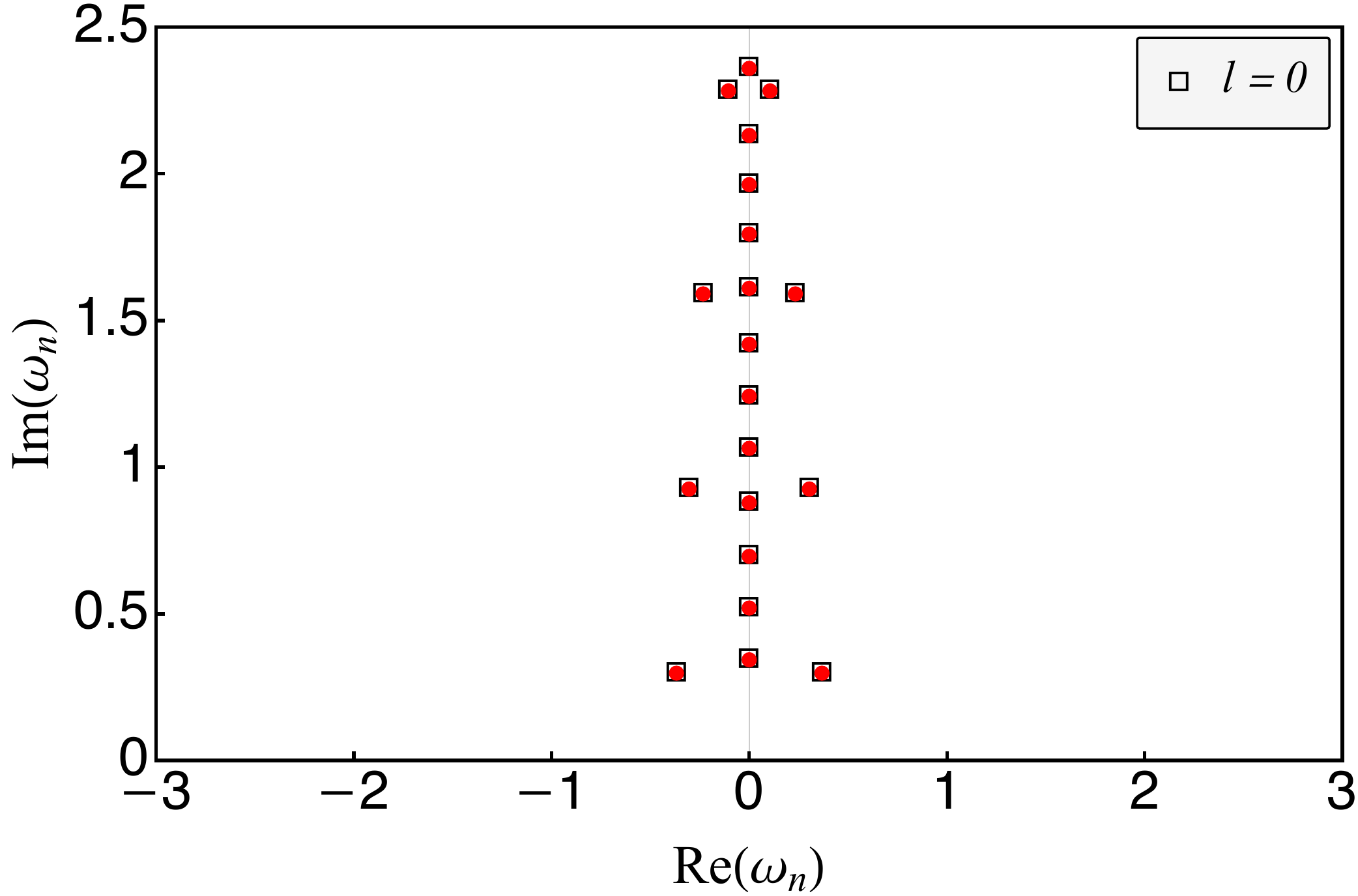}
	\endminipage\hfill
	\minipage{0.33\textwidth}
	\includegraphics[width=\linewidth]{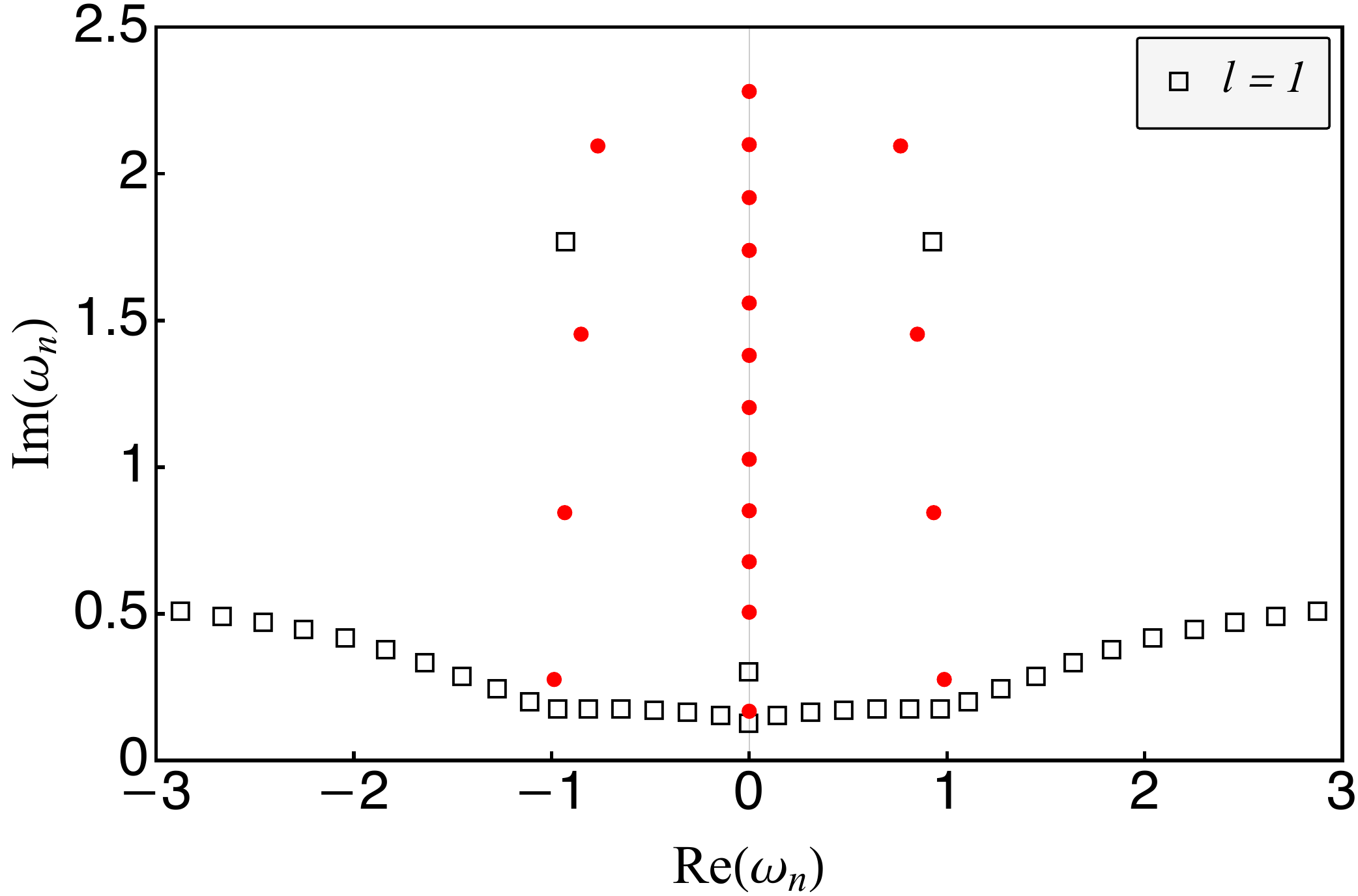}
	\endminipage\hfill
	\minipage{0.33\textwidth}
	\includegraphics[width=\linewidth]{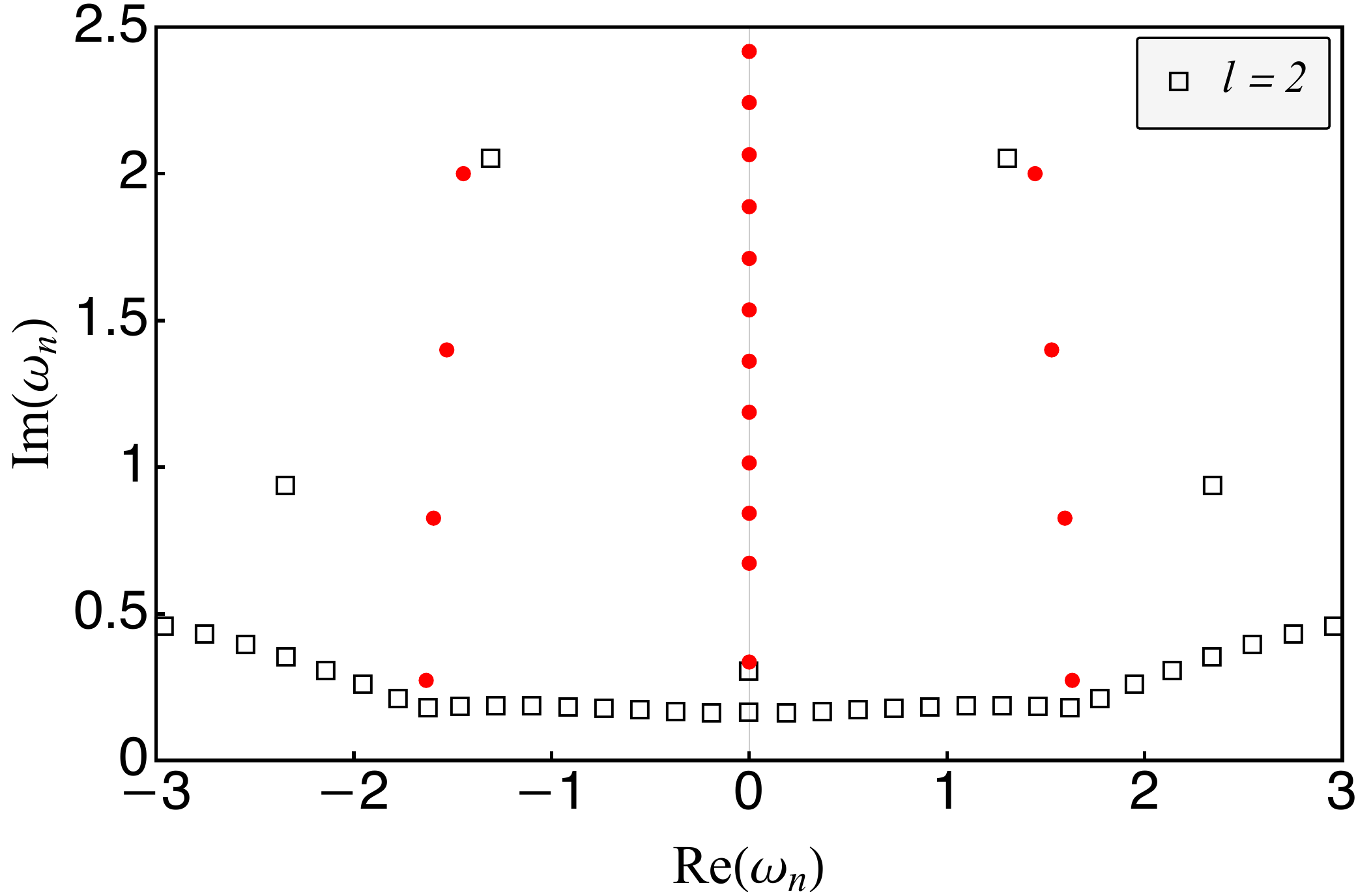}
	\endminipage
\caption{We have plotted the QNM frequencies of the perturbed scattering potential for deterministic perturbations of norm $||\delta V_d||={10^{-1}}$ with $k=60$ for various choices of the angular number associated with \emph{scalar} modes. In particular, we consider three possible choices of the angular number --- $\ell=0$ (left panel), $\ell=1$ (middle panel) and $\ell=2$ (right panel), in the background a \RN\-de Sitter black hole with $M = 1$, $Q=0.9$ and $\Lambda = 0.01$. We have also superimposed the unperturbed QNM frequencies (indicated in red) over the perturbed ones to explicitly demonstrate the drifting. The characteristic length scale has been set to be $\lambda = 2 r_+$.}
\label{fig_RNdS_lamb_Q_vary_l}
\end{figure}

Finally, we also present how the nature of the instability varies with the angular number $\ell$, again for scalar perturbation modes in \ref{fig_RNdS_lamb_Q_vary_l}. We consider fixed values for all the black hole hairs, namely mass, charge, and cosmological constant, but vary the angular number $\ell$. We have also fixed the strength of the high-frequency perturbation to be large ($||\delta V_{d}||=10^{-1}$)  Once again, we find that for $\ell=0$, the QNM spectrum is stable against large high-frequency deterministic perturbations. The stability is not just restricted to the fundamental modes, but it is rather remarkable that all the higher overtones are also stable under such a large perturbation, an aspect worth exploring further. However, for $\ell>0$, all the modes are unstable under such a large perturbation. This behavior is similar to what we have seen in the case of the Schwarzschild de Sitter black holes. Most importantly, except for the $\ell=0$ case, the perturbed QNM frequencies have lower values for their imaginary parts, even when the black hole is electrically charged, which implies that the violation of the strong cosmic censorship conjecture could be avoided. Since perturbations to the scattering potential arising from the neighboring gravitating systems are generic, possibly the imaginary part of the fundamental QNM frequency will be smaller than the unperturbed ones. This will reduce the regularity of the fundamental mode at the Cauchy horizon, possibly ensuring that Christodoulou's version of the strong cosmic censorship conjecture is respected. Hence the validity of the strong cosmic censorship conjecture may not require a quantum field to exist, rather it demands a large perturbation by the neighboring gravitating system near the event horizon. We hope to address this question in more detail in the future. 

\section{Discussion and concluding remarks}
\label{sec:disscusion_conclusion}

Probing the ringdown spectrums from the merger of binary black holes have been one of the key observational frontiers in gravitational wave astrophysics. The timescale associated with the characteristic exponential decay along with the frequency of sinusoidal oscillations in the ringdown phase is a unique testimonial regarding the nature of the central supermassive compact object. An understanding of the fundamental QNM in the ringdown spectrum can be used to test the no-hair theorems, as well as it may provide us hint for the existence of physics beyond general relativity in the vicinity of the photon region. Detecting higher overtones open up further avenues of exploration, ranging from the nature of the central compact object (e.g., whether the object is a classical black hole, a black hole with quantum corrections, or, an exotic compact object) to imposing tight constraints on the physics beyond general relativity. Therefore the importance of studying the spectrum of the QNM frequencies cannot be over-emphasized.      

On the other hand, the stability of the QNM spectrum itself is an intriguing avenue to explore. Normally one would expect the QNM spectrum to be stable owing to the spectral theorem. However, the spectral theorem applies only to self-adjoint operators. Even though the perturbations in the black hole background are governed by self-adjoint operators, the boundary conditions satisfied by these perturbations are dissipative. As a consequence, the spectral problem associated with finding the QNMs of a black hole cannot be described by a self-adjoint operator and hence the spectral theorem is no longer applicable. Therefore, the study regarding the stability of the QNM spectrum is performed by invoking the notion of the black hole's pseudospectrum, which we have elaborated on in the previous sections. These ideas have been applied to studying the stability of the QNM spectra of asymptotically flat spacetimes, and it was demonstrated that for large perturbations most of the overtones are unstable, however, the fundamental mode remains stable. Motivated by this result, we have studied the corresponding situation for asymptotically de Sitter spacetime in the present work. We summarize below the main results obtained in our analysis regarding the stability of QNMs in asymptotically de Sitter spacetimes.

\begin{itemize}

\item The construction of a hyperboloidal coordinate system for asymptotically de Sitter spacetimes, which neatly captures the boundary conditions satisfied by the perturbations near the horizon and the cosmological horizon, has been worked out, within the minimal gauge prescription \cite{PanossoMacedo:2018hab}. There is one crucial difference between the hyperboloidal coordinate system of an asymptotically de Sitter spacetime from that of an asymptotically flat spacetime --- after the minimal gauge choice has been imposed, there is no residual gauge freedom in the hyperboloidal coordinate system for asymptotically de Sitter spacetimes, and hence the transformation from the $(t,r,\theta,\phi)$ to $(\tau,\sigma,\theta,\phi)$ coordinate system does not involve any free parameter and is unique. 

\item The differential operator determining the behavior of the perturbing field between the event horizon and the cosmological horizon of an asymptotically de Sitter spacetime, in the hyperboloidal coordinate system, contains a Sturm–Liouville operator which is singular at the boundaries. However, the nature of singularity is different from that of asymptotically flat spacetimes, since in the present context, the singularities at the boundaries are removable. While, in asymptotically flat spacetimes, the null infinity is an essential singularity of the differential operator. Therefore, the structure of the differential operator governing the perturbation is different in the presence of a positive cosmological constant and hence one expects the pseudospectrum to also depart from that of the asymptotically flat black holes.  

\item One of the most striking results, in the context of asymptotically de Sitter spacetime is that, unlike the case of asymptotically flat spacetime, for large enough perturbation with $||\delta V_{d}||=10^{-1}$ and for $k=60$, the fundamental mode itself gets dislodged for both gravitational and scalar perturbations provided that the cosmological constant has a small value. Though, for scalar perturbation, the migration of the fundamental QNM is of the same order as that of the perturbing potential, and hence is probably stable, but the migration for gravitation perturbation is at least an order of magnitude larger, and thus unstable. This result is seemingly true for both Schwarzschild de Sitter and \RN de Sitter spacetimes. Thus, the fundamental QNM associated with the gravitational perturbation is unstable in asymptotic de Sitter spacetimes with a small cosmological constant. Moreover, even for random perturbations with magnitude $||\delta V_{r}||=10^{-1}$, the fundamental QNM is unstable in the presence of a small positive cosmological constant.

\item We must emphasize another intriguing result derived here --- a singular behavior when the cosmological constant goes to zero. For the case depicting $\Lambda=0$, which corresponds to the asymptotically flat Schwarzschild spacetime, \cite{Jaramillo:2020tuu} has demonstrated the stability of the fundamental mode, whereas in the asymptotically de Sitter spacetime, we find an instability for the gravitational quasi-normal modes, which becomes worse for smaller cosmological constant, i.e., as $\Lambda\to 0$. This singular behaviour of the $\Lambda\to0$ limit arises due to the emergence of the additional de Sitter modes in the asymptotically de Sitter spacetimes, which are absent in the asymptotically flat case. The above singular limit can also be understood in the following manner: it is not the same, on the one hand, to take first the limit $\Lambda \to 0$ for the spacetime and then consider the QNM spectrum (leading to stability of the fundamental QNM \cite{Jaramillo:2020tuu}) and, on the other hand, to calculate the QNM spectrum for small, but non-zero $\Lambda$ and \emph{then} take the limit $\Lambda \to 0$ (leading to an enhanced instability of the fundamental QNM). 

\item For asymptotically de Sitter spacetime, it turns out that deterministic and large perturbations with $k=1$ can affect the QNM spectrum significantly, except for the fundamental mode. While for asymptotically flat spacetime, deterministic perturbation with $k=1$ has very little to no influence on the QNM spectrum --- another crucial difference between the pseudo spectrum of asymptotically flat and asymptotically de Sitter spacetimes. 

\item The behavior of the $\ell=0$ mode for scalar perturbation is another intriguing result that has been churned out by our analysis. It turns out that, for both Schwarzschild de Sitter and \RN de Sitter spacetimes, there is very little effect on the unperturbed QNM spectrum even when the scattering potential is perturbed by $\mathcal{O}(10^{-1})$ terms. Thus it seems that in the presence of a positive cosmological constant, the $\ell=0$ mode of the scalar perturbation becomes highly stable against external perturbations.  

\item Finally in the presence of a small postive cosmological constant, it turns out that for $\ell>0$, large enough perturbations, whether they are deterministic or, random, can dislodge the fundamental QNM, and the imaginary part of the perturbed QNM becomes smaller. Thus the decay timescale of the perturbed fundamental QNM is larger. This holds for \RN de Sitter spacetime as well and possibly provides a way to reinforce the strong cosmic censorship conjecture purely classically. This is because, the regularity at the Cauchy horizon is determined by the ratio $(\textrm{Im}~\omega_{\rm min}/\kappa_{\rm Cauchy})$, where $\omega_{\rm min}$ is the fundamental QNM frequency and $\kappa_{\rm Cauchy}$ is the surface gravity at the Cauchy horizon. Since under external perturbation, $\textrm{Im}~\omega_{\rm min}$ becomes smaller, the above ratio decreases, thereby increasing the irregularity at the Cauchy horizon. This in turn can rescue the strong cosmic censorship conjecture, classically, for a \RN de Sitter black hole.

\end{itemize}

Having summarized the results derived in this paper, let us briefly touch upon the future directions to explore. First of all, our results provide a preliminary indication that the violation of the strong cosmic censorship conjecture for \RN de Sitter black hole can be avoided, classically, for gravitational perturbation. However, one needs to present a more thorough analysis, e.g., the following questions need to be answered --- what is the change in the imaginary part of the fundamental QNM frequency, will the change be enough to restore the strong cosmic censorship conjecture, etc.? It will be worth exploring whether the destabilization of the fundamental mode holds for exotic compact objects as well in an asymptotically de Sitter spacetime and implications for their stability. The reason behind the surprising stable nature of the $\ell=0$ QNM modes for scalar perturbation of asymptotically de Sitter spacetime needs further investigation. Moreover, it will be interesting to explore how generic the destabilization of the fundamental QNM is, with respect to the nature of the deterministic perturbation potential. We hope to return to these issues in the future.


\section*{Acknowledgement}
The authors thank the referee for various useful and insightful comments and suggestions. The authors also acknowledge useful discussions with An{\i}l Zengino\u{g}lu. 
S.S. thanks Srijit Bhattacharjee for several fruitful and illuminating discussions on various topics related to the pseudospectra and quasi-normal modes of black holes during the course of this work. S.S. would also like to thank IACS, Kolkata for the hospitality during his academic visit, that led to the initiation of this project. S.S. also expresses his gratitude to Divyesh N. Solanki for his assistance and coordination in helping S.S. to set up and operate some of the computational resources used in this work. S.S. acknowledges funding from SERB, DST, Government of India through a Junior Research Fellowship (JRF) under research grant no: CRG/2020/004347. The computational results reported in this work were performed on the Central Computing Facility (CCF) of IIIT, Allahabad. The research of M.R. is funded by the National Post-Doctoral Fellowship (N-PDF) from SERB, DST, Government of India (Reg. No. PDF/2021/001234). Research of S.C. is funded by the INSPIRE Faculty fellowship from DST, Government of India (Reg. No. DST/INSPIRE/04/2018/000893) and by the Start-Up Research Grant from SERB, DST, Government of India (Reg. No. SRG/2020/000409). 

\providecommand{\href}[2]{#2}\begingroup\raggedright\endgroup

\end{document}